\documentclass[a4paper,11pt]{article}

\usepackage{jheppub_mod} 
\usepackage[T1]{fontenc} 

\usepackage{hyperref}
\usepackage{graphicx}
\usepackage{amsmath,amssymb,slashed}
\usepackage{booktabs,tabulary}
\usepackage{dsfont}
\usepackage{color}
\usepackage{multirow}
\usepackage{subcaption}

\usepackage{braket} 
\usepackage{bbold} 
\usepackage{csquotes} 
\usepackage{tikz} 
\usepackage[compat=1.0.0]{tikz-feynman} 

\tikzfeynmanset{momentum/arrow distance=2mm, momentum/arrow shorten=0.25} 

\newcommand{\LN}[1]{\textsc{#1}} 
\newcommand{\Lat}[1]{\textit{#1}} 

\newcommand{\beq}{\begin{equation}}
\newcommand{\eeq}{\end{equation}}

\newcommand{\Ax}{A} 

\newcommand{\Maxial}{m_{\Ax}}
\newcommand{\Mf}{m_{f_1}}
\newcommand{\Mfprime}{m_{f_1'}}
\newcommand{\Ma}{m_{a_1}}

\newcommand{\Mrho}{M_{\rho}}
\newcommand{\Mrhoprime}{M_{\rho'}}
\newcommand{\Gammarho}{\Gamma_{\rho}}
\newcommand{\Gammarhoprime}{\Gamma_{\rho'}}

\newcommand{\Mpi}{M_{\pi}}
\newcommand{\Momega}{M_{\omega}}

\newcommand{\Mphi}{M_{\phi}}

\newcommand{\Momegaprime}{M_{\omega'}}
\newcommand{\Mphiprime}{M_{\phi'}}

\newcommand{\M}{\mathcal{M}}

\newcommand{\F}{\mathcal{F}}

\newcommand{\Q}{\mathcal{Q}}

\newcommand{\Lagrangian}{\mathcal{L}}

\newcommand{\GeV}{\,\text{GeV}}
\newcommand{\MeV}{\,\text{MeV}}
\newcommand{\keV}{\,\text{keV}}

\newcommand{\perc}{\%}

\newcommand{\iu}{i}

\newcommand{\tior}{\mathrm{T}}

\newcommand{\unity}{\mathds{1}}

\newcommand{\Tr}{\text{Tr}}
\newcommand{\diag}{\text{diag}}

\newcommand{\Order}{\mathcal{O}}

\newcommand{\dfx}{\mathrm{d}^4x\,}
\newcommand{\dfk}{\mathrm{d}^4k\,}
\newcommand{\du}{\mathrm{d}u\,}
\newcommand{\dv}{\mathrm{d}v\,}
\newcommand{\dx}{\mathrm{d}x\,}
\newcommand{\dy}{\mathrm{d}y\,}
\newcommand{\dz}{\mathrm{d}z\,}

\renewcommand{\Re}{\text{Re}\,}
\renewcommand{\Im}{\text{Im}\,}

\newcommand{\SUthree}{SU(3)}

\newcommand{\Uthree}{U(3)}

\newcommand{\eps}{\epsilon}

\newcommand{\ex}{\mathrm{e}}

\newcommand{\aT}{\text{a}}
\newcommand{\sT}{\text{s}}

\newcommand{\disp}{\text{disp}}
\newcommand{\sthr}{s_\text{thr}}

\newcommand{\asym}{\text{asym}}
\newcommand{\eff}{\text{eff}}

\newcommand{\sm}{s_\text{m}}
\newcommand{\vm}{v_\text{m}}

\newcommand{\Rud}{\text{\cite{Rudenko:2017bel}}}

\allowdisplaybreaks[1]

\title{On the transition form factors of the
axial-vector resonance $\boldsymbol{f_1(1285)}$ and its decay
into $\boldsymbol{e^+e^-}$}

\author[a]{Marvin Zanke,}
\author[b]{Martin Hoferichter,}
\author[a]{and Bastian Kubis}

\affiliation[a]{
Helmholtz-Institut f\"ur Strahlen- und Kernphysik (Theorie) and \\
Bethe Center for Theoretical Physics, Universit\"at Bonn, 53115 Bonn, Germany}

\affiliation[b]{
Albert Einstein Center for Fundamental Physics, Institute for Theoretical Physics, University of Bern, Sidlerstrasse 5, 3012 Bern, Switzerland}

\emailAdd{zanke@hiskp.uni-bonn.de}
\emailAdd{hoferichter@itp.unibe.ch}
\emailAdd{kubis@hiskp.uni-bonn.de}

\abstract{
Estimating the contribution from axial-vector intermediate states to hadronic light-by-light scattering requires input on their transition form factors (TFFs). Due to the \LN{Landau}--\LN{Yang} theorem, any experiment sensitive to these TFFs needs to involve at least one virtual photon, which complicates their measurement. Phenomenologically, the situation is best for the $f_1(1285)$ resonance, for which 
information is available from $e^+e^-\to e^+e^- f_1$, $f_1\to 4\pi$, $f_1\to \rho \gamma$, $f_1\to \phi \gamma$, and $f_1\to e^+e^-$. We provide a comprehensive analysis of the $f_1$ TFFs in the framework of vector meson dominance, including short-distance constraints, to
determine to which extent the three independent TFFs can be constrained from the available experimental input---a prerequisite for improved calculations of the axial-vector contribution to hadronic light-by-light scattering. In particular, we focus on the process $f_1\to e^+e^-$, evidence for which has been reported recently by SND for the first time, and discuss the impact that future improved measurements will have on the determination of the $f_1$ TFFs.    
}

\begin{document} 
\renewcommand{\figureautorefname}{Fig.}
\renewcommand{\tableautorefname}{Table}
\renewcommand{\chapterautorefname}{Ch.}
\renewcommand{\sectionautorefname}{Sec.}
\renewcommand{\subsectionautorefname}{Sec.}
\renewcommand{\appendixautorefname}{App.}
\def\equationautorefname~#1\null{Eq.~(#1)\null}
\maketitle

\section{Introduction}
\label{sec:intro}

The interaction of an axial-vector resonance $A$ with two electromagnetic currents is subject to the venerable \LN{Landau}--\LN{Yang} theorem~\cite{Landau:1948kw,Yang:1950rg}, which states that a spin-$1$ particle cannot decay into two on-shell photons. Accordingly, the decay $A\to\gamma\gamma$ is forbidden, and the simplest process from which information on the general $A\to\gamma^*\gamma^*$ matrix element can be extracted is the singly-virtual process. Such measurements are available from the (space-like) reaction $e^+e^-\to e^+e^- A$ for $A=f_1(1285)$ and 
$A=f_1(1420)$~\cite{Gidal:1987bn,Gidal:1987bm,Aihara:1988bw,Aihara:1988uh,Achard:2001uu,Achard:2007hm}, providing results for the so-called equivalent two-photon decay width $\tilde\Gamma_{\gamma\gamma}$ as well as some constraints on the momentum dependence of the process. Assuming $\Uthree$ symmetry then allows some inference for $A=a_1(1260)$, but other direct phenomenological input is scarce.  

Recently, renewed interest in the electromagnetic properties of axial-vector resonances has been triggered by hadronic corrections to the anomalous magnetic moment of the muon, with the current Standard-Model prediction~\cite{Aoyama:2020ynm,Aoyama:2012wk,Aoyama:2019ryr,Czarnecki:2002nt,Gnendiger:2013pva,Davier:2017zfy,Keshavarzi:2018mgv,Colangelo:2018mtw,Hoferichter:2019gzf,Davier:2019can,Keshavarzi:2019abf,Hoid:2020xjs,Kurz:2014wya,Melnikov:2003xd,Masjuan:2017tvw,Colangelo:2017qdm,Colangelo:2017fiz,Hoferichter:2018dmo,Hoferichter:2018kwz,Gerardin:2019vio,Bijnens:2019ghy,Colangelo:2019lpu,Colangelo:2019uex,Blum:2019ugy,Colangelo:2014qya},
\beq
\label{amuSM}
a_\mu^\text{SM}=116\,591\,810(43)\times 10^{-11},
\eeq
differing from experiment~\cite{Bennett:2006fi,Abi:2021gix,Albahri:2021ixb,Albahri:2021kmg,Albahri:2021mtf},
\beq
\label{exp}
a_\mu^\text{exp}=116\,592\,061(41)\times 10^{-11},
\eeq
by $4.2\sigma$. While at present the uncertainty is dominated by hadronic vacuum polarization, with an emerging tension between the determination from $e^+e^-$ data~\cite{Aoyama:2020ynm,Davier:2017zfy,Keshavarzi:2018mgv,Colangelo:2018mtw,Hoferichter:2019gzf,Davier:2019can,Keshavarzi:2019abf,Hoid:2020xjs} and lattice QCD~\cite{Aoyama:2020ynm,Chakraborty:2017tqp,Borsanyi:2017zdw,Blum:2018mom,Giusti:2019xct,Shintani:2019wai,Davies:2019efs,Gerardin:2019rua,Aubin:2019usy,Giusti:2019hkz,Borsanyi:2020mff}, see Refs.~\cite{Crivellin:2020zul,Keshavarzi:2020bfy,Malaescu:2020zuc,Colangelo:2020lcg}, the ultimate precision expected from the Fermilab~\cite{Grange:2015fou} and J-PARC~\cite{Abe:2019thb} experiments demands that also the second-most-uncertain contribution, hadronic light-by-light (HLbL) scattering, be further improved. The uncertainty of the current phenomenological estimate, $a_\mu^\text{HLbL}=92(19)\times
10^{-11}$~\cite{Aoyama:2020ynm,Melnikov:2003xd,Masjuan:2017tvw,Colangelo:2017qdm,Colangelo:2017fiz,Hoferichter:2018dmo,Hoferichter:2018kwz,Gerardin:2019vio,Bijnens:2019ghy,Colangelo:2019lpu,Colangelo:2019uex,Pauk:2014rta,Danilkin:2016hnh,Jegerlehner:2017gek,Knecht:2018sci,Eichmann:2019bqf,Roig:2019reh}, is dominated by the intermediate- and high-energy regions of the loop integral. In fact, while at low energies the few dominant hadronic channels can be taken into account explicitly in a dispersive approach~\cite{Hoferichter:2013ama,Colangelo:2014dfa,Colangelo:2014pva,Colangelo:2015ama,Danilkin:2021icn}---in terms of pseudoscalar TFFs and partial-wave amplitudes for $\gamma^*\gamma^*\to\pi\pi$~\cite{GarciaMartin:2010cw,Hoferichter:2011wk,Moussallam:2013una,Danilkin:2018qfn,Hoferichter:2019nlq,Danilkin:2019opj}---between $(1\text{--}2)\GeV$ multi-hadron channels become relevant, which ultimately need to be matched to short-distance constraints for the HLbL amplitude~\cite{Melnikov:2003xd,Bijnens:2019ghy,Colangelo:2019lpu,Colangelo:2019uex,Knecht:2020xyr,Ludtke:2020moa,Bijnens:2020xnl,Bijnens:2021jqo,Colangelo:2021nkr}. At these intermediate energies, though, the potentially most sizable contribution originates from hadronic channels that include axial-vector resonances, especially given the role they may play in the transition to the asymptotic constraints~\cite{Melnikov:2003xd,Jegerlehner:2017gek,Roig:2019reh,Leutgeb:2019gbz,Cappiello:2019hwh,Masjuan:2020jsf}. So far, however, the available estimates of axial-vector contributions are model dependent, both because evaluated with a Lagrangian model for the HLbL tensor itself and because of uncertainties in the interaction with the electromagnetic currents, as parameterized in terms of their TFFs.

As a first step to improving this situation, a systematic analysis of the axial-vector TFFs has been presented recently in Ref.~\cite{Hoferichter:2020lap}, including the decomposition into \LN{Lorentz} structures that guarantee the absence of kinematic singularities in the TFFs, following the recipe of \LN{Bardeen}, \LN{Tung}, and \LN{Tarrach} (\LN{BTT})~\cite{Bardeen:1969aw,Tarrach:1975tu}, and the derivation of short-distance constraints in analogy to the light-cone expansion of \LN{Brodsky} and \LN{Lepage} (\LN{BL})~\cite{Lepage:1979zb,Lepage:1980fj,Brodsky:1981rp}. Here, we provide a comprehensive analysis of the TFFs of the $f_1(1285)$, for which the most phenomenological input is available. In addition to $e^+e^-\to e^+e^-f_1$~\cite{Aihara:1988bw,Aihara:1988uh,Achard:2001uu}, there are data for $f_1\to 4\pi$~\cite{Zyla:2020zbs}, $f_1\to \rho\gamma$~\cite{Zyla:2020zbs,Amelin:1994ii}, $f_1\to\phi\gamma$~\cite{Zyla:2020zbs,Bityukov:1987bj}, and, most recently, $f_1\to e^+e^-$~\cite{Achasov:2019wtd}, all of which probe different aspects of the TFFs, as we will study in detail in this paper. 

Given that there are three independent TFFs, in contrast to just one in the case of pseudoscalar mesons, a full dispersive reconstruction as in Refs.~\cite{Hoferichter:2018dmo,Hoferichter:2018kwz,Niecknig:2012sj,Schneider:2012ez,Hoferichter:2012pm,Hoferichter:2014vra,Hoferichter:2021lct} for the $\pi^0$ or in progress for $\eta$, $\eta'$~\cite{Stollenwerk:2011zz,Hanhart:2013vba,Kubis:2015sga,Xiao:2015uva,Gan:2020aco} appears not feasible given the available data. Accordingly, we will study the simplest vector-meson-dominance (VMD) ansatz, to elucidate which parameters can presently be determined from experiment. In contrast to previous work~\cite{Rudenko:2017bel,Milstein:2019yvz}, our parameterization ensures the absence of kinematic singularities, includes short-distance constraints, and accounts for the spectral function of the isovector resonances. In particular, we critically examine which of the processes listed above do allow for an unambiguous extraction of TFF properties. We focus on the $f_1\to e^+e^-$ decay, evidence for which has been observed only recently by the SND collaboration~\cite{Achasov:2019wtd}, with future improvements possible in the context of the ongoing program to measure $e^+e^-\to\text{hadrons}$ cross sections. Further, since this process involves a loop integration that depends on all three TFFs, it should provide some sensitivity also to the doubly-virtual TFFs, which are particularly difficult to measure otherwise.

The outline of this article is as follows:   in \autoref{sec:lorentz_decomposition}, we review the \LN{BTT} decomposition of the $A\to\gamma^*\gamma^*$ matrix element as well as the asymptotic constraints. In \autoref{sec:vmd}, we then construct a minimal VMD ansatz, an extended version,
and study their asymptotic behavior. 
The tree-level processes $e^+e^-\to e^+e^-f_1$, $f_1\to 4\pi$, 
and $f_1\to V\gamma$ ($V=\rho,\phi,\omega$) used to constrain the parameters are discussed in \autoref{sec:tree_level}, 
followed by the $f_1\to e^+e^-$ decay in \autoref{sec:f1ee}.
The full phenomenological analysis is provided
in \autoref{sec:pheno}, before we summarize our findings in \autoref{sec:summary}. Further details are provided in the appendices. 

\section{\LN{Lorentz} decomposition and \LN{Brodsky}--\LN{Lepage} limit}
\label{sec:lorentz_decomposition}

The matrix element for the decay of an axial-vector meson into two virtual photons, $\Ax(P,\lambda_\Ax) \to \gamma^*(q_1,\lambda_1) \gamma^*(q_2,\lambda_2)$, is given by~\cite{Hoferichter:2020lap} 
\beq
  \braket{\gamma^*(q_1,\lambda_1)\gamma^*(q_2,\lambda_2)|\Ax(P,\lambda_\Ax)} = \iu (2\pi)^4 \delta^{(4)}(q_1 + q_2 - P) \, 
  \M\big(\{\Ax,\lambda_\Ax\} \to \{\gamma^*,\lambda_1\} \{\gamma^*,\lambda_2\}\big)  
\eeq
in terms of helicity amplitudes
\beq \label{eq:amplitude}
\M\big(\{\Ax,\lambda_\Ax\} \to \{\gamma^*,\lambda_1\} \{\gamma^*,\lambda_2\}\big) =
  e^2 {\eps_\mu^{\lambda_1}}^*(q_1) {\eps_\nu^{\lambda_2}}^*(q_2) \eps_\alpha^{\lambda_\Ax}(P) \M^{\mu \nu \alpha}(q_1,q_2),
\eeq
where we introduced the tensor matrix element $\M^{\mu \nu \alpha}(q_1,q_2)$ by means of
\begin{align}
	\M^{\mu \nu}(\{P,\lambda_\Ax\} \to q_1, q_2) &= \eps_\alpha^{\lambda_\Ax}(P) \M^{\mu \nu \alpha}(q_1,q_2) \notag \\
	&= \iu \int{\dfx \ex^{\iu q_1 \cdot x} \braket{0|\tior\{j_\text{em}^\mu(x) j_\text{em}^\nu(0)\}|\Ax(P,\lambda_\Ax)}}.
\end{align}
In deriving these relations, the axial-vector meson is treated as an asymptotic state in the narrow-width approximation; furthermore, the electromagnetic quark current is given by
\begin{align}
	j_\text{em}^\mu(x) &= \bar{q}(x) \Q \gamma^\mu q(x), & q(x) &= (u(x),d(x),s(x))^\intercal, & \Q = \frac13 \diag(2,-1,-1).
\end{align}

\subsection{\LN{Lorentz} structures}

Following the \LN{BTT} approach~\cite{Bardeen:1969aw,Tarrach:1975tu}, the tensor matrix element $\M^{\mu \nu \alpha}(q_1,q_2)$ can be decomposed into three independent \LN{Lorentz} structures and scalar functions $\F_i(q_1^2,q_2^2)$ that are free of kinematic singularities, with the result~\cite{Hoferichter:2020lap}
\beq\label{eq:MDecomposition}
	\M^{\mu \nu \alpha}(q_1,q_2) = \frac{\iu}{\Maxial^2} \sum_{i=1}^3{T_i^{\mu \nu \alpha}(q_1,q_2) \F_i(q_1^2,q_2^2)},
\eeq
where $\Maxial$ is the mass of the respective axial-vector meson and
\begin{align}\label{eq:structures}
	T_1^{\mu \nu \alpha}(q_1,q_2) &= \eps^{\mu \nu \beta \gamma} {q_1}_\beta {q_2}_\gamma (q_1^\alpha - q_2^\alpha), \notag \\
	T_2^{\mu \nu \alpha}(q_1,q_2) &= \eps^{\alpha \nu \beta \gamma} {q_1}_\beta {q_2}_\gamma q_1^\mu + \eps^{\alpha \mu \nu \beta} {q_2}_\beta q_1^2, \notag \\
	T_3^{\mu \nu \alpha}(q_1,q_2) &= \eps^{\alpha \mu \beta \gamma} {q_1}_\beta {q_2}_\gamma q_2^\nu + \eps^{\alpha \mu \nu \beta} {q_1}_\beta q_2^2,
\end{align}
with the convention $\eps^{0123} = +1$.
Under photon crossing ($\mu \leftrightarrow \nu$ and $q_1 \leftrightarrow q_2$), the structures transform according to $T_1^{\nu \mu \alpha}(q_2,q_1) = -T_1^{\mu \nu \alpha}(q_1,q_2)$ and $T_2^{\nu \mu \alpha}(q_2,q_1) = -T_3^{\mu \nu \alpha}(q_1,q_2)$, so that for the form factors we find $\F_1(q_2^2,q_1^2) = -\F_1(q_1^2,q_2^2)$ and $\F_2(q_2^2,q_1^2) = -\F_3(q_1^2,q_2^2)$ on account of \LN{Bose} symmetry, $\M^{\mu \nu \alpha}(q_1,q_2) = \M^{\nu \mu \alpha}(q_2,q_1)$.
The prefactor $\iu/\Maxial^2$ in \autoref{eq:MDecomposition}
has been chosen to obtain dimensionless TFFs $\F_i(q_1^2,q_2^2)$ with real-valued normalization.

The \LN{Landau}--\LN{Yang} theorem~\cite{Landau:1948kw,Yang:1950rg} forbids the decay into two on-shell photons, \Lat{i.e.}, at least one photon has to be virtual. In particular, the decay width\footnote{This expression includes a factor $1/2$ due to the indistinguishability of the two on-shell photons.}
\beq\label{eq:regularDecayWidth}
	\Gamma(\Ax \to \gamma \gamma) = \frac{1}{32\pi \Maxial} \lvert \M(\Ax \to \gamma \gamma) \rvert^2
\eeq
vanishes~\cite{Hoferichter:2020lap}, where $\lvert \M(\Ax \to \gamma \gamma) \rvert^2$ is the squared spin-average of the helicity amplitudes, \autoref{eq:amplitude},
for on-shell photons. Instead, the so-called equivalent two-photon decay width is defined as~\cite{Aihara:1988bw}\footnote{The equivalent two-photon decay width is sometimes defined without the factor of $1/2$, see Ref.~\cite{Olsson:1987jk}.}
\beq
 	\widetilde{\Gamma}_{\gamma \gamma} = \lim_{q_1^2 \to 0} \frac{1}{2} \frac{\Maxial^2}{q_1^2} \Gamma(\Ax \to \gamma_\text{L}^* \gamma_\text{T}),
\eeq
where the spin-averaged---longitudinal-transversal ($\text{LT}$)---width is given by
\beq
	\Gamma(\Ax \to \gamma_\text{L}^* \gamma_\text{T}) = \frac13 \sum_{\substack{\lambda_\Ax = \{0,\pm\} \\ \lambda_2 = \pm}} \int{\mathrm{d}\Gamma_{\Ax \to \gamma^* \gamma^*}^{0 \lambda_2 | \lambda_\Ax}} \Big\rvert_{q_2^2 = 0},
\eeq
and the differential decay width for fixed polarization reads
\beq
	\mathrm{d}\Gamma_{\Ax \to \gamma^* \gamma^*}^{\lambda_1 \lambda_2 | \lambda_\Ax} = \frac{1}{32 \pi^2 \Maxial^2} \frac{\sqrt{\lambda(\Maxial^2, q_1^2, q_2^2)}}{2\Maxial} \lvert \M(\{\Ax,\lambda_\Ax\} \to \{\gamma^*,\lambda_1\} \{\gamma^*,\lambda_2\}) \rvert^2 \mathrm{d}\Omega,
\eeq
with center-of-mass solid angle $\Omega$ and the \LN{K\"all\'en} function $\lambda(a,b,c) = a^2 + b^2 + c^2 - 2ab -2ac -2bc$. In terms of the $\F_i(q_1^2,q_2^2)$ one has~\cite{Hoferichter:2020lap}
\beq\label{eq:twoPhotonDecayWidth}
	\widetilde{\Gamma}_{\gamma \gamma} = \frac{\pi \alpha^2}{12} \Maxial \lvert \F_2(0,0) \rvert^2 = \frac{\pi \alpha^2}{12} \Maxial \lvert \F_3(0,0) \rvert^2,
\eeq
where $\alpha = e^2/(4\pi)$ is the fine-structure constant.

\subsection{Asymptotic constraints}
\label{sec:asymptotic_constraints}

In analogy to the asymptotic limits of the pseudoscalar TFF derived in Refs.~\cite{Lepage:1979zb,Lepage:1980fj,Brodsky:1981rp}, one can use a light-cone expansion to obtain the asymptotic behavior of the axial-vector TFFs. Using the distribution amplitudes from Refs.~\cite{Yang:2005gk,Yang:2007zt}, the asymptotic behavior is given by~\cite{Hoferichter:2020lap} 
\begin{align}\label{eq:FFAsymptotic}
	\F_1(q_1^2,q_2^2) &= \Order(1/q_i^6), \notag \\
	\F_2(q_1^2,q_2^2) &= F_\Ax^\eff \Maxial^3 \int_0^1{\du \frac{u \phi(u)}{(u q_1^2 + (1-u) q_2^2 - u (1-u) \Maxial^2)^2}} + \Order(1/q_i^6), \notag \\
	\F_3(q_1^2,q_2^2) &= -F_\Ax^\eff \Maxial^3 \int_0^1{\du \frac{(1-u) \phi(u)}{(u q_1^2 + (1-u) q_2^2 - u (1-u) \Maxial^2)^2}} + \Order(1/q_i^6),
\end{align}
where we generically denoted powers of asymptotic momenta by $q_i = q_1,q_2$ and the wave function $\phi(u) = 6 u (1-u)$ 
is the asymptotic form that already contributes to the pseudoscalar case. 
In writing \autoref{eq:FFAsymptotic}, we furthermore defined an effective decay constant
\beq\label{eq:effectiveDecayConstant}
	F_\Ax^\eff = 4 \sum_a{C_a F_\Ax^a},
\eeq
where the decay constants $F_\Ax^a$ are defined via
\beq
	\braket{0|\bar{q}(0) \gamma_\mu \gamma_5 \frac{\lambda^a}{2} q(0)|\Ax(P,\lambda_\Ax)} = F_\Ax^a \Maxial \eps_\mu.
\eeq
The \LN{Gell-Mann} matrices $\lambda_a$ and the conveniently normalized unit matrix $\lambda_0 = \sqrt{2/3} \, \unity$ determine the flavor decomposition, with the flavor weights $C_a$ in the effective decay constant given by $C_a = 1/2 \, \Tr(\Q^2 \lambda^a)$, \Lat{i.e.}, $C_0 = 2/(3 \sqrt{6})$, $C_3 = 1/6$, and $C_8 = 1/(6 \sqrt{3})$.

\begin{figure}[t]
	\centering
	\includegraphics[width=0.95\textwidth]{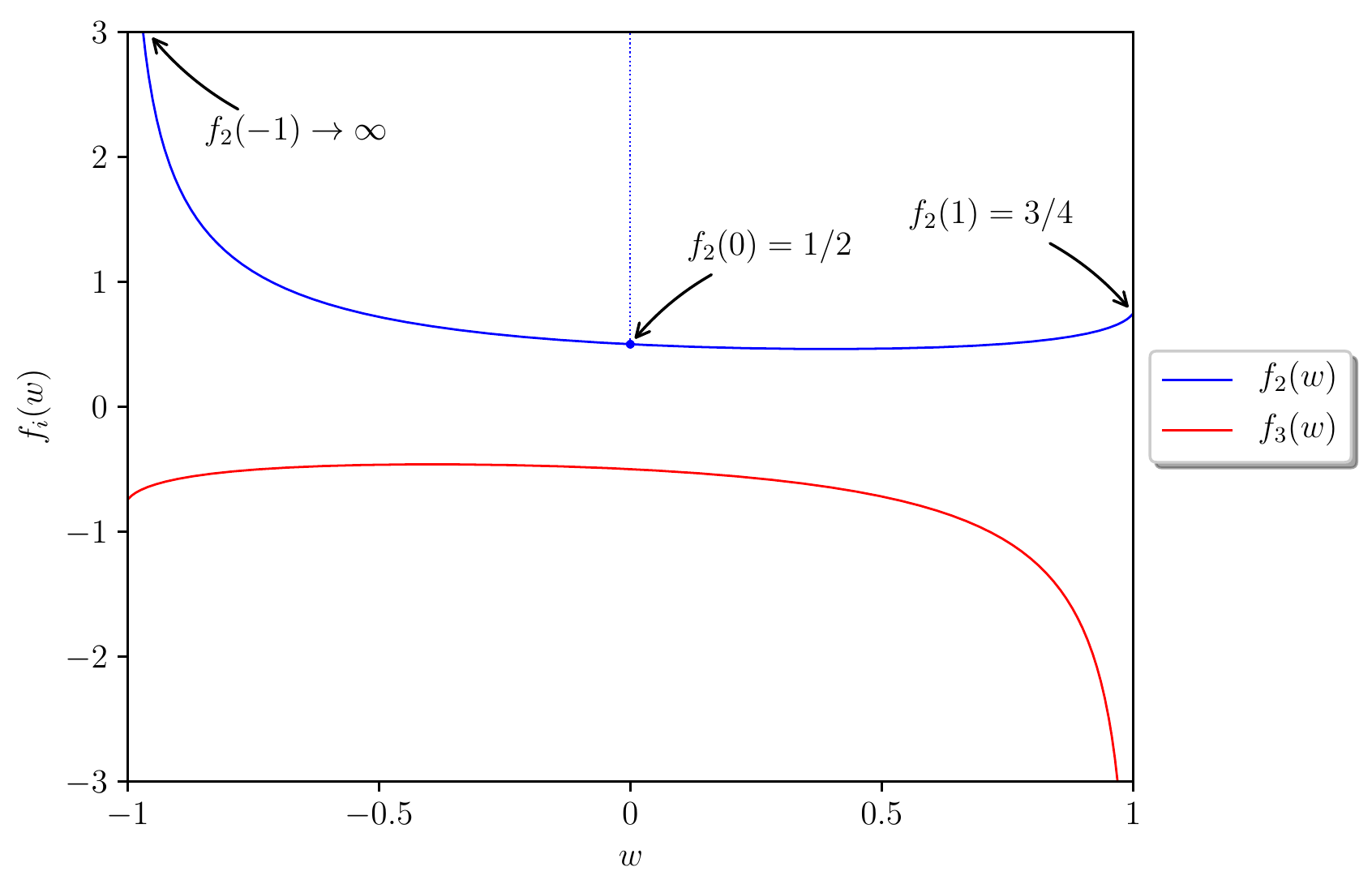}
	\caption{Asymmetry functions $f_2(w)$ and $f_3(w)$, \autoref{eq:fw}, with values for the limiting cases $w\in\{-1,0,1\}$ of $f_2(w)$, corresponding to $q_1^2 = 0$, $q_1^2 = q_2^2$, and $q_2^2 = 0$, respectively. 
	The analogous limits for $f_3(w)$ follow from $f_2(-w) = -f_3(w)$.}
	\label{fig:asymmetryFunctions}
\end{figure}

In \autoref{eq:FFAsymptotic} we retained the leading mass effects in the denominator, but stress that this does not suffice for a consistent treatment of such corrections. We will thus mostly set $\Maxial = 0$ in the denominators when implementing the short-distance constraints, but address the treatment of the leading mass effects in \autoref{appx:asymptotics}. Rewriting the results in terms of 
the average photon virtuality $Q^2$ and the asymmetry parameter $w$,
\begin{align}
	Q^2 &= \frac{q_1^2 + q_2^2}{2} \in [0,\infty), & w &= \frac{q_1^2 - q_2^2}{q_1^2 + q_2^2} \in [-1,1],
\end{align}
one finds the scaling~\cite{Hoferichter:2020lap}
\begin{align}\label{eq:FFLimits}
	\F_1(q_1^2,q_2^2) &= \Order(1/Q^6), \notag \\
	\F_i(q_1^2,q_2^2) &= \frac{F_\Ax^\eff \Maxial^3}{Q^4}f_i(w) +\Order(1/Q^6), \qquad i = 2,3,
\end{align}
with
\begin{align}
\label{eq:fw}
 f_{2/3}(w) &= \frac{3}{4w^3}\left(3 \mp 2w + \frac{(3\pm w)(1\mp w)}{2w} \log\frac{1-w}{1+w}\right).
\end{align}
The asymmetry functions $f_{2/3}(w)$ are shown in \autoref{fig:asymmetryFunctions}, where we also illustrate the values of the function $f_2(w)$ for the limiting cases $w=-1$ ($q_1^2 = 0$), $w=0$ ($q_1^2 = q_2^2$), and $w=1$ ($q_2^2 = 0$); since $f_2(-w) = -f_3(w)$, the analogous limits for $f_3(w)$ follow accordingly.

More specifically, the symmetric doubly-virtual and singly-virtual asymptotic limits of the TFFs---the latter often being referred to as the \LN{BL} limit---become
\begin{align}\label{eq:FFDVSV}
	\F_2(q^2,q^2) &= \frac{F_\Ax^\eff \Maxial^3}{2q^4}+\Order(1/q^6), & \F_2(q^2,0) &= \frac{3 F_\Ax^\eff \Maxial^3}{q^4}+\Order(1/q^6), \notag \\
	\F_3(q^2,q^2) &= -\frac{F_\Ax^\eff \Maxial^3}{2q^4}+\Order(1/q^6), & \F_3(0,q^2) &= -\frac{3 F_\Ax^\eff \Maxial^3}{q^4}+\Order(1/q^6),
\end{align}
while the expressions for $\F_2(0,q^2)$ and $\F_3(q^2,0)$ diverge. Given that the derivation of \autoref{eq:FFAsymptotic} can only be justified from the operator product expansion for $|w|<1/2$~\cite {Gorsky:1987,Manohar:1990hu}, the singly-virtual limits need to be treated with care.\footnote{In soft-collinear effective theory (SCET) the \LN{BL} factorization can be derived with the kernel corresponding to the perturbatively calculable
SCET \LN{Wilson} coefficient and the wave function 
to the non-perturbative matrix element of a SCET operator~\cite{Bauer:2002nz,Rothstein:2003wh,Grossmann:2015lea}. The asymptotic result as given in \autoref{eq:FFAsymptotic} follows in the limit of conformal symmetry of QCD~\cite{Braun:2003rp}.} However, physical helicity amplitudes only depend on the well-defined limits in \autoref{eq:FFDVSV}, in such a way that the problematic limits $\F_2(0,q^2)$ and $\F_3(q^2,0)$ do not contribute to observables. We will return to this point in the context of the $f_1\to e^+e^-$ loop integral.

\section{Vector meson dominance}
\label{sec:vmd}

Given the scarcity of data for axial-vector resonances, we will perform our phenomenological analysis in the context of a VMD description, which has proven to provide successful approximations for a host of low-energy hadron--photon processes~\cite{Sakurai:1960ju,Sakurai:1969,Landsberg:1986fd,Meissner:1987ge,Klingl:1996by,Fang:2021wes}. Most notably, the underlying assumption that the interaction is dominated by the exchange of vector mesons predicts the charge radius of the pion at the level of $10\perc$. Even though the ensuing model dependence is hard to estimate \Lat{a priori}, this approach allows us to analyze all experimental constraints simultaneously in a common framework, which could be refined as soon as improved data become available.

To construct VMD representations of the TFFs as defined in \autoref{sec:lorentz_decomposition}, it is convenient to recast them in terms of their symmetric (s) and antisymmetric (a) combinations
\begin{align}\label{eq:FFNewBasis}
	\F_{\aT_1}(q_1^2,q_2^2) &= \F_1(q_1^2,q_2^2), \notag \\
	\F_{\aT_2}(q_1^2,q_2^2) &= \F_2(q_1^2,q_2^2) + \F_3(q_1^2,q_2^2), \notag \\
	\F_{\sT}(q_1^2,q_2^2) &= \F_2(q_1^2,q_2^2) - \F_3(q_1^2,q_2^2),
\end{align}
with the indicated symmetry properties 
under the exchange of momenta, $q_1^2 \leftrightarrow q_2^2$. Consequently, the basis of structures transforms according to
\begin{align}
	T_{\aT_1}^{\mu \nu \alpha}(q_1,q_2) &= T_1^{\mu \nu \alpha}(q_1,q_2) \notag \\
	&= \eps^{\mu \nu \beta \gamma} {q_1}_\beta {q_2}_\gamma (q_1^\alpha - q_2^\alpha), \notag \\
	T_{\aT_2}^{\mu \nu \alpha}(q_1,q_2) &= \frac12 \big[T_2^{\mu \nu \alpha}(q_1,q_2) + T_3^{\mu \nu \alpha}(q_1,q_2)\big] \notag \\
	&= \frac12 {q_1}_\beta {q_2}_\gamma \left(\eps^{\alpha \nu \beta \gamma} q_1^\mu + \eps^{\alpha \mu \beta \gamma} q_2^\nu\right) + \frac12 \eps^{\alpha \mu \nu \beta} ({q_2}_\beta q_1^2 + {q_1}_\beta q_2^2), \notag \\
	T_{\sT}^{\mu \nu \alpha}(q_1,q_2) &= \frac12 \big[T_2^{\mu \nu \alpha}(q_1,q_2) - T_3^{\mu \nu \alpha}(q_1,q_2)\big] \notag \\
	&= \frac12 {q_1}_\beta {q_2}_\gamma \left(\eps^{\alpha \nu \beta \gamma} q_1^\mu - \eps^{\alpha \mu \beta \gamma} q_2^\nu\right) + \frac12 \eps^{\alpha \mu \nu \beta} ({q_2}_\beta q_1^2 - {q_1}_\beta q_2^2),
\end{align}
where these functions fulfill the same symmetry properties under photon crossing. 
Given this alternative basis, the equivalent two-photon decay width, \autoref{eq:twoPhotonDecayWidth}, becomes
\beq\label{eq:twoPhotonDecayWidthNewBasis}
	\widetilde{\Gamma}_{\gamma \gamma} = \frac{\pi \alpha^2}{48} \Maxial \lvert \F_\sT(0,0) \rvert^2
\eeq
and the tensor matrix element of \autoref{eq:MDecomposition} takes the form
\beq\label{eq:MDecompositionNewBasis}
	\M^{\mu \nu \alpha}(q_1,q_2) = \frac{\iu}{\Maxial^2} \sum_{i=\aT_1,\aT_2,\sT}{T_i^{\mu \nu \alpha}(q_1,q_2) \F_i(q_1^2,q_2^2)}.
\eeq

\subsection{Quantum numbers and mixing effects}
\label{sec:mixing_effects}

Since by far the best phenomenological information is available for the $f_1 \equiv f_1(1285)$, we will focus on this resonance in the remainder of this work, but remark that information on the $f_1' \equiv f_1(1420)$ and the $a_1(1260)$ can be derived when assuming $\Uthree$ flavor symmetry. As a first step towards constructing our VMD ansatz for the TFFs,\footnote{Related models for the $f_1$ have previously been constructed in the literature~\cite{Rudenko:2017bel,Milstein:2019yvz}, see \autoref{appx:literature} for a more detailed comparison.} we review the relevant quantum numbers and mixing patterns. 
From the $G$-parity $G=+$ of the $f_1$, it is immediately clear that both photons have to be either in their isoscalar or isovector state when neglecting isospin-breaking effects. Hence, the VMD coupling can only proceed via $\rho \rho$-like or via some combination of an $\omega$- and $\phi$-like vector meson, 
each of which will be discussed in turn in \autoref{sec:vmd_isovector} and \autoref{sec:vmd_isoscalar}, respectively. As we will show in the following, it is the isovector channel that dominates, with isoscalar corrections typically at the level of $5\perc$.

To this end, we have to take into account mixing effects between the (physical) mesons of the corresponding $J^{PC} = 1^{++}$ axial-vector nonet, \Lat{i.e.}, the mixing pattern~\cite{Zyla:2020zbs}
\beq\label{eq:mixingAngle}
	\begin{pmatrix} f_1 \\ f_1' \end{pmatrix} = \begin{pmatrix} \cos\theta_\Ax & \sin\theta_\Ax \\ -\sin\theta_\Ax & \cos\theta_\Ax \end{pmatrix} \begin{pmatrix} f^0 \\ f^8 \end{pmatrix},
\eeq
where $f^0$ and $f^8$ denote the isoscalar singlet and octet states of the $J^{PC} = 1^{++}$ nonet and $\theta_\Ax$ is the corresponding mixing angle.
Pure octet/singlet mixing is reproduced for $\theta_\Ax = \pi/2$, whereas ideal mixing is obtained for $\theta_\Ax = \arctan(1/\sqrt{2})$. 

Including only the two resonances $f_1$ and $f_1'$, the $\Uthree$ parameterization of the $J^{PC} = 1^{++}$ axial vectors reads
\beq
\label{axial_vector_SU3}
	\Phi_\mu^\Ax = \begin{pmatrix} \sqrt{\frac{2}{3}} f^0 + \frac{1}{\sqrt{3}} f^8 & 0 & 0 \\ 0 & \sqrt{\frac{2}{3}} f^0 + \frac{1}{\sqrt{3}} f^8 & 0 \\ 0 & 0 & \sqrt{\frac{2}{3}} f^0 -\frac{2}{\sqrt{3}} f^8 \end{pmatrix}_\mu,
\eeq
and when splitting the charge matrix into isovector and isoscalar components according to $\Q = \Q_3 + \Q_8$,
\begin{align}
\label{charge_SU3}
	\Q_3 &= \frac{1}{2} \diag(1, -1, 0), & \Q_8 &= \frac{1}{6} \diag(1, 1, -2),
\end{align}
one finds
\begin{align}\label{eq:axialIsoCouplings}
	\Tr[\Phi_\mu^\Ax \Q_3 \Q_3] &= \frac{{f_1}_\mu(\sqrt{2} \cos\theta_\Ax + \sin\theta_\Ax) + {f'_1}_\mu(\cos\theta_\Ax - \sqrt{2} \sin\theta_\Ax)}{2\sqrt{3}}, \notag \\
	\Tr[\Phi_\mu^\Ax \Q_8 \Q_8] &= \frac{{f_1}_\mu(\sqrt{2} \cos\theta_\Ax - \sin\theta_\Ax) - {f'_1}_\mu(\cos\theta_\Ax + \sqrt{2} \sin\theta_\Ax)}{6\sqrt{3}}.
\end{align}
Using the mixing angle $\theta_\Ax = 62(5)^\circ$ as determined by the L3 collaboration~\cite{Achard:2001uu,Achard:2007hm}, see \autoref{sec:L3}, one thus finds that the ratio $R_{\text{S/V}}$ of isoscalar to isovector contributions for the $f_1 \gamma \gamma$ coupling is given by 
\beq\label{eq:ratioISIVContribution}
	R_{\text{S/V}} =\frac{\sqrt{2}-\tan\theta_\Ax}{3(\sqrt{2}+\tan\theta_\Ax)} = -4.7(3.4)\perc.
\eeq

\subsection{Isovector contributions}
\label{sec:vmd_isovector}

For the isovector contributions to the TFFs in \autoref{eq:FFNewBasis} we include the $\rho\equiv\rho(770)$ and the $\rho'\equiv\rho(1450)$, since this is the minimal particle content that produces a non-vanishing contribution for the antisymmetric TFFs. 
We propose the minimal parameterizations
\begin{align}\label{eq:VMDParametrization}
	\F_{\aT_{1/2}}^{I=1}(q_1^2,q_2^2) &= \frac{C_{\aT_{1/2}} \Mrho^2 \Mrhoprime^2}{(q_1^2 - \Mrho^2 + \iu \sqrt{q_1^2} \, \Gammarho(q_1^2))(q_2^2 - \Mrhoprime^2 + \iu \sqrt{q_2^2} \, \Gammarhoprime(q_2^2))} - (q_1 \leftrightarrow q_2), \notag \\
	\F_{\sT}^{I=1}(q_1^2,q_2^2) &= \frac{C_{\sT} \Mrho^4}{(q_1^2 - \Mrho^2 + \iu \sqrt{q_1^2} \, \Gammarho(q_1^2))(q_2^2 - \Mrho^2 + \iu \sqrt{q_2^2} \, \Gammarho(q_2^2))},
\end{align}
where $\Gammarho(q^2)$ and $\Gammarhoprime(q^2)$ are yet to be specified energy-dependent widths.\footnote{In writing the propagator poles of our VMD model with energy-dependent widths, we stick to the convention of Ref.~\cite{Hoferichter:2014vra}.}
Moreover, $\rho\rho'$ and $\rho'\rho'$ terms will be added to $\F_{\sT}(q_1^2,q_2^2)$ below, to help incorporate the asymptotic constraints from \autoref{sec:asymptotic_constraints}.  
We adopt the dispersion-theoretical point of view to model the
singularities of the TFFs based on vector-meson poles, and refrain from
constructing these using effective Lagrangians in order to facilitate
the implementation of high-energy constraints.

Concerning the energy-dependent width $\Gammarho(q^2)$, the decay $\rho \to \pi \pi$ is described by
\begin{align}\label{eq:energyDependentWidthRho}
	\Gammarho(q^2) &= \theta\big(q^2 - 4\Mpi^2\big) \frac{\gamma_{\rho \to \pi \pi}(q^2)}{\gamma_{\rho \to \pi \pi}(\Mrho^2)} \Gammarho, & \gamma_{\rho \to \pi \pi}(q^2) &= \frac{(q^2 - 4\Mpi^2)^{3/2}}{q^2},
\end{align}
where $\gamma_{\rho \to \pi \pi}(q^2)$ is constructed to be in accord with the behavior of the decay width for variable $\Mrho^2 = q^2$, see \autoref{eq:decayWidthRhoPiPi}, and $\Gammarho$ is the total width of the $\rho$ meson.
For the energy-dependent width $\Gammarhoprime(q^2)$ on the other hand, we will consider two different parameterizations.  First, we assume the decay channel $\rho' \to 4\pi$ to be dominant and thus adopt the near-threshold behavior of the four-pion phase space~\cite{Leutwyler:2002hm,Hanhart:2012wi}. Second, we construct a spectral shape from the decay channels
 $\rho' \to \omega \pi$ ($\omega \to 3\pi$) and $\rho' \to \pi \pi$, neglecting, however, another significant contribution from $\rho' \to a_1 \pi$ ($a_1 \to 3 \pi$)~\cite{Zyla:2020zbs}. 
These parameterizations read
\begin{align}\label{eq:energyDependentWidthRhoPrime}
	\Gammarhoprime^{(4\pi)}(q^2) &= \theta\big(q^2 - 16\Mpi^2\big) \frac{\gamma_{\rho' \to 4\pi}(q^2)}{\gamma_{\rho' \to 4\pi}(\Mrhoprime^2)} \Gammarhoprime, & \gamma_{\rho' \to 4\pi}(q^2) &= \frac{(q^2 - 16\Mpi^2)^{9/2}}{(q^2)^2},
\end{align}
where $\gamma_{\rho' \to 4\pi}(q^2)$ is taken from Refs.~\cite{Leutwyler:2002hm,Hanhart:2012wi} and $\Gammarhoprime$ is the total decay width of the $\rho'$ meson, and
\begin{align}\label{eq:energyDependentWidthRhoPrimeAlternative}
	\Gammarhoprime^{(\omega \pi,\pi \pi)}(q^2) &= \theta\big(q^2 - (\Momega + \Mpi)^2\big) \frac{\gamma_{\rho' \to \omega \pi}(q^2)}{\gamma_{\rho' \to \omega \pi}(\Mrhoprime^2)} \Gamma_{\rho' \to \omega \pi}\notag\\
	&+ \theta\big(q^2 - 4\Mpi^2\big) \frac{\gamma_{\rho' \to \pi \pi}(q^2)}{\gamma_{\rho' \to \pi \pi}(\Mrhoprime^2)} \Gamma_{\rho' \to \pi \pi},
\end{align}
where
\begin{align}\label{eq:gammaRhoPrimeOmegaPi}
	\gamma_{\rho' \to \omega \pi}(q^2) &= \frac{\lambda(q^2, \Momega^2, \Mpi^2)^{3/2}}{(q^2)^{3/2}},\qquad \gamma_{\rho' \to \pi \pi}(q^2) = \frac{(q^2 - 4\Mpi^2)^{3/2}}{q^2}.
\end{align}
Estimates for the branching fractions required to evaluate these expressions are provided in \autoref{appx:SU3}. Finally, the standard form of the $\rho\to\pi\pi$ spectral function in \autoref{eq:energyDependentWidthRho} proves disadvantageous for the evaluation of superconvergence relations in \autoref{sec:vmd_asmyptotics} due to its high-energy behavior.  
We thus follow Refs.~\cite{Adolph:2015tqa,VonHippel:1972fg} and introduce barrier factors according to
\begin{align}\label{eq:energyDependentWidthRho_barrier_factors}
	\Gamma^{(1)}_\rho(q^2)&=\Gammarho(q^2) \frac{\Mrho^2-4\Mpi^2+4p_R^2}{q^2-4\Mpi^2+4p_R^2},\qquad p_R=202.4\MeV,\notag\\
	\Gamma^{(2)}_\rho(q^2)&=\Gamma^{(1)}_\rho(q^2)\frac{\sqrt{q^2}}{\Mrho},
\end{align}
where concurrent adjustments to the $\rho' \to \pi \pi$ channel of $\Gammarhoprime^{(\omega \pi, \pi \pi)}(q^2)$, \autoref{eq:energyDependentWidthRhoPrimeAlternative}, are implied. In the end, the numerical impact of the choice of the $\rho$  spectral function is subdominant, and our results will be shown for $\Gamma^{(2)}_\rho(q^2)$ (both for the $\rho$ and the $2\pi$ component of $\Gammarhoprime^{(\omega \pi,\pi \pi)}(q^2)$),  which is identified as the best phenomenological description for the $\rho$ meson in Ref.~\cite{Adolph:2015tqa}.

For the one-loop process $f_1 \to e^+ e^-$ discussed in \autoref{sec:f1ee} we will use dispersively improved variants of the isovector form factors
to ensure the correct analyticity properties when inserting the TFFs into the loop integral. The corresponding spectral representations are constructed from the energy-dependent widths, \Lat{i.e.},
\begin{align}\label{eq:FormFactorsSpectralRepresentation}
	\widehat{\F}^{I=1}_{\aT_{1/2}}(q_1^2,q_2^2) &= \frac{C_{\aT_{1/2}} \Mrho^2 \Mrhoprime^2}{N_\aT} \bigg[P_\rho^\disp(q_1^2) P_{\rho'}^\disp(q_2^2) - P_{\rho'}^\disp(q_1^2) P_\rho^\disp(q_2^2)\bigg], \notag \\
	\widehat{\F}^{I=1}_{\sT}(q_1^2,q_2^2) &= \frac{C_{\sT} \Mrho^4}{N_\sT} P_\rho^\disp(q_1^2) P_\rho^\disp(q_2^2),
\end{align}
where the dispersive $\rho$ and $\rho'$ propagators are given by
\begin{align}\label{eq:dispersivePropagators}
	P_\rho^\disp(q^2) &= \frac{1}{\pi} \int_{4\Mpi^2}^\infty{\dx \, \frac{\Im\big[P^\text{BW}_\rho(x)\big]}{q^2 - x + i\eps}}, \notag \\
	P_{\rho'}^\disp(q^2) &= \frac{1}{\pi} \int_{\sthr}^\infty{\dy \, \frac{\Im\big[P^\text{BW}_{\rho'}(y)\big]}{q^2 - y + i\eps}}.
\end{align}
The spectral functions are
\begin{align}\label{eq:FormFactorsSpectralFunctions}
	\Im\big[P^\text{BW}_\rho(x)\big] &= \frac{-\sqrt{x} \, \Gammarho(x)}{(x-\Mrho^2)^2 + x \Gammarho(x)^2}, \notag \\
	\Im\big[P^\text{BW}_{\rho'}(y)\big] &= \frac{-\sqrt{y} \, \Gammarhoprime(y)}{(y-\Mrhoprime^2)^2 + y \Gammarhoprime(y)^2},
\end{align}
and the threshold $\sthr \in \{16\Mpi^2,4\Mpi^2\}$ depends on the choice of $\Gammarhoprime(q^2)$, \autoref{eq:energyDependentWidthRhoPrime} or \autoref{eq:energyDependentWidthRhoPrimeAlternative}.
The normalization constants $N_\aT$ and $N_\sT$ are introduced in order to retain the form factor normalizations $C_{\aT_{1/2}}$ and $C_{\sT}$ from \autoref{eq:VMDParametrization},
\begin{align}\label{eq:renormalizationsConstantsVMD}
	N_\aT &= \Mrho^2 \Mrhoprime^2 P_\rho^\disp(0) P_{\rho'}^\disp(0), \notag \\
	N_\sT &= \Mrho^4 P_\rho^\disp(0) P_\rho^\disp(0),
\end{align}
\Lat{i.e.}, to ensure that the constants $C_{\aT_{1/2}}$ and $C_{\sT}$ carry the same meaning in the original and the dispersively improved VMD parameterizations, see \autoref{tab:renormalizations}.
With these conventions, we will drop the distinction between $\F_i(q_1^2,q_2^2)$ and $\widehat\F_i(q_1^2,q_2^2)$ in the following, the understanding being that $f_1\to e^+e^-$ is evaluated with the dispersively improved variants.

\begin{table}[t]
	\centering
	\begin{tabular}{c  c  c  c}
	\toprule
		$\Gammarhoprime(q^2)$ & $N_\aT$ & $N_\sT$ & $\widetilde{N}_\sT$\\
		\midrule
		$\Gammarhoprime^{(4\pi)}(q^2)$ & $0.577^{-0.037}_{+0.045}$ & $0.805$ & $0.805 (1 - \eps_1 - \eps_2) + 0.577^{-0.037}_{+0.045} \eps_1 + 0.414^{-0.051}_{+0.067} \eps_2$\\
		$\Gammarhoprime^{(\omega \pi,\pi \pi)}(q^2)$ & $0.642^{-0.039}_{+0.046}$ & $0.805$ & $0.805 (1 - \eps_1 - \eps_2) + 0.642^{-0.039}_{+0.046} \eps_1 + 0.512^{-0.060}_{+0.076} \eps_2$\\\bottomrule
	\end{tabular}
	\caption{Numerical values of the normalization constants given in \autoref{eq:renormalizationsConstantsVMD} and \autoref{eq:normalizationConstantExtendedVMD}. The uncertainties refer to the variation $\Gammarhoprime=(400\pm 60)\MeV$, see \autoref{appx:constants}.}
	\label{tab:renormalizations}
\end{table}

Given that excited $\rho$ mesons need to be introduced for the antisymmetric TFFs, it is natural to consider an extended VMD parameterization of the symmetric form factor including $\rho\rho'$ and $\rho'\rho'$ terms,  
\begin{align}
\label{extended_VMD}
	\widetilde{\F}_{\sT}^{I=1}(q_1^2,q_2^2) &= C_{\sT} \Bigg[\frac{(1 - \eps_1 - \eps_2) \Mrho^4}{(q_1^2 - \Mrho^2 + \iu \sqrt{q_1^2} \, \Gammarho(q_1^2))(q_2^2 - \Mrho^2 + \iu \sqrt{q_2^2} \, \Gammarho(q_2^2))} \notag \\
	&\qquad \, \, + \frac{(\eps_1/2) \Mrho^2 \Mrhoprime^2}{(q_1^2 - \Mrho^2 + \iu \sqrt{q_1^2} \, \Gammarho(q_1^2))(q_2^2 - \Mrhoprime^2 + \iu \sqrt{q_2^2} \, \Gammarhoprime(q_2^2))} \notag \\
	&\qquad \, \, + \frac{(\eps_1/2) \Mrhoprime^2 \Mrho^2}{(q_1^2 - \Mrhoprime^2 + \iu \sqrt{q_1^2} \, \Gammarhoprime(q_1^2))(q_2^2 - \Mrho^2 + \iu \sqrt{q_2^2} \, \Gammarho(q_2^2))} \notag \\
	&\qquad \, \, + \frac{\eps_2 \Mrhoprime^4}{(q_1^2 - \Mrhoprime^2 + \iu \sqrt{q_1^2} \, \Gammarhoprime(q_1^2))(q_2^2 - \Mrhoprime^2 + \iu \sqrt{q_2^2} \, \Gammarhoprime(q_2^2))}\Bigg],
\end{align}
which is normalized in such a way that $\widetilde{\F}_{\sT}^{I=1}(0,0) = C_{\sT} = \F_{\sT}^{I=1}(0,0)$. Here, $\eps_1$ and $\eps_2$ could be treated as additional free parameters, but instead we will use this freedom to match to the asymptotic constraints in \autoref{sec:vmd_asmyptotics}.
Similarly to \autoref{eq:FormFactorsSpectralRepresentation}, the spectral representation for $\widetilde{\F}_{\sT}^{I=1}(q_1^2,q_2^2)$ is given by
\begin{align}\label{extended_VMD_Spectral_Representation}
	\widetilde{\F}_{\sT}^{I=1}(q_1^2,q_2^2) 
	&= \frac{C_{\sT}}{\widetilde{N}_\sT} \Bigg[(1 - \eps_1 - \eps_2)\Mrho^4 P_\rho^\disp(q_1^2) P_\rho^\disp(q_2^2) + \frac{\eps_1 \Mrho^2 \Mrhoprime^2}{2} P_\rho^\disp(q_1^2) P_{\rho'}^\disp(q_2^2) \notag \\
	&\qquad \quad \, + \frac{\eps_1 \Mrhoprime^2 \Mrho^2}{2} P_{\rho'}^\disp(q_1^2) P_\rho^\disp(q_2^2) + \eps_2 \Mrhoprime^4 P_{\rho'}^\disp(q_1^2) P_{\rho'}^\disp(q_2^2)\Bigg],
\end{align}
with normalization
\begin{align}\label{eq:normalizationConstantExtendedVMD}
	\widetilde{N}_\sT &= (1 - \eps_1 - \eps_2)\Mrho^4 P_\rho^\disp(0) P_\rho^\disp(0) \notag\\
	&+ \eps_1 \Mrho^2 \Mrhoprime^2 P_\rho^\disp(0) P_{\rho'}^\disp(0) + \eps_2 \Mrhoprime^4 P_{\rho'}^\disp(0) P_{\rho'}^\disp(0),
\end{align}
see \autoref{tab:renormalizations}.

\subsection{Isoscalar contributions}
\label{sec:vmd_isoscalar}

In the following, we estimate the isoscalar contributions to the TFFs of \autoref{eq:FFNewBasis} under the assumption of $\Uthree$ flavor symmetry, where we will include the resonances $\omega \equiv \omega(782)$ and $\phi \equiv \phi(1020)$ as well as their excited equivalents $\omega'\equiv\omega(1420)$ and $\phi'\equiv\phi(1680)$ into our parameterization. Mixing effects between the (physical) mesons of the corresponding $J^{PC} = 1^{--}$ vector-meson nonets are taken into account via the pattern~\cite{Zyla:2020zbs}
\beq\label{eq:mixingAngleVectors}
	\begin{pmatrix} \omega^{(')} \\ \phi^{(')} \end{pmatrix} = \begin{pmatrix} \cos\theta_{V^{(')}} & \sin\theta_{V^{(')}} \\ -\sin\theta_{V^{(')}} & \cos\theta_{V^{(')}} \end{pmatrix} \begin{pmatrix} \omega^{0(')} \\ \omega^{8(')} \end{pmatrix},
\eeq
where $\omega^{0(')}$ and $\omega^{8(')}$ denote the isoscalar singlet and octet states of the respective vector-meson nonet with mixing angle $\theta_{V^{(')}}$. For our considerations, we assume both nonets to be ideally mixed, \Lat{i.e.}, $\theta_{V} =  \arctan(1/\sqrt{2}) = \theta_{V'}$. Finally, we need the $\Uthree$ parameterization of the $J^{PC} = 1^{--}$ vector mesons, which reads
\beq
\label{vector_SU3}
	\Phi_\mu^{V^{(')}} = \begin{pmatrix} \rho^{0(')} + \omega^{(')} & 0 & 0 \\ 0 & -\rho^{0(')} + \omega^{(')}  & 0 \\ 0 & 0 &  -\sqrt{2} \phi^{(')}\end{pmatrix}_\mu
\eeq
when including only the aforementioned resonances.

Since the $\Uthree$ couplings $f_1 \omega \phi$, $f_1 \omega' \phi$, and $f_1 \omega \phi'$ vanish for ideally mixed vector mesons, we propose the minimal parameterizations
\begin{align}\label{eq:VMDParametrizationIsoscalar}
	\F_{\aT_{1/2}}^{I=0}(q_1^2,q_2^2) &= \frac{C^{\omega \omega'}_{\aT_{1/2}} \Momega^2 \Momegaprime^2}{(q_1^2 - \Momega^2)(q_2^2 - \Momegaprime^2)} + \frac{C^{\phi \phi'}_{\aT_{1/2}} \Mphi^2 \Mphiprime^2}{(q_1^2 - \Mphi^2)(q_2^2 - \Mphiprime^2)} - (q_1 \leftrightarrow q_2), \notag \\
	\F_{\sT}^{I=0}(q_1^2,q_2^2) &= \frac{C^{\omega \omega}_{\sT} \Momega^4}{(q_1^2 - \Momega^2)(q_2^2 - \Momega^2)} + \frac{C^{\phi \phi}_{\sT} \Mphi^4}{(q_1^2 - \Mphi^2)(q_2^2 - \Mphi^2)}.
\end{align}
 The resonances $\omega$ and $\phi$ should be well described by a narrow-resonance approximation---with $M_V^2\to M_V^2-i\eps$ for time-like applications---while for a realistic description of the excited-state isoscalar resonances their widths would need to be taken into account.    
Due to the expected smallness of the isoscalar contributions, see \autoref{eq:ratioISIVContribution}, we refrain from giving an extended VMD parameterization analogous to \autoref{extended_VMD}. 

With the $\Uthree$ parameterization of the axial-vector mesons, $\Phi_\mu^\Ax$, and the charge matrix $\Q$ from \autoref{sec:mixing_effects}, the ratios of isoscalar to isovector couplings are found to be\footnote{The notation is to be understood in such a way that for each term the prefactor of the fields indicated as a subscript is taken, with the $\Uthree$ parameterizations from \autoref{axial_vector_SU3},   \autoref{charge_SU3}, and \autoref{vector_SU3}. In the ratios only the traces are relevant, as the common Lagrangian parameters cancel. }
\begin{align}
	\frac{C^{\omega \omega'}_{\aT_{1/2}}}{C_{\aT_{1/2}}} &= \frac{C^{\omega \omega}_{\sT}}{C_{\sT}} = \frac{\Tr[\Phi_\mu^\Ax \Phi_\nu^V \Phi_\kappa^{V^{(')}}]\rvert_{{f_1}_\mu \omega_\nu \omega^{(')}_\kappa} \Tr[\Phi_\alpha^V \Q]\rvert_{\omega_\alpha} \Tr[\Phi_\beta^{V^{(')}} \Q]\rvert_{\omega^{(')}_\beta}}{\Tr[\Phi_\mu^\Ax \Phi_\nu^V \Phi_\kappa^{V^{(')}}]\rvert_{{f_1}_\mu \rho_\nu \rho^{(')}_\kappa} \Tr[\Phi_\alpha^V \Q]\rvert_{\rho_\alpha} \Tr[\Phi_\beta^{V^{(')}} \Q]\rvert_{\rho^{(')}_\beta}} = \frac{1}{9}, \\
	\frac{C^{\phi \phi'}_{\aT_{1/2}}}{C_{\aT_{1/2}}} &= \frac{C^{\phi \phi}_{\sT}}{C_{\sT}} = \frac{\Tr[\Phi_\mu^\Ax \Phi_\nu^V \Phi_\kappa^{V^{(')}}]\rvert_{{f_1}_\mu \phi_\nu \phi^{(')}_\kappa} \Tr[\Phi_\alpha^V \Q]\rvert_{\phi_\alpha} \Tr[\Phi_\beta^{V^{(')}} \Q]\rvert_{\phi^{(')}_\beta}}{\Tr[\Phi_\mu^\Ax \Phi_\nu^V \Phi_\kappa^{V^{(')}}]\rvert_{{f_1}_\mu \rho_\nu \rho^{(')}_\kappa} \Tr[\Phi_\alpha^V \Q]\rvert_{\rho_\alpha} \Tr[\Phi_\beta^{V^{(')}} \Q]\rvert_{\rho^{(')}_\beta}} = \frac{2(\sqrt{2} - 2\tan\theta_\Ax)}{9(\sqrt{2} + \tan\theta_\Ax)},\notag
\end{align}
which, using the mixing angle $\theta_\Ax = 62(5)^\circ$ as determined by the L3 collaboration~\cite{Achard:2001uu,Achard:2007hm}, see \autoref{sec:L3}, implies 
\begin{align}\label{eq:SU3RatiosCouplings}
	R^\omega &= \frac{C^{\omega \omega'}_{\aT_{1/2}}}{C_{\aT_{1/2}}} = \frac{C^{\omega \omega}_{\sT}}{C_{\sT}} = \frac{1}{9}, & R^\phi &= \frac{C^{\phi \phi'}_{\aT_{1/2}}}{C_{\aT_{1/2}}} = \frac{C^{\phi \phi}_{\sT}}{C_{\sT}} = -0.158(34).
\end{align}
The additional suppression in \autoref{eq:ratioISIVContribution} then results from a cancellation between $\omega$ and $\phi$ contributions
\begin{align}
\label{eq:isoscalar_cancellation}
 R_{\text{S/V}} = R^\omega + R^\phi= 11.1\perc-15.8(3.4)\perc = -4.7(3.4)\perc.
\end{align}
In practice, we will restrict the analysis of isoscalar contributions to the symmetric TFF. First, $\F_{\sT}(q_1^2,q_2^2)$ gives the dominant contribution to the observables, so that the most important isoscalar correction is expected from there. In addition, for the antisymmetric TFFs we would need to include the excited $\omega'$ and $\phi'$ states, incurring significant uncertainties from their spectral functions and, especially for the $f_1\to e^+e^-$ application, the asymptotic matching due to their large masses. Alternatively, isoscalar antisymmetric TFFs could be produced via deviations from ideal $\phi$--$\omega$ mixing, but again the uncertainties would be difficult to control. For these reasons we conclude that the isoscalar contributions to the antisymmetric TFFs should be irrelevant at present, with potential future refinements once better data become available.

\subsection{Asymptotics}
\label{sec:vmd_asmyptotics}

\begin{figure}[t]
	\centering
	\includegraphics[width=0.95\textwidth]{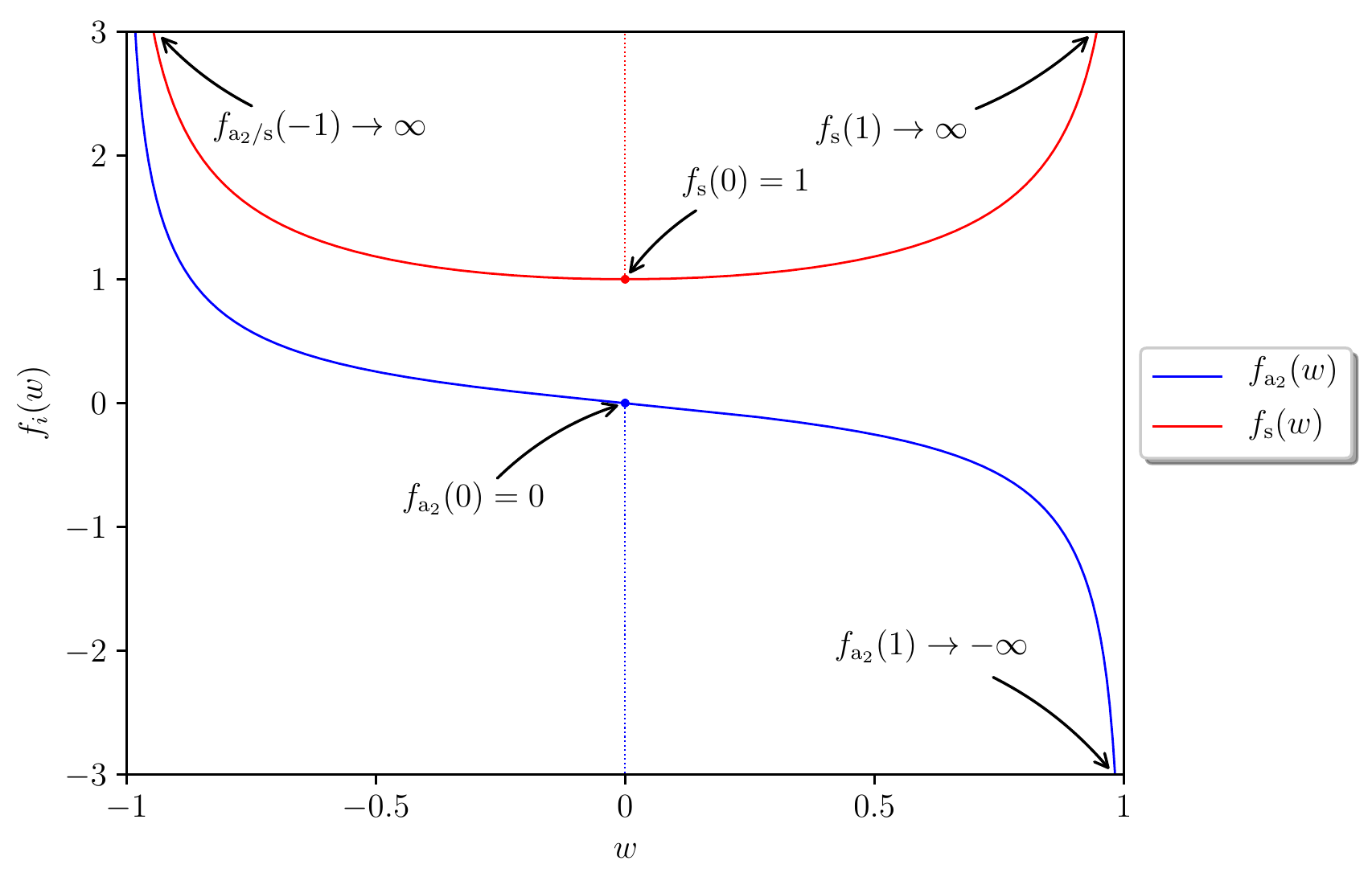}
	\caption{Asymmetry functions $f_{\aT_2}(w)$ and $f_{\sT}(w)$, \autoref{eq:FFLimitsNewBasis}, with values for the limiting cases $w\in\{-1,0,1\}$, corresponding to $q_1^2 = 0$, $q_1^2 = q_2^2$, and $q_2^2 = 0$ respectively.}
	\label{fig:asymmetryFunctionsNewBasis}
\end{figure}

The VMD representations for the TFFs should comply with the asymptotic constraints reviewed in \autoref{sec:asymptotic_constraints}, mainly to ensure that the $f_1\to e^+e^-$ loop integral does not receive unphysical contributions in the high-energy region. We will focus on the isovector amplitudes, given the strong suppression of the isoscalar contributions. 
Translated to the basis of (anti-)symmetric TFFs, we have 
\begin{align}\label{eq:FFLimitsNewBasis}
	\F_{\aT_1}(q_1^2,q_2^2) &= \Order(1/Q^6), \notag \\
	\F_{\aT_2}(q_1^2,q_2^2) &= \frac{F_{f_1}^\eff \Mf^3}{Q^4} f_{\aT_2}(w) + \Order(1/Q^6), & f_{\aT_2}(w) &= \frac{3}{4w^3}\left(6 + \frac{3-w^2}{w} \log\frac{1-w}{1+w}\right), \notag \\
	\F_{\sT}(q_1^2,q_2^2) &= \frac{F_{f_1}^\eff \Mf^3}{Q^4} f_{\sT}(w)  + \Order(1/Q^6), & f_{\sT}(w) &= -\frac{3}{2w^3}\left(2w + \log\frac{1-w}{1+w}\right),
\end{align}
see \autoref{fig:asymmetryFunctionsNewBasis}. The symmetrical doubly-virtual limits become ($\lambda \approx 1$)
\begin{align}\label{eq:FFDVSVNewBasis}
	\F_{\aT_2}(q^2,\lambda q^2) &= - \frac{6 F_{f_1}^\eff \Mf^3}{q^4} k(\lambda) + \Order(1/q^6),  \qquad
	\F_{\sT}(q^2,q^2) = \frac{F_{f_1}^\eff \Mf^3}{q^4}  + \Order(1/q^6),\notag\\
	k(\lambda) &= \frac{3\lambda^2 - (\lambda^2 + 4 \lambda + 1) \log\lambda -3}{(\lambda - 1)^4}=\Order(\lambda-1),
\end{align}
 but upon symmetrization all singly-virtual limits of $\F_{\aT_2/\sT}(q_1^2,q_2^2)$ diverge. For this reason, the asymptotic limits for $\F_{\aT_2/\sT}(q_1^2,q_2^2)$ cannot be considered in isolation, but need to be implemented in such a way as to reproduce the physical behavior of $\F_{2/3}(q_1^2,q_2^2)$.

We first consider the asymptotic behavior of the minimal VMD
parameterization, \autoref{eq:VMDParametrization},
\begin{align}\label{eq:FFDVSVVMD}
	\F_{\aT_{1/2}}^{I=1}(q^2, \lambda q^2) &\propto \frac{\lambda - 1}{\lambda^2} \frac{1}{q^6}, & \F_{\aT_{1/2}}^{I=1}(q^2, 0) &\propto \frac{1}{q^2}, \notag \\
	\F_{\sT}^{I=1}(q^2, q^2) &\propto \frac{1}{q^4}, & \F_{\sT}^{I=1}(q^2, 0) &\propto \frac{1}{q^2}.
\end{align}
In this case, the scaling is correct in the doubly-virtual direction of 
$\F_{\aT_1/\sT}^{I=1}(q_1^2,q_2^2)$, while $\F_{\aT_2}^{I=1}(q_1^2,q_2^2)$ drops too fast and the singly-virtual limits too slowly, see \autoref{tab:FFLimits}. Phenomenologically, the symmetric TFF gives the dominant contribution to $f_1\to e^+e^-$, see \autoref{sec:f1ee}, so that here also the coefficient deserves some attention. Comparing the asymptotic limit of \autoref{eq:VMDParametrization} with \autoref{eq:FFDVSVNewBasis}, the VMD ansatz for $\F_\sT(q_1^2,q_2^2)$ implies the following estimate for the effective decay constant defined in \autoref{eq:effectiveDecayConstant}:
\beq
\label{eq:effectiveDecayConstantVMDIsoscalar}
	F_{f_1}^\eff\Big|_\text{VMD} = \frac{C_\sT \Mrho^4}{\Mf^3} = 159(19) \MeV,
\eeq
where we already used the L3 result for $C_\sT$ including the isoscalar contribution, see  \autoref{eq:CsIsoscalar} below. Within uncertainties, this value agrees with the result from light-cone sum rules (LCSRs)~\cite{Yang:2007zt,Hoferichter:2020lap}
\beq\label{eq:effectiveDecayConstantLCSR}
	F_{f_1}^\eff\Big|_\text{LCSRs} = 146(7)_\text{LCSRs}(12)_{\theta_\Ax} \MeV,
\eeq
so that even the minimal VMD ansatz should display a reasonable asymptotic behavior.  

\begin{table}[t]
	\centering
	\begin{tabular}{ c  c  c  c  c  c}
	\toprule
		 & \multicolumn{2}{c }{$\F_{\aT_1}(q_1^2,q_2^2)$} & $\F_{\aT_2}(q_1^2,q_2^2)$ & $\F_{\sT}(q_1^2,q_2^2)$ & $\F_2(q_1^2,q_2^2)$\\
		& $q_{1/2}^2 \approx q^2$ & $q_2^2 = 0$ &$q_{1/2}^2 \approx q^2$ & $q_{1/2}^2 = q^2$ & $q_{2}^2 = 0$\\
		\midrule
		Light-cone expansion & $1/q^6$ & $1/q_1^6$ & $1/q^4$  & $1/q^4$ & $1/q_1^4$\\
		VMD (isovector) & $1/q^6$ & $1/q_1^2$ & $1/q^6$  & $1/q^4$ & $1/q_1^2$\\
		$\widetilde{\text{VMD}}$ (isovector) & $1/q^6$  & $1/q_1^2$  & $1/q^6$ & $1/q^6$  & $1/q_1^4$\\\bottomrule
	\end{tabular}
	\caption{Comparison of the asymptotic behavior of the TFFs as predicted by the light-cone expansion, \autoref{eq:FFLimitsNewBasis} and \autoref{eq:FFDVSVNewBasis}, with the implementation in the VMD representations, \autoref{eq:VMDParametrization} and \autoref{extended_VMD}. The doubly-virtual limits of $\widetilde{\text{VMD}}$ are tailored to decrease as $1/q^6$, so that the behavior of the light-cone expansion is reproduced   
	by adding \autoref{eq:asym_double_spectral}.}
	\label{tab:FFLimits}
\end{table}

To go beyond this minimal implementation, we now turn to the extended VMD ansatz for $\F_\sT(q_1^2,q_2^2)$. We follow the strategy from Refs.~\cite{Hoferichter:2018dmo,Hoferichter:2018kwz} and add an explicit asymptotic term that incorporates the correct doubly-virtual behavior, obtained by rewriting \autoref{eq:FFAsymptotic} in terms of a dispersion relation; see also Ref.~\cite{Khodjamirian:1997tk}. Accordingly, we need to ensure that the isovector VMD contribution to $\F_\sT(q^2,q^2)$ behaves $\propto 1/q^6$, resulting in 
\beq\label{eq:eps2VMD}
	\eps_2 = \frac{(1-\eps_1)\Mrho^4 + \eps_1 \Mrho^2 \Mrhoprime^2}{\Mrho^4 - \Mrhoprime^4}.
\eeq
This leaves the freedom to choose $\eps_1$, which we use to implement the physical singly-virtual scaling of $\F_2^{I=1}(q^2,0)=[\F_{\aT_2}^{I=1}(q^2,0)+\widetilde{\F}_{\sT}^{I=1}(q^2,0)]/2\propto 1/q^4$, leading to
\beq\label{eq:eps1VMD}
	\eps_1 = -2 \frac{C_{\aT_2} (\Mrho^4 -\Mrhoprime^4) + C_\sT \Mrho^2 \Mrhoprime^2}{C_\sT (\Mrho^2 - \Mrhoprime^2)^2}.
\eeq
Further, the coefficient of $1/q^4$ in the resulting $\F_2^{I=1}(q^2,0)$ only depends on $C_\sT$, and matching to \autoref{eq:FFDVSV} implies
\beq\label{eq:effectiveDecayConstantExtendedVMDIsoscalar}
	F_{f_1}^\eff\Big|_{\widetilde{\text{VMD}}} = \frac{C_\sT \Mrho^2\Mrhoprime^2}{6\Mf^3} = 95(12) \MeV,
\eeq
reasonably close to the LCSR estimate of \autoref{eq:effectiveDecayConstantLCSR}. In general, the choice for $\eps_1$ in \autoref{eq:eps1VMD} enforces the expected singly-virtual behavior at the expense of a large coefficient, \Lat{e.g.}, for $C_{\aT_2}=0$ one has $\eps_1=-1.08$, so that a better low-energy phenomenology might be achieved when considering $\eps_1$ a free parameter instead. We will continue to use \autoref{eq:eps1VMD} as a benchmark scenario in comparison to the minimal VMD ansatz, keeping this caveat regarding $\eps_1$ in mind.

In choosing the above $\eps_{1/2}$, we did not take the spectral representations of \autoref{eq:FormFactorsSpectralRepresentation} and \autoref{extended_VMD_Spectral_Representation} into account, which would lead to a set of superconvergence relations that need to be fulfilled, but instead made an approximate choice in terms of \autoref{extended_VMD} and \autoref{eq:VMDParametrization}. More specifically, these superconvergence relations read
\begin{align}
	\Order(1/q^6) &= \frac{C_\sT}{\widetilde{N}_\sT q^4}\Big[(1-\eps_1-\eps_2) \Mrho^4 P^0_\rho P^0_\rho + \eps_1 \Mrho^2 \Mrhoprime^2 P^0_\rho P^0_{\rho'} + \eps_2 \Mrhoprime^4 P^0_{\rho'} P^0_{\rho'}\Big], \\
	\Order(1/q^4) &= -\frac{C_{\aT_2} \Mrho^2 \Mrhoprime^2}{2 N_\aT q^2}\Big[P^0_\rho \bar{P}^0_{\rho'} - P^0_{\rho'} \bar{P}^0_\rho\Big] \notag \\
	&- \frac{C_\sT}{2 N_\sT q^2} \Big[(1-\eps_1-\eps_2) \Mrho^4 P^0_\rho \bar{P}^0_\rho + \frac{\eps_1 \Mrho^2 \Mrhoprime^2}{2} \Big(P^0_\rho \bar{P}^0_{\rho'} + P^0_{\rho'} \bar{P}^0_\rho\Big) + \eps_2 \Mrhoprime^4 P^0_{\rho'} \bar{P}^0_{\rho'}\Big], \notag
\end{align}
where we defined
\begin{align}\label{eq:superconvergence_integrals}
	P^0_\rho &= -\frac{1}{\pi} \int_{4\Mpi^2}^\infty{\dx \, \Im\big[P^\text{BW}_\rho(x)\big]}, & \bar{P}^0_\rho &= -\frac{1}{\pi} \int_{4\Mpi^2}^\infty{\dx \, \frac{\Im\big[P^\text{BW}_\rho(x)\big]}{x}}, \notag \\
	P^0_{\rho'} &= -\frac{1}{\pi} \int_{\sthr}^\infty{\dy \, \Im\big[P^\text{BW}_{\rho'}(y)\big]}, & \bar{P}^0_{\rho'} &= -\frac{1}{\pi} \int_{\sthr}^\infty{\dy \, \frac{\Im\big[P^\text{BW}_{\rho'}(y)\big]}{y}}.
\end{align}
Solving this for $\eps_2$ and $\eps_1$, we find
\begin{align}\label{eq:epsSCR}
	\eps_2 &= \frac{(1-\eps_1) \big(\Mrho^2P^0_\rho\big)^2 + \eps_1 \Mrho^2 P^0_\rho \Mrhoprime^2 P^0_{\rho'}}{ \big(\Mrho^2P^0_\rho\big)^2  -  \big(\Mrhoprime^2P^0_{\rho'}\big)^2}, \notag \\
	\eps_1 &= -2 \frac{\frac{C_{\aT_2}}{N_\aT} \big[\big(\Mrho^2 P^0_\rho\big)^2 -\big(\Mrhoprime^2 P^0_{\rho'}\big)^2\big] + \frac{C_\sT}{N_\sT} \Mrho^2 P^0_\rho \Mrhoprime^2 P^0_{\rho'}}{\frac{C_\sT}{N_\sT} (\Mrho^2 P^0_\rho - \Mrhoprime^2 P^0_{\rho'})^2},
\end{align}
in accordance with \autoref{eq:eps2VMD} and \autoref{eq:eps1VMD} upon the replacements
\begin{align}\label{eq:SCR_replacements}
	\Mrho^2 &\to \Mrho^2 P^0_\rho, & \Mrhoprime^2 &\to \Mrhoprime^2 P^0_{\rho'}, \notag \\
	C_{\aT_2} &\to \frac{C_{\aT_2}}{N_\aT}, & C_\sT &\to \frac{C_\sT}{N_\sT}.
\end{align}
Numerical values for $P^0_{\rho}$ and $P^0_{\rho'}$ are collected in \autoref{tab:superconvergence_relations_result}. These results show that most correction factors are close to unity, in which case the only potentially significant correction arises from the different normalizations $N_{\aT}$ and $N_{\sT}$ 
for $\eps_1$, see \autoref{tab:renormalizations}. However, 
our central results will employ $\Gammarhoprime^{(\omega \pi,\pi \pi)}(q^2)$,
and given the abovementioned caveats in the choice of $\eps_1$, we conclude that at the current level of accuracy the naive VMD expressions \autoref{eq:eps2VMD} and \autoref{eq:eps1VMD} are sufficient.

\begin{table}[t]
	\centering
	\begin{tabular}{c c c c}
	\toprule
		$\Gamma_{\rho^{(\prime)}}(q^2)$ & $\Gammarho^{(2)}(q^2)$ & $\Gammarhoprime^{(4\pi)}(q^2)$ & $\Gammarhoprime^{(\omega \pi,\pi \pi)}(q^2)$\\\midrule
		$P^0_{\rho^{(\prime)}}$  & $1.023$ & $0.718^{-0.057}_{+0.070}$ & $0.918^{-0.073}_{+0.087}$\\\bottomrule
	\end{tabular}
	\caption{Numerical values of $P^0_\rho$ and $P^0_{\rho'}$, \autoref{eq:superconvergence_integrals}, as obtained with the parameterizations $\Gammarho^{(2)}(q^2)$, $\Gammarhoprime^{(4\pi)}(q^2)$, and $\Gammarhoprime^{(\omega \pi,\pi \pi)}(q^2)$, \autoref{eq:energyDependentWidthRho_barrier_factors}, \autoref{eq:energyDependentWidthRhoPrime}, and \autoref{eq:energyDependentWidthRhoPrimeAlternative}, needed for \autoref{eq:epsSCR}. The uncertainties refer to the variation $\Gammarhoprime=(400\pm 60)\MeV$, see \autoref{appx:constants}.}
	\label{tab:superconvergence_relations_result}
\end{table}

The doubly-virtual behavior is implemented as follows~\cite{Hoferichter:2018dmo,Hoferichter:2018kwz}: 
first, we rewrite the asymptotic form factors $\F_2(q_1^2,q_2^2)$ and $\F_3(q_1^2,q_2^2)$ from \autoref{eq:FFAsymptotic} into a double-spectral representation, which allows us to isolate the different energy regions, in particular those that give rise to the correct asymptotic limits. Setting 
$\Maxial = 0$ in the respective integrands of \autoref{eq:FFAsymptotic}, we observe that 
\begin{align}\label{eq:FFDispersiveIntegral}
	\F_2(q_1^2,q_2^2) &= - F_\Ax^\eff \Maxial^3 \frac{\partial}{\partial q_1^2} \int_0^1{\du \frac{\phi(u)}{u q_1^2 + (1-u) q_2^2}} + \Order(1/q_i^6), \notag \\
	\F_3(q_1^2,q_2^2) &= F_\Ax^\eff \Maxial^3 \frac{\partial}{\partial q_2^2} \int_0^1{\du \frac{\phi(u)}{u q_1^2 + (1-u) q_2^2}} + \Order(1/q_i^6) 
\end{align}
 take exactly the same form as for the pseudoscalar case, except for the partial derivatives with respect to $q_i^2$. Accordingly, the same arguments as in Refs.~\cite{Khodjamirian:1997tk,Hoferichter:2018dmo,Hoferichter:2018kwz} apply, and the integral over the wave function can be formally expressed by a double-spectral representation
\beq\label{eq:double_spectral_integral}
	I(q_1^2,q_2^2) = \int_0^1{\du \frac{\phi(u)}{u q_1^2 + (1-u) q_2^2}} 
	= \frac{1}{\pi^2} \int_0^\infty{\dx \int_0^\infty{\dy \frac{\rho^\asym(x,y)}{(x-q_1^2)(y-q_2^2)}}},
\eeq
with double-spectral density
\beq
\label{eq:double_spectral}
	\rho^\asym(x,y) = 3 \pi^2 x y\delta''(x-y).
\eeq
The asymptotic form arises from the high-energy part of these integrals, so that, to avoid overlap with the VMD contribution at low energies, we impose a lower cutoff $\sm$, which, in the language of LCSRs, could be identified with the continuum threshold. Evaluating the partial derivatives and dropping surface terms in the evaluation of the $\delta$ distribution~\cite{Hoferichter:2018dmo,Hoferichter:2018kwz}, we find
\begin{align}
	\F_2^\asym(q_1^2,q_2^2) &= - F_\Ax^\eff \Maxial^3 \frac{\partial}{\partial q_1^2} \left[ \frac{1}{\pi^2} \int_{\sm}^\infty{\dx \int_{\sm}^\infty{\dy \frac{\rho^\asym(x,y)}{(x-q_1^2)(y-q_2^2)}}} \right] + \Order(1/q_i^6) \notag \\
	&= 3 F_\Ax^\eff \Maxial^3 \int_{\sm}^\infty{\dx \frac{q_2^2 (x + q_1^2)}{(x - q_1^2)^3 (x - q_2^2)^2}} + \Order(1/q_i^6),\notag\\
	\F_3^\asym(q_1^2,q_2^2) &= -3 F_\Ax^\eff \Maxial^3 \int_{\sm}^\infty{\dy \frac{q_1^2 (y + q_2^2)}{(y - q_1^2)^2 (y - q_2^2)^3}} + \Order(1/q_i^6).
	\label{eq:asym_double_spectral}
\end{align}
By construction, the asymptotic contributions in this form saturate the doubly-virtual limits of \autoref{eq:FFDVSVNewBasis}, while not affecting the singly-virtual contributions $\F_2(q^2,0)$, $\F_3(0,q^2)$ already taken into account via the extended VMD representation. The opposite---unphysical---cases $\F_2(0,q^2)$, $\F_3(q^2,0)$, which do not contribute to helicity amplitudes, are equally suppressed in the $f_1\to e^+e^-$ loop integral, see \autoref{sec:f1ee}. 
Given that $\Maxial > 1\GeV$, it is also worthwhile to consider the potential impact of mass corrections to the asymptotic constraints. A formulation in terms of a generalized double-spectral density is given in \autoref{appx:asymptotics}.

In conclusion, the extended VMD ansatz together with the asymptotic contribution of \autoref{eq:asym_double_spectral} complies with the short-distance constraints of \autoref{eq:FFAsymptotic}, apart from the singly-virtual behavior of $\F_{\aT_1}(q_1^2,q_2^2)$ and small violations due to the isoscalar contributions of the form factors, see \autoref{eq:ratioISIVContribution}. As we will demonstrate below that $\F_{\aT_1}(q_1^2,q_2^2)$ gives the smallest contribution to the $f_1\to e^+e^-$ loop integral, see \autoref{eq:f1e+e-AmplitudeFinal}, the resulting VMD representation should provide a decent approximation to its high-energy part. In particular, the sensitivity to the high-energy assumptions can be monitored by comparing the two VMD variants constructed in this section.

\section{Tree-level processes}
\label{sec:tree_level}

The VMD parameterizations constructed in the previous section involve the free parameters $C_{\aT_1}$, $C_{\aT_2}$, and $C_{\sT}$ (and, for the extended variant, the onset of the asymptotic contributions $\sm$). In the following, we collect the available data that can, in principle, determine these parameters, starting with the processes in which the TFFs appear at tree level:
\begin{enumerate}
 \item $e^+e^-\to e^+e^- f_1$, which mainly determines the equivalent two-photon decay width $\widetilde{\Gamma}_{\gamma \gamma}^{f_1}$, see \autoref{sec:L3};
 \item $f_1\to 4\pi$, sensitive to the TFFs via $f_1\to\rho\rho\to 4\pi$, see \autoref{sec:4pi};
 \item $f_1\to \rho\gamma$, whose branching fraction and helicity components encode information on the TFFs, see \autoref{sec:rhogamma}.
\end{enumerate}
In a more rigorous, dispersive, reconstruction of the TFFs, the
(partially) hadronic final states would serve as input to a
determination of their discontinuities.
The strategy to investigate the impact of these reactions on a determination of the various TFFs has already been followed in Refs.~\cite{Rudenko:2017bel,Milstein:2019yvz}, albeit with rather different form factor parameterizations.  
Moreover, we investigate the following tree-level decays:
\begin{enumerate}\setcounter{enumi}{3}
 \item $f_1\to \phi\gamma$ and $f_1\to \omega\gamma$, where the measured branching fraction of the former allows for a consistency check of our $\Uthree$ assumption for the isoscalar TFFs and the latter predicts a branching ratio that can be confronted with potential future measurements, see \autoref{sec:phigamma_omegagamma}.
 \end{enumerate}

\subsection[$e^+e^-\to e^+e^- f_1$]{$\boldsymbol{e^+e^-\to e^+e^- f_1}$}
\label{sec:L3}

\begin{figure}[t]
	\centering
	\includegraphics{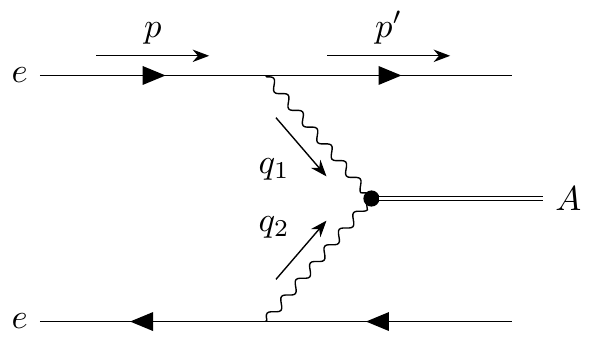}
	\caption{\LN{Feynman} diagram for two-photon hadron formation in electron--positron scattering.}
	\label{fig:twoPhotonHadronFormation}
\end{figure}

In contrast to (pseudo-)scalar or tensor resonances, axial-vector resonances are only visible in $e^+e^-$ collisions, see \autoref{fig:twoPhotonHadronFormation}, as long as at least one of the photons is off shell, a direct consequence of the 
\LN{Landau}--\LN{Yang} theorem~\cite{Landau:1948kw,Yang:1950rg}.
The required challenging measurements have been performed 
for the $f_1$ and $f_1'$, by the MARK II~\cite{Gidal:1987bn,Gidal:1987bm}, the TPC/Two-Gamma~\cite{Aihara:1988bw,Aihara:1988uh}, and, more recently, by the L3~\cite{Achard:2001uu,Achard:2007hm} collaborations. With both measurements required to constrain the mixing angle $\theta_\Ax$ from the data, we will restrict our analysis to the L3 data, given that they are more accurate than the results from the preceding experiments.
The L3 analyses are based on the model of Ref.~\cite{Schuler:1997yw}, 
which assumes $\F_1(q_1^2,q_2^2) = 0$ for the first form factor from \autoref{eq:MDecomposition} and uses a dipole ansatz for $\F_2(q^2,0) = -\F_3(0,q^2)$, with
\beq\label{eq:FFSchuler}
	\F_\text{D}(q^2,0) = \frac{\F_\text{D}(0,0)}{(1-q^2/\Lambda_\text{D}^2)^2}.
\eeq
Under the assumption $B(f_1' \to K\bar{K}\pi) = 1$---which appears justified in light of the smallness of the other available channels~\cite{Zyla:2020zbs}---the measured parameters are
\begin{align}
	\widetilde{\Gamma}_{\gamma \gamma}^{f_1} &= 3.5(6)(5) \keV, & \Lambda_{f_1} &= 1.04(6)(5) \GeV, \notag \\
	\widetilde{\Gamma}_{\gamma \gamma}^{f_1'} &= 3.2(6)(7) \keV, & \Lambda_{f_1'} &= 0.926(72)(32) \GeV,
	\label{eq:L3}
\end{align}
where the quoted uncertainties are statistical and systematic, respectively. 
Employing the two-photon decay widths of the $f_1$ and $f_1'$, 
the mixing angle of the $J^{PC} = 1^{++}$ axial-vector nonet as defined in \autoref{eq:mixingAngle} can be extracted as follows: 
one calculates the coupling of the axial-vector mesons $f_1$ and $f_1'$ to two photons in analogy to \autoref{eq:axialIsoCouplings}, yielding
\beq\label{eq:axialPhotonCoupling}
	\Tr[\Phi_\mu^\Ax \Q \Q] = \frac{{f_1}_\mu(2 \sqrt{2} \cos\theta_\Ax + \sin\theta_\Ax) + {f'_1}_\mu(\cos\theta_\Ax - 2 \sqrt{2} \sin\theta_\Ax)}{3\sqrt{3}},
\eeq
so that using the formula for the equivalent two-photon decay width $\widetilde{\Gamma}_{\gamma \gamma}$, \autoref{eq:twoPhotonDecayWidth}, one finds
\beq
	\frac{\widetilde{\Gamma}_{\gamma \gamma}^{f_1}}{\widetilde{\Gamma}_{\gamma \gamma}^{f_1'}} = \frac{\Mf}{\Mfprime}  \left\lvert \frac{2 \sqrt{2} + \tan\theta_\Ax}{1 - 2\sqrt{2} \tan\theta_\Ax} \right\rvert^2
	=\frac{\Mf}{\Mfprime} \cot^2(\theta_\Ax - \theta_0),
\eeq
where $\theta_0 = \arcsin(1/3)$. 
Solving for $\theta_\Ax$ and inserting the above values for $\widetilde{\Gamma}_{\gamma \gamma}^{f_1}$ and $\widetilde{\Gamma}_{\gamma \gamma}^{f_1'}$, one finds the result of Refs.~\cite{Achard:2001uu,Achard:2007hm},
\beq
	\theta_\Ax = 62(5)^\circ,
\eeq
where the statistical and systematic uncertainties have been added in quadrature.

Next, the measurement of $\widetilde{\Gamma}_{\gamma \gamma}^{f_1}$ determines the normalization of the symmetric TFF, $\lvert C_{\sT}\rvert = \lvert \F_{\sT}^{I=1}(0,0)\rvert$ when neglecting the isoscalar contributions, according to \autoref{eq:twoPhotonDecayWidthNewBasis},
\beq\label{eq:Cs}
	\lvert C_{\sT}\rvert = 0.89(10).
\eeq
Taking into account the isoscalar contributions and, in particular, the ratios $R^\omega$ and $R^\phi$ of isoscalar to isovector couplings, \autoref{eq:SU3RatiosCouplings}, the normalization of the symmetric TFF becomes $\lvert \F_{\sT}^{I=1}(0,0) + \F_{\sT}^{I=0}(0,0)\rvert = (1 + R^\omega + R^\phi) \lvert C_{\sT}\rvert = 0.953(34) \lvert C_{\sT}\rvert$, resulting in
\beq\label{eq:CsIsoscalar}
	\lvert C_{\sT}\rvert = 0.93(11),
\eeq
which is slightly larger than \autoref{eq:Cs}, as expected from the negative ratio found in the estimate of \autoref{eq:ratioISIVContribution}. In the following, we will use \autoref{eq:CsIsoscalar} for the normalization of the symmetric TFF.

In addition, \autoref{eq:L3} determines the slope of $\F_2(q^2,0)$, based on the assumption of a dipole form. The asymptotic behavior matches onto 
\autoref{eq:FFDVSV} with~\cite{Hoferichter:2020lap}
\beq
\label{eq:scale_dipole}
 	F_{f_1}^\eff\Big|_\text{L3}=\frac{C_\sT \Lambda_{f_1}^4}{6\Mf^3}=86(28)\MeV,
\eeq
below both the LCSR estimate, \autoref{eq:effectiveDecayConstantLCSR}, and the effective decay constant implied by VMD, \autoref{eq:effectiveDecayConstantVMDIsoscalar}, and close to the scale derived from the singly-virtual behavior of the extended VMD representation, \autoref{eq:effectiveDecayConstantExtendedVMDIsoscalar}.\footnote{Matching the effective decay constant in the doubly-virtual direction to the 
quark model of Ref.~\cite{Schuler:1997yw} instead, one would obtain 
$F_{f_1}^\eff\big|_\text{L3}=C_\sT \Lambda_{f_1}^4/(4\Mf^3)=129(42)\MeV$, closer to \autoref{eq:effectiveDecayConstantVMDIsoscalar}
and \autoref{eq:effectiveDecayConstantLCSR}. This reflects the factor $3/2$ by which the relative coefficients of the singly- and doubly-virtual limits differ between the quark model and the \LN{BL} prediction~\cite{Hoferichter:2020lap}.}
The  uncertainty in \autoref{eq:scale_dipole} is mainly driven by the dipole parameter $\Lambda_\text{D}$.
 In fact, most of the data points measured by the L3 collaboration lie well below the obtained dipole scale, in such a way that the data should be similarly well described by a monopole ansatz,
\beq
	\F_\text{M}(q^2,0) = \frac{\F_\text{M}(0,0)}{1-q^2/\Lambda_\text{M}^2},
\eeq
when adjusting the slopes of the parameterizations to coincide at $q^2 = 0$. The corresponding monopole scale 
becomes
\beq
	\Lambda_\text{M} = \frac{\Lambda_\text{D}}{\sqrt{2}} = 0.74(6) \GeV \approx \Mrho,
\eeq
thus providing strong motivation for the VMD representation constructed in \autoref{sec:vmd}.

To constrain the singly-virtual VMD limits further, we need to match the L3 parameterization onto the full description of the $e^+e^-\to e^+e^- f_1$ cross section, which depends on the combination~\cite{Hoferichter:2020lap}
\begin{align}
\label{L3_matching}
 \bigg|\bigg(1-\frac{q^2}{\Mf^2}\bigg)\F_1(q^2,0)-\frac{q^2}{\Mf^2}\F_2(q^2,0)\bigg|^2-\frac{2q^2}{\Mf^2}\big|\F_2(q^2,0)\big|^2
 =\frac{-q^2}{\Mf^2}\bigg(2-\frac{q^2}{\Mf^2}\bigg)|\F_\text{D}(q^2,0)|^2.
\end{align}
The normalization agrees by construction, while matching the slopes at $q^2 = 0$ leads to 
\begin{align}
 \frac{2}{\Lambda_\text{D}^2}&=\frac{1}{N_{\omega\phi}}\Bigg[\frac{1}{\Mrho^2}+\frac{R^\omega}{\Momega^2}+\frac{R^\phi}{\Mphi^2}+\frac{\Mrhoprime^2-\Mrho^2}{\Mrho^2\Mrhoprime^2}\frac{C_{\aT_1}+C_{\aT_2}}{C_\sT}-\frac{\Mf^2(\Mrhoprime^2-\Mrho^2)^2}{\Mrho^4\Mrhoprime^4 N_{\omega\phi}}\bigg(\frac{C_{\aT_1}}{C_{\sT}}\bigg)^2\Bigg]
\end{align}
for the minimal VMD representation, and 
\begin{align}
 \frac{2}{\Lambda_\text{D}^2}&=\frac{1}{N_{\omega\phi}}\Bigg[\frac{1}{\Mrho^2}+\frac{1}{\Mrhoprime^2}+\frac{R^\omega}{\Momega^2}+\frac{R^\phi}{\Mphi^2}+\frac{\Mrhoprime^2-\Mrho^2}{\Mrho^2\Mrhoprime^2}\frac{C_{\aT_1}}{C_\sT}-\frac{\Mf^2(\Mrhoprime^2-\Mrho^2)^2}{\Mrho^4\Mrhoprime^4 N_{\omega\phi}}\bigg(\frac{C_{\aT_1}}{C_{\sT}}\bigg)^2\Bigg]
\end{align}
for the extended one. The factor $N_{\omega\phi}=1+R^\omega+R^\phi$ arises from accounting for the isoscalar terms in the normalization, see \autoref{eq:CsIsoscalar}.

\subsection[$f_1\to 4\pi$]{$\boldsymbol{f_1\to 4\pi}$}
\label{sec:4pi}

\begin{figure}[t]
	\centering
	\includegraphics{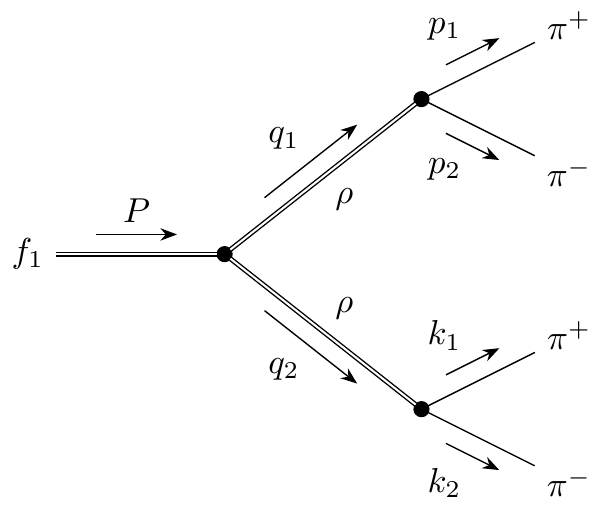}
	\qquad \qquad
	\includegraphics{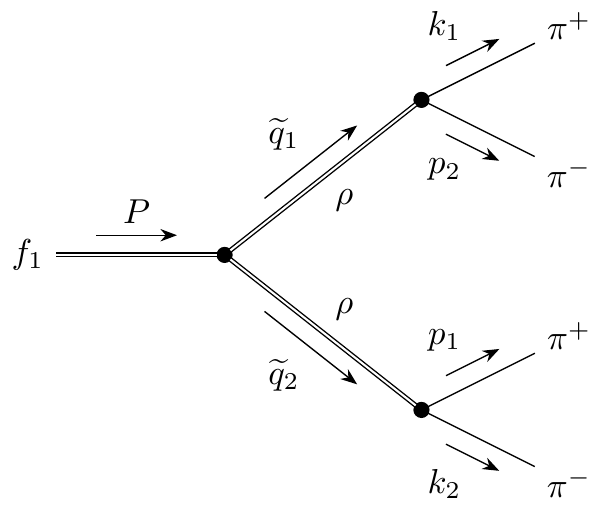}
	\caption{\LN{Feynman} diagrams for $f_1 \to \pi^+ \pi^- \pi^+ \pi^-$ via two $\rho$ mesons. Since the two $\pi^+$ and $\pi^-$ are respectively indistinguishable, there exist two contributions (\textit{left} and \textit{right}).}
	\label{fig:f4pi}
\end{figure}

In addition to $e^+e^-\to e^+e^-f_1$, the normalization of the symmetric TFF would be accessible in the process $f_1\to\rho\rho\to4\pi$ if the $\rho$ intermediate states largely saturated the decay within regions of the phase space reasonably close to their mass shell. In fact, up to corrections due to the two-pion channel $\rho'\to\pi^+\pi^-$, such an identification appears natural within the VMD approach.
In constructing an amplitude $\M(f_1 \to \pi^+ \pi^- \pi^+ \pi^-)$, which can be obtained by means of $\M(f_1 \to {\rho^0}^* {\rho^0}^*)$ and the $\rho \pi \pi$ coupling dictated by \autoref{eq:lagrangianRhoPiPi}, 
only the symmetric form factor $\F_{\sT}^{I=1}(q_1^2,q_2^2)$ and the symmetric \LN{Lorentz} structure $T_{\sT}^{\mu \nu \alpha}(q_1,q_2)$ are relevant under the above assumptions and when restricting to the minimal VMD parameterization.
More specifically, we use the amplitude $\M(f_1 \to \gamma^* \gamma^*)$, in the decomposition of \autoref{eq:MDecompositionNewBasis}, and remove the external photons by dropping the relevant $\rho$-meson propagator poles and the factors of $e$, at the same time dividing by the $\rho \gamma$ coupling $\widetilde{g}_{\rho \gamma}$, \autoref{eq:couplingRhoGammaVMD}, for each cut photon. 
In doing so, we arrive at
\begin{align}\label{eq:f1RhoRhoAmplitude}
	\M(f_1 \to {\rho^0}^* {\rho^0}^*)
	&= \frac{C_{f \rho \rho}}{2} \eps^*_\mu(q_1) \eps^*_\nu(q_2) \eps_\alpha(P)\notag\\ &\times\left[{q_1}_\beta {q_2}_\gamma\left(\eps^{\alpha \nu \beta \gamma} q_1^\mu - \eps^{\alpha \mu \beta \gamma} q_2^\nu \right) + \eps^{\alpha \mu \nu \beta} ({q_2}_\beta q_1^2 - {q_1}_\beta q_2^2)\right],
\end{align}
where we defined $C_{f \rho \rho} = C_\sT \Mrho^4/(\Mf^2 \widetilde{g}_{\rho \gamma}^2)$. 
Observing that there exist two diagrams for $f_1 \to \pi^+ \pi^- \pi^+ \pi^-$ due to the indistinguishability of the two $\pi^+$ and $\pi^-$---see \autoref{fig:f4pi}---we use the $\rho \pi \pi$ coupling as prescribed by \autoref{eq:lagrangianRhoPiPi} to deduce
\begin{align}\label{eq:f14piAmplitude}
	\M(f_1 \to \pi^+ \pi^- \pi^+ \pi^-)
	&= \frac{2 C_{f \rho \rho} g_{\rho \pi \pi}^2}{\big(q_1^2 - \Mrho^2 + \iu \sqrt{q_1^2} \, \Gammarho(q_1^2)\big)\big(q_2^2 - \Mrho^2 + \iu \sqrt{q_2^2} \, \Gammarho(q_2^2)\big)} \eps_\alpha(P) \eps^{\alpha \mu \nu \beta} \notag \\
	&\hspace{-1.2cm}\times \Big[\left(\Mpi^2 + (p_1 \cdot p_2)\right) {k_1}_\beta {k_2}_\nu (p_2 - p_1)_\mu - \left(\Mpi^2 + (k_1 \cdot k_2)\right) {p_1}_\beta {p_2}_\mu (k_2 - k_1)_\nu\Big]\notag\\
	&+ (p_1 \leftrightarrow k_1).
\end{align}
Here, the momenta are defined as in \autoref{fig:f4pi} and the pions are on shell, $p_{1/2}^2 = \Mpi^2 = k_{1/2}^2$.

Given this amplitude, one can calculate the decay width and thus branching ratio via the four-body phase-space integration of
\beq
	\mathrm{d}\Gamma(f_1 \to \pi^+\pi^-\pi^+\pi^-) = \frac{1}{2\Mf}\left\lvert\M(f_1 \to \pi^+\pi^-\pi^+\pi^-)\right\rvert^2 \mathrm{d}\Phi_4(P,p_1,p_2,k_1,k_2).
\eeq
We use the differential four-body phase space $\mathrm{d}\Phi_4(P,p_1,p_2,k_1,k_2)$ in the form~\cite{Zyla:2020zbs}
\beq\label{eq:phase_space_recursion}
	\mathrm{d}\Phi_4(P,p_1,p_2,k_1,k_2) = \mathrm{d}\Phi_2(q_1;p_1,p_2) \mathrm{d}\Phi_2(q_2;k_1,k_2) \mathrm{d}\Phi_2(P;q_1,q_2) \frac{\mathrm{d}q_1^2}{2\pi} \frac{\mathrm{d}q_2^2}{2\pi},
\eeq
where $\mathrm{d}\Phi_2(P;q_1,q_2)$, $\mathrm{d}\Phi_2(q_1;p_1,p_2)$, and $\mathrm{d}\Phi_2(q_2;k_1,k_2)$ are the respective two-body phase spaces of the subsystems $\{\rho(q_1)\rho(q_2)\}$, $\{\pi^+(p_1)\pi^-(p_2)\}$, and $\{\pi^+(k_1)\pi^-(k_2)\}$. Since the integration volumes of the phase spaces are \LN{Lorentz} invariant, each two-body phase space can be evaluated in the corresponding center-of-mass frame and we have to perform an explicit \LN{Lorentz} transformation from the center-of-mass frames of $\{\pi^+(p_1)\pi^-(p_2)\}$ and $\{\pi^+(k_1)\pi^-(k_2)\}$ into the one of $\{\rho(q_1)\rho(q_2)\}$ in order to evaluate scalar products of the kind $(p_i \cdot k_j)$, $i,j \in \{1,2\}$, appearing in $\lvert\M(f_1 \to \pi^+\pi^-\pi^+\pi^-)\rvert^2$---see, \Lat{e.g.}, Ref.~\cite{Guo:2011ir} for more details.\footnote{While two diagrams contribute, as shown in \autoref{fig:f4pi}, the decay rate involves an additional symmetry factor of $S=1/(2!)^2$ because of the two pairs of indistinguishable particles in the final state.}
We perform the phase space integration numerically with the \Lat{Cuhre} algorithm from the \Lat{Cuba} library~\cite{Hahn:2004fe}, where the energy-dependent width $\Gammarho(q^2)$ is as specified in \autoref{eq:energyDependentWidthRho_barrier_factors}, and obtain~\cite{Zanke:thesis}   
\beq
	\Gamma(f_1 \to \pi^+\pi^-\pi^+\pi^-) = \lvert C_\sT\rvert^2 \lvert g_{\rho \gamma}\rvert^4 \lvert g_{\rho \pi \pi}\rvert^4 \times 0.63\times 10^{-10}\GeV.
\eeq
Combining the above result with the values $\lvert g_{\rho \gamma}\rvert = 4.96$ and $\lvert g_{\rho \pi \pi} \rvert = 5.98$, \autoref{eq:couplingRhoGamma} and \autoref{eq:couplingRhoPiPi}, we find the branching ratio to be given by
\beq
	B(f_1 \to \pi^+ \pi^- \pi^+ \pi^-) = |C_\sT|^2 \times 0.215(10) \perc.
\eeq
The comparison with the experimental ratio $B(f_1 \to \pi^+ \pi^- \pi^+ \pi^-) = 10.9(6) \perc$~\cite{Zyla:2020zbs} yields
\beq
	\lvert C_\sT \rvert = 7.1(3), 
\eeq
in serious disagreement with \autoref{eq:CsIsoscalar}.

Including $\rho'$ contributions within the minimal VMD representation, there are four additional diagrams as compared to \autoref{fig:f4pi} and the corresponding master formula  takes the form
\begin{align}
\label{f14pi_minimal}
	 \Gamma(f_1 \to \pi^+\pi^-\pi^+\pi^-) = \lvert g_{\rho\gamma}\rvert^4 \lvert g_{\rho\pi\pi}\rvert^4 &\Big[C_{\aT_1}^2 \kappa^2 \Gamma_{\aT_1} + C_{\aT_2}^2 \kappa^2 \Gamma_{\aT_2} + C_{\sT}^2 \Gamma_{\sT}^{(1)} + C_{\aT_1} C_{\aT_2} \kappa^2 \Gamma_{\aT_1,\aT_2} \notag\\
 	& + C_{\aT_1} C_{\sT} \kappa \Gamma_{\aT_1,\sT}^{(1)} + C_{\aT_2} C_{\sT} \kappa \Gamma_{\aT_2,\sT}^{(1)} \Big],
\end{align}
where 
\beq
\label{rho'_kappa}
	\kappa = \frac{\Mrhoprime^2}{\Mrho^2} \frac{\widetilde{g}_{\rho \gamma}}{\widetilde{g}_{\rho' \gamma}} \frac{g_{\rho' \pi \pi}}{g_{\rho \pi \pi}} = \frac{g_{\rho'\gamma}g_{\rho'\pi\pi}}{g_{\rho\gamma}g_{\rho\pi\pi}} \approx -0.7,
\eeq
see \autoref{def_kappa}, and 
the numerical values of the defined decay rates are collected in \autoref{tab:Constants4pi}.
For the extended VMD representation, yet two additional diagrams have to be taken into account, resulting in the master formula
\begin{align}
\label{f14pi_extended} 
 	\Gamma(f_1 \to \pi^+\pi^-\pi^+\pi^-) &= \lvert g_{\rho\gamma}\rvert^4 \lvert g_{\rho\pi\pi}\rvert^4 \notag \\
	&\hspace{-0.8cm} \times \bigg[C_{\aT_1}^2 \kappa^2 \Gamma_{\aT_1} + C_{\aT_2}^2 \kappa^2 \Gamma_{\aT_2} + C_{\sT}^2 \Big[(1-\eps_1-\eps_2)^2 \Gamma_{\sT}^{(1)} + \eps_1^2 \kappa^2 \Gamma_{\sT}^{(2)} + \eps_2^2 \kappa^4 \Gamma_{\sT}^{(3)} \notag\\
	&\hspace{-0.3cm} + (1-\eps_1-\eps_2) \eps_1 \kappa \Gamma_{\sT}^{(4)} + (1-\eps_1-\eps_2) \eps_2 \kappa^2 \Gamma_{\sT}^{(5)} + \eps_1 \eps_2 \kappa^3 \Gamma_{\sT}^{(6)}\Big] \notag \\
	&\hspace{-0.3cm} + C_{\aT_1} C_{\aT_2} \kappa^2 \Gamma_{\aT_1,\aT_2} + C_{\aT_1} C_{\sT} \Big[(1-\eps_1-\eps_2) \kappa \Gamma_{\aT_1,\sT}^{(1)} + \eps_1 \kappa^2 \Gamma_{\aT_1,\sT}^{(2)} + \eps_2 \kappa^3 \Gamma_{\aT_1,\sT}^{(3)} \Big] \notag \\
	&\hspace{-0.3cm} + C_{\aT_2} C_{\sT} \Big[ (1-\eps_1-\eps_2) \kappa \Gamma_{\aT_2,\sT}^{(1)} + \eps_1 \kappa^2 \Gamma_{\aT_2,\sT}^{(2)} + \eps_2 \kappa^3 \Gamma_{\aT_2,\sT}^{(3)} \Big] \bigg],
\end{align}
see \autoref{tab:Constants4pi} for the numerical values of the decay rates. The numerical pattern shows that even though the coupling $\kappa$ itself is $\Order(1)$, $\rho'$ contributions are significantly suppressed, both due to the propagators in \autoref{eq:f14piAmplitude} and because the $\rho'$ can never be on shell in the available phase space. For the solutions of the global phenomenological analysis in \autoref{sec:pheno}, we find that the interference effects tend to even slightly reduce the branching ratio in the minimal VMD case, while the large values of $(1-\eps_1-\eps_2)$ in the extended VMD fits can increase $B(f_1\to\pi^+\pi^-\pi^+\pi^-)$ to the level of $1\perc$, still far below the experimental value. 

\begin{table}[t]
	\centering
	\begin{tabular}{ r  r  r  r r r}
	\toprule
	$\Gamma_{\sT}^{(1)}$ & $\Gamma_{\sT}^{(2)}$ & $\Gamma_{\sT}^{(3)}$ & $\Gamma_{\sT}^{(4)}$
	& $\Gamma_{\sT}^{(5)}$ & $\Gamma_{\sT}^{(6)}$\\
	$0.63$ & $0.01$ & $0.00$ & $0.16$ & $0.01$ & $0.00$\\\midrule
	$\Gamma_{\aT_1}$ & $\Gamma_{\aT_2}$ & $\Gamma_{\aT_1,\aT_2}$ & & & \\
	$0.02$ & $0.18$ & $-0.06$ & & &\\\midrule
	$\Gamma_{\aT_1,\sT}^{(1)}$ & $\Gamma_{\aT_2,\sT}^{(1)}$ &
	$\Gamma_{\aT_1,\sT}^{(2)}$ & $\Gamma_{\aT_2,\sT}^{(2)}$ &
	$\Gamma_{\aT_1,\sT}^{(3)}$ & $\Gamma_{\aT_2,\sT}^{(3)}$\\
	$-0.12$ & $0.54$ & $-0.01$ & $0.05$ & $0.00$ & $0.00$\\
	\bottomrule
	\end{tabular}
	\caption{Decay rates needed for the evaluation of \autoref{f14pi_minimal} and \autoref{f14pi_extended}, all in units of $10^{-10}\GeV$. The $\rho$ and $\rho'$ spectral functions are evaluated with \autoref{eq:energyDependentWidthRho_barrier_factors} and \autoref{eq:energyDependentWidthRhoPrimeAlternative}, respectively. The latter variant is chosen for consistency with the estimate of the $\rho'\to\pi\pi$ coupling via \autoref{rho'_kappa}, see \autoref{appx:SU3}.}
	\label{tab:Constants4pi} 
\end{table}

The reason for this incompatibility can be understood as follows. The available phase space prohibits the two $\rho$ mesons from being simultaneously on-shell, and the corresponding loss of resonance enhancement for two intermediate $\rho$ mesons implies that other decay mechanisms become more important. A candidate for such a mechanism is given by the decay
$f_1 \to a_1 \pi \to \rho \pi \pi \to 4\pi$, see \autoref{appx:f14pi} for an estimate of this decay channel. From this analysis, we indeed infer that the intermediate state $a_1 \pi$ likely saturates the decay width to a large extent, so that we have to conclude that the decay $f_1\to 4\pi$ does not allow one to extract further information on the $f_1$ TFFs.
We will thus disregard this input entirely and adopt \autoref{eq:CsIsoscalar} for the symmetric normalization. With $C_{\aT_1}$, $C_{\aT_2}$, and $C_{\sT}$ all real couplings, we will further fix the global sign by demanding that $C_{\sT}$ be positive,
\beq\label{eq:CouplingCs}
	C_{\sT} = 0.93(11). 
\eeq


\subsection[$f_1\to \rho\gamma$]{$\boldsymbol{f_1\to \rho\gamma}$}
\label{sec:rhogamma}

\begin{figure}[t]
	\centering
	\includegraphics{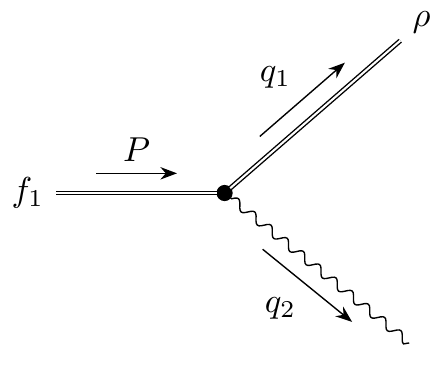}
	\caption{\LN{Feynman} diagram for $f_1 \to \rho \gamma$ consistent with $\M(f_1 \to \gamma^* \gamma^*)$.}
	\label{fig:fRhoPhoton}
\end{figure}

The construction of the amplitude for $f_1 \to \rho \gamma$ proceeds along the same lines as for $f_1\to 4\pi$, via
 $\M(f_1 \to \gamma^* \gamma^*)$, either by using the minimal or the extended VMD parameterization. By definition, this decay channel only probes the isovector contribution, up to negligible isospin-breaking effects. 

\begin{figure}[t]
	\centering
	\includegraphics[width=0.5\textwidth]{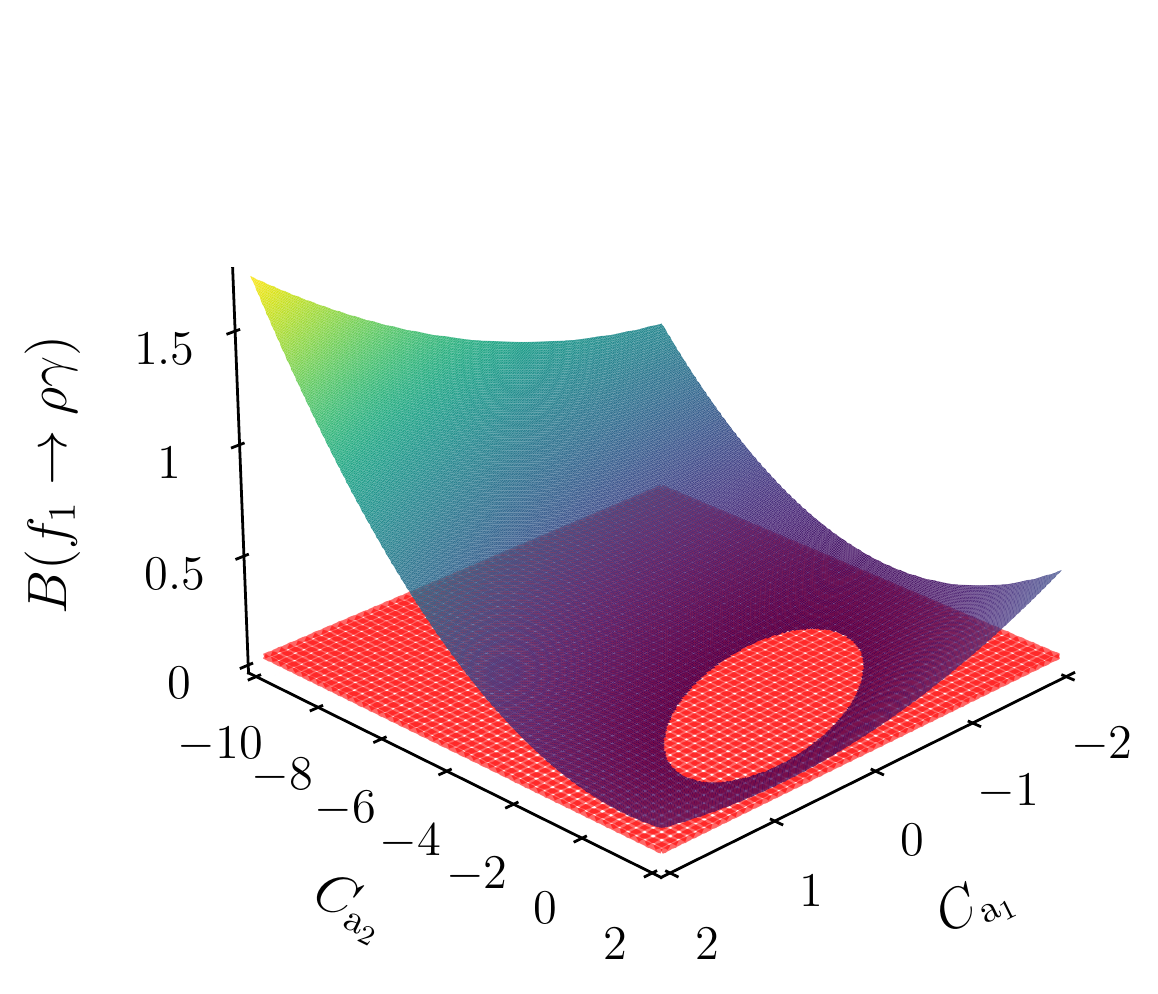}
	\caption{Surface plot of $B(f_1 \to \rho \gamma)$ (\textit{blue-yellow textured}), \autoref{eq:BfRhoGamma}, using the central value of $C_\sT = 0.93(11)$, \autoref{eq:CouplingCs}, together with the central value of $B(f_1 \to \rho \gamma) = 4.2(1.0) \perc$ (\textit{red}), see \autoref{sec:pheno}.  } 
	\label{fig:B_f_rho_gamma}
\end{figure}

For the amplitude $\M(f_1 \to \rho \gamma)$, we then proceed as stated above, starting with the minimal VMD ansatz, and consider the $\rho$ meson and photon on shell, $q_1^2 = \Mrho^2$, $q_2^2 = 0$, and $\eps^*(q_1) \cdot q_1 = 0 = \eps^*(q_2) \cdot q_2$, which also implies $\Gammarho(q_2^2=0) = 0 = \Gammarhoprime(q_2^2=0)$ according to \autoref{eq:energyDependentWidthRho}--\autoref{eq:energyDependentWidthRho_barrier_factors}. The corresponding diagram is depicted in \autoref{fig:fRhoPhoton} and we find 
\begin{align}\label{eq:f1RhoPhotonAmplitude}
	\M(f_1 \to \rho \gamma) &= C_{f \rho \gamma} \eps_\mu^*(q_1) \eps_\nu^*(q_2) \eps_\alpha(P) \notag \\
	&\times \left[C_{\aT_1} \eps^{\mu \nu \beta \gamma} {q_1}_\beta {q_2}_\gamma (q_1^\alpha - q_2^\alpha) + \frac{\Mrho^2}{2} C_{\aT_2} \eps^{\alpha \mu \nu \beta} {q_2}_\beta + \frac{\Mrho^2}{2} C_{\sT} \eps^{\alpha \mu \nu \beta} {q_2}_\beta \right],
\end{align}
where we introduced $C_{f \rho \gamma} = e \Mrho^2/(\widetilde{g}_{\rho \gamma} \Mf^2)$.
The branching ratio of the decay is given by
\beq\label{eq:BfRhoGamma}
	B(f_1 \to \rho \gamma) = \frac{B_1 C_{\aT_1}^2 + B_2 \big(C_{\aT_2}^2 + C_{\sT}^2 + 2 C_{\aT_2} C_{\sT}\big) - B_3 \big(C_{\aT_1} C_{\aT_2} + C_{\aT_1} C_{\sT}\big)}{\Gamma_f},
\eeq
where---as throughout this work---the coupling constants are assumed to be purely real and we defined the coefficients
\begin{align}
	B_1 &= \frac{\alpha |g_{\rho \gamma}|^2 \big(\Mf^2 - \Mrho^2\big)^5}{24 \Mf^9}, &B_2 &= \frac{\alpha |g_{\rho \gamma}|^2 \Mrho^2 \big(\Mf^2 - \Mrho^2\big)^3 \big(\Mf^2 + \Mrho^2\big)}{96 \Mf^9}, \notag \\
	B_3 &= \frac{\alpha |g_{\rho \gamma}|^2 \Mrho^2 \big(\Mf^2 - \Mrho^2\big)^4}{24 \Mf^9}.
\end{align}
As depicted in \autoref{fig:B_f_rho_gamma}, the solution of \autoref{eq:BfRhoGamma} in terms of the unknown couplings $C_{\aT_1}$ and $C_{\aT_2}$ represents an ellipse, where we used the central values of $C_\sT = 0.93(11)$ and $B(f_1 \to \rho \gamma) = 4.2(1.0) \perc$, see \autoref{sec:pheno}, to illustrate the cut surfaces. Although it is straightforward to actually solve \autoref{eq:BfRhoGamma} for such an equation, we refrain from doing so here since there is no unique solution anyway without further input. 

The equivalent amplitude in the extended VMD representation reads
\begin{align}\label{eq:f1RhoPhotonAmplitudeExtendedVMD}
	\widetilde{\M}(f_1 \to \rho \gamma) &= C_{f \rho \gamma} \eps_\mu^*(q_1) \eps_\nu^*(q_2) \eps_\alpha(P) \\
	&\hspace{-0.8cm}\times \left[C_{\aT_1} \eps^{\mu \nu \beta \gamma} {q_1}_\beta {q_2}_\gamma (q_1^\alpha - q_2^\alpha) + \frac{\Mrho^2}{2} C_{\aT_2} \eps^{\alpha \mu \nu \beta} {q_2}_\beta + \frac{\Mrho^2}{2} C_{\sT}\Big(1- \frac{\eps_1}{2} - \eps_2\Big) \eps^{\alpha \mu \nu \beta} {q_2}_\beta \right], \notag
\end{align}
the only difference compared to the minimal VMD parameterization being that $C_\sT \to \widetilde{C}_\sT = (1- \eps_1/2 - \eps_2) C_\sT$. Hence, the branching ratio given in \autoref{eq:BfRhoGamma} becomes
\beq\label{eq:BfRhoGammaExtendedVMD}
	\widetilde{B}(f_1 \to \rho \gamma) = \frac{B_1 C_{\aT_1}^2 + B_2 \big(C_{\aT_2}^2 + \widetilde{C}_{\sT}^2 + 2 C_{\aT_2} \widetilde{C}_{\sT}\big) - B_3 \big(C_{\aT_1} C_{\aT_2} + C_{\aT_1} \widetilde{C}_{\sT}\big)}{\Gamma_f},
\eeq
which, when inserting $\eps_1$ and $\eps_2$ from \autoref{sec:vmd_asmyptotics}, simplifies to
\beq
	\widetilde{B}(f_1 \to \rho \gamma) = \frac{B_1 C_{\aT_1}^2 + \widetilde{B}_2 C_{\sT}^2 - \widetilde{B}_3 C_{\aT_1} C_{\sT}}{\Gamma_f},
\eeq
where we defined the coefficients
\begin{align}
	\widetilde{B}_2 &= \frac{\Mrhoprime^4}{(\Mrhoprime^2 - \Mrho^2)^2} B_2, &
	\widetilde{B}_3 &= \frac{\Mrhoprime^2}{\Mrhoprime^2 - \Mrho^2} B_3.
\end{align}
In this variant, the dependence on $C_{\aT_2}$ thus disappears from the branching fraction,
which is a subtle consequence of the correlation between $C_{\aT_2}$ and $C_{\sT}$ imposed via
the singly-virtual high-energy behavior, see \autoref{eq:eps1VMD}.

\begin{figure}[t]
	\centering
	\includegraphics{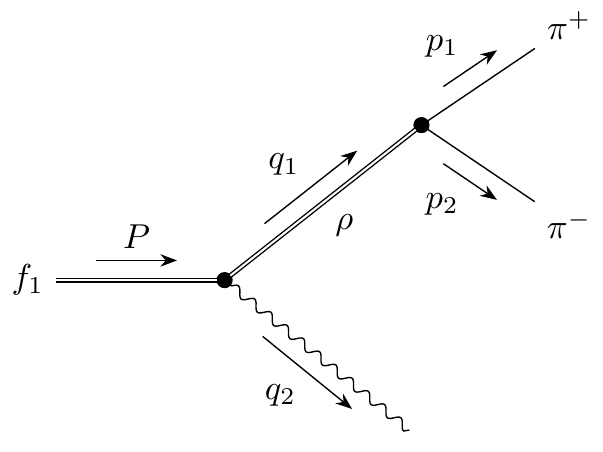}
	\caption{\LN{Feynman} diagram for $f_1 \to \rho \gamma \to \pi^+ \pi^- \gamma$ consistent with $\M(f_1 \to \gamma^* \gamma^*)$.}
	\label{fig:fPiPiPhoton}
\end{figure}

Another measured quantity of interest with regard to $f_1 \to \rho \gamma$ is the ratio of the $\rho$-meson's helicity amplitudes in its rest frame, which is accessible through the subsequent decay $\rho \to \pi^+ \pi^-$. 
In a similar manner to how we obtained the $f_1 \to \rho \gamma$ amplitudes in \autoref{eq:f1RhoPhotonAmplitude} and \autoref{eq:f1RhoPhotonAmplitudeExtendedVMD},
we can construct an amplitude for $f_1 \to \rho \gamma \to \pi^+ \pi^- \gamma$, where we indeed consider the subsequent decay of an on-shell $\rho$ meson and furthermore use the $\rho \pi \pi$ coupling given by \autoref{eq:lagrangianRhoPiPi}; the process is depicted in \autoref{fig:fPiPiPhoton}.

Imposing $q_1^2 = \Mrho^2$, thus also $\Gammarho(q_1^2 = \Mrho^2) = \Gammarho$ according to \autoref{eq:energyDependentWidthRho_barrier_factors}, $q_2^2 = 0 = \eps^*(q_2) \cdot q_2$, and $p_1^2 = \Mpi^2 = p_2^2$, we find
\begin{align}
	\M(f_1 \to \rho \gamma \to \pi^+ \pi^- \gamma) &= \frac{C_{f \rho \gamma} g_{\rho \pi \pi}}{\Mrho \Gammarho} \eps^*_\nu(q_2) \eps_\alpha(P) (p_2 - p_1)_\mu \\
	&\times \Big[C_{\aT_1} \eps^{\mu \nu \beta \gamma} {q_1}_\beta {q_2}_\gamma (q_1^\alpha - q_2^\alpha) + \frac{\Mrho^2}{2}C_{\aT_2}\eps^{\alpha \mu \nu \beta} {q_2}_\beta + \frac{\Mrho^2}{2}C_{\sT} \eps^{\alpha \mu \nu \beta} {q_2}_\beta \Big]\notag
\end{align}
with the minimal VMD parameterization, where the constant $C_{f \rho \gamma} = e \Mrho^2/(\widetilde{g}_{\rho \gamma} \Mf^2)$ is defined as in \autoref{eq:f1RhoPhotonAmplitude}. The equivalent expression $\widetilde{\M}(f_1 \to \rho \gamma \to \pi^+ \pi^- \gamma)$ in the extended VMD variant is obtained for $C_\sT \to \widetilde{C}_\sT = (1 - \eps_1/2 - \eps_2)C_\sT$. Transforming into the rest frame of the $\rho$ meson, one finds the spin-averaged amplitude squared to be of the form
\beq
	\left|\M(f_1 \to \rho \gamma \to \pi^+ \pi^- \gamma)\right|^2 = \mathrm{M}_{\text{TT}} \sin^2{\theta_{\pi^+\gamma}} + \mathrm{M}_{\text{LL}} \cos^2{\theta_{\pi^+\gamma}},
\eeq
where $\theta_{\pi^+\gamma}$ is the angle between the final-state $\pi^+$ and photon and
\beq\label{eq:ratioHelicityAmplitudes}
	r_{\rho \gamma} = \frac{\mathrm{M}_{\text{LL}}}{\mathrm{M}_{\text{TT}}} = \frac{2 \Mf^2 \Mrho^2}{\big[\Mrho^2 - 2 \big(\Mf^2 - \Mrho^2\big) C_{\aT_1}/\big(C_{\aT_2} + C_{\sT}\big)\big]^2}
\eeq
is the corresponding ratio of the longitudinal and transversal $\rho$-meson helicity amplitudes. In the extended VMD case, one again needs to replace $C_\sT \to \widetilde{C}_\sT = (1 - \eps_1/2 - \eps_2)C_\sT$, which then further simplifies to
\beq
	\widetilde{r}_{\rho \gamma} = \frac{\widetilde{\mathrm{M}}_{\text{LL}}}{\widetilde{\mathrm{M}}_{\text{TT}}} = \frac{2 \Mf^2 \Mrho^2 \Mrhoprime^4}{\big[\Mrho^2 \Mrhoprime^2 - 2 \big(\Mf^2 - \Mrho^2\big) (\Mrhoprime^2 - \Mrho^2) C_{\aT_1}/C_{\sT}\big]^2}
\eeq
when inserting $\eps_1$ and $\eps_2$ from \autoref{sec:vmd_asmyptotics}. The coupling $C_{\aT_2}$ therefore does not contribute to either $f_1\to\rho\gamma$ observable in the extended VMD ansatz. 

\begin{figure}[t]
	\centering
	\includegraphics[width=0.495\textwidth]{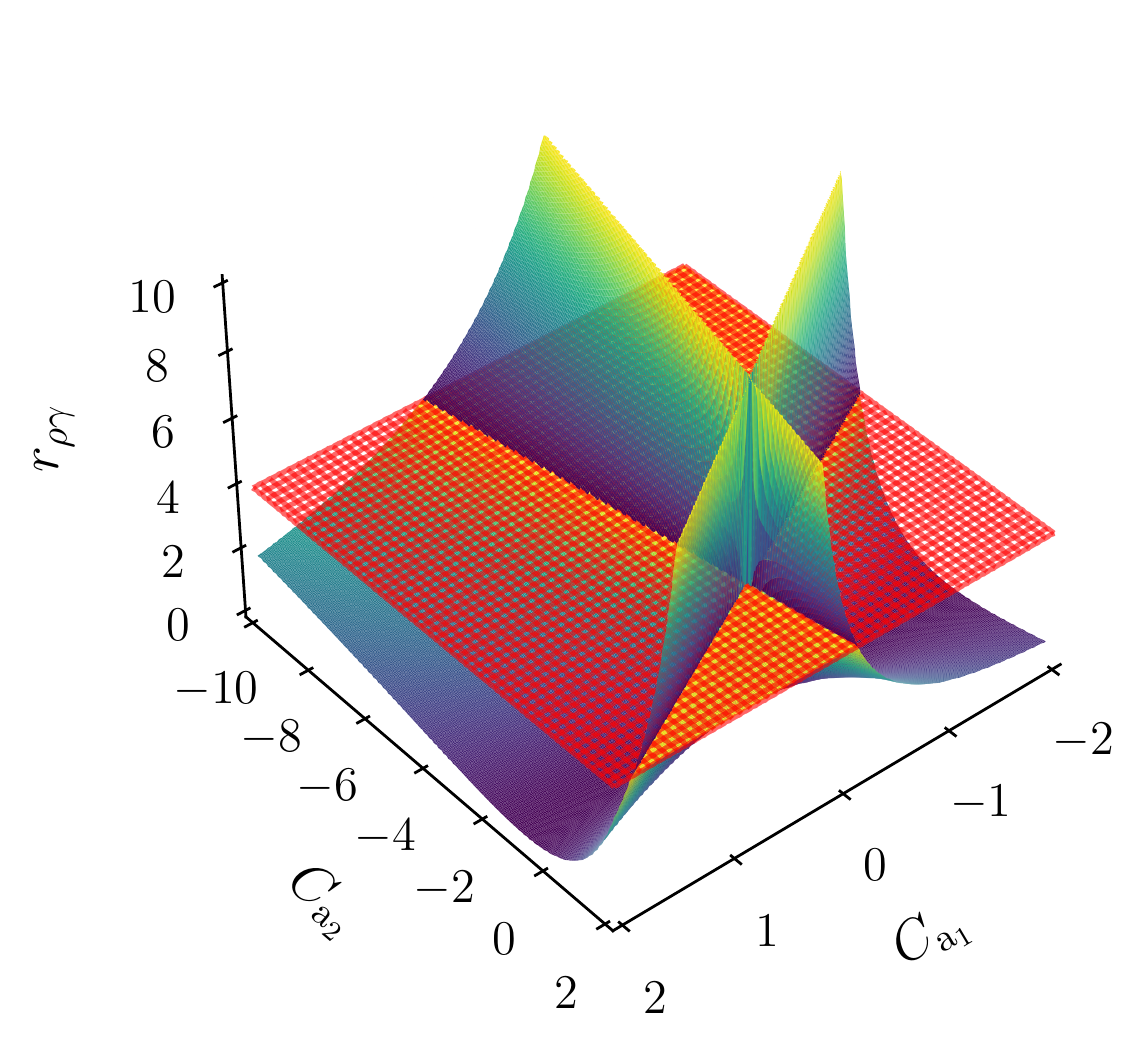}
	\hfill
	\includegraphics[width=0.495\textwidth]{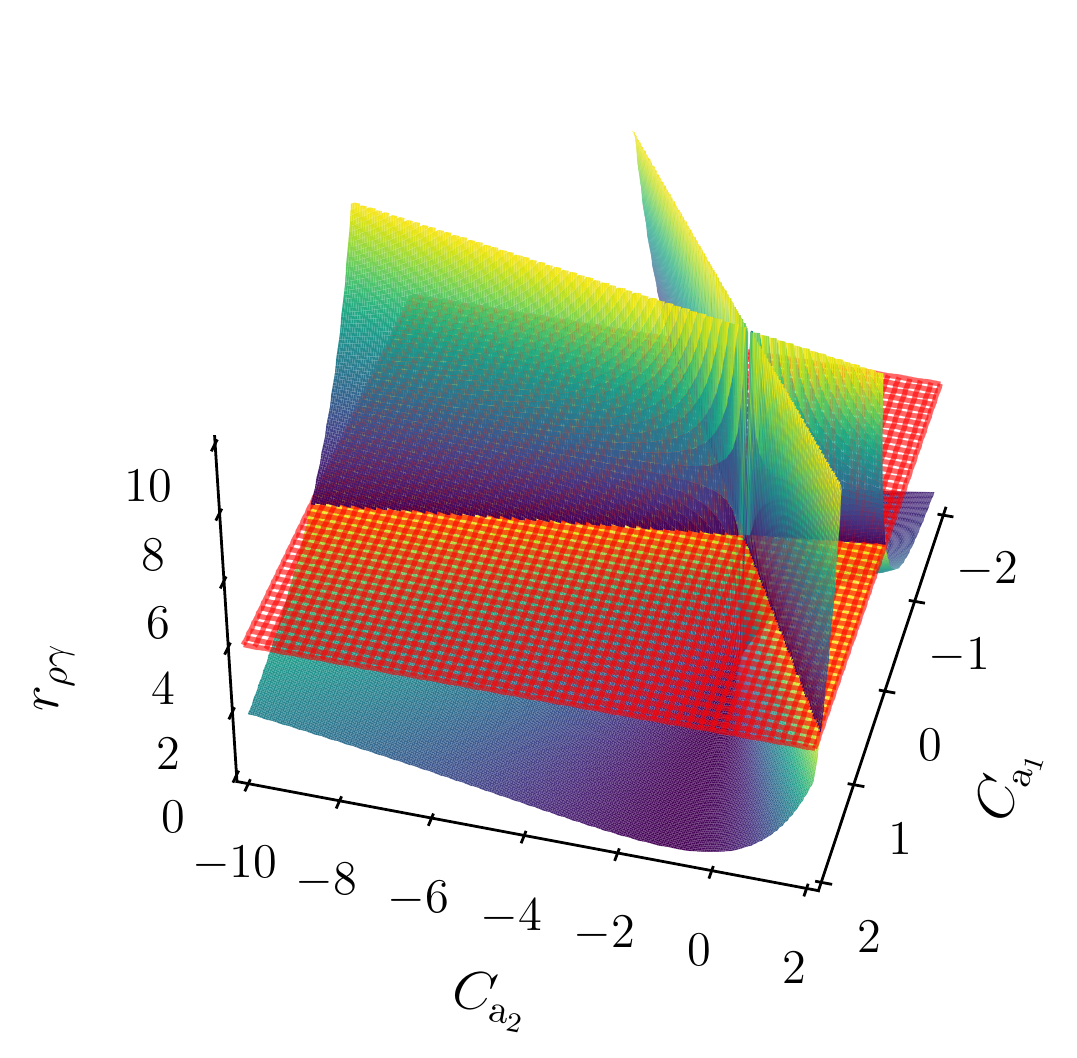}
	\caption{Surface plots of $r_{\rho \gamma}$ (\textit{blue-yellow textured}), \autoref{eq:ratioHelicityAmplitudes}, using the central value of $C_\sT = 0.93(11)$, \autoref{eq:CouplingCs}, together with the central value of the measurement $r_{\rho \gamma} = 3.9(1.3)$~\cite{Amelin:1994ii} (\textit{red}) from two different perspectives (\textit{left} and \textit{right}).}
	\label{fig:r_rho_gamma}
\end{figure}

The solution of \autoref{eq:ratioHelicityAmplitudes} in terms of the unknown couplings $C_{\aT_1}$ and $C_{\aT_2}$ is given by four unconnected straight lines, as apparent from \autoref{fig:r_rho_gamma}, where we used the central values of $C_\sT = 0.93(11)$, \autoref{eq:CouplingCs}, and the measurement $r_{\rho \gamma} = 3.9(0.9)(1.0) = 3.9(1.3)$~\cite{Amelin:1994ii} for illustration. Similar to the discussion regarding $B(f_1 \to \rho \gamma)$, we refrain from giving the explicit form of the solution here and postpone the phenomenological analysis to \autoref{sec:pheno}.

\subsection[$f_1\to \phi\gamma$ and $f_1\to \omega\gamma$]{$\boldsymbol{f_1\to \phi\gamma}$ and $\boldsymbol{f_1\to \omega\gamma}$}
\label{sec:phigamma_omegagamma}

The branching ratio of $f_1 \to \phi \gamma$ has been measured experimentally, $B(f_1 \to \phi \gamma) = 0.74(26)\times 10^{-3}$~\cite{Bityukov:1987bj,Zyla:2020zbs}, and thus allows for another consistency check of our VMD representations, in particular,  the
$\Uthree$ assumptions for the isoscalar TFFs. Similarly, we can predict the branching fraction for  
$f_1 \to \omega \gamma$ once all the parameters are determined, which could 
be confronted with potential future measurements.

In complete analogy to \autoref{sec:rhogamma}, we construct amplitudes for $f_1 \to V \gamma$, $V = \phi, \omega$, \Lat{i.e.},
\begin{align}\label{eq:f1PhiPhotonAmplitude}
	\M(f_1 \to V \gamma) &= C_{f V \gamma} \eps_\mu^*(q_1) \eps_\nu^*(q_2) \eps_\alpha(P) \\
	&\times \left[C^{V V'}_{\aT_1} \eps^{\mu \nu \beta \gamma} {q_1}_\beta {q_2}_\gamma (q_1^\alpha - q_2^\alpha) + \frac{M_V^2}{2} C^{V V'}_{\aT_2} \eps^{\alpha \mu \nu \beta} {q_2}_\beta + \frac{M_V^2}{2} C^{V V}_{\sT} \eps^{\alpha \mu \nu \beta} {q_2}_\beta \right],\notag
\end{align}
where we defined $C_{f V \gamma} = e M_V^2/(\widetilde{g}_{V \gamma} \Mf^2)$.
In terms of the ratio $R^V = R^\phi, R^\omega$ of isoscalar to isovector couplings, \autoref{eq:SU3RatiosCouplings}, the branching ratio of the decay is given by
\beq\label{eq:BfPhiGamma}
	B(f_1 \to V \gamma) = (R^V)^2 \frac{B^V_1 C_{\aT_1}^2 + B^V_2 \big(C_{\aT_2}^2 + C_{\sT}^2 + 2 C_{\aT_2} C_{\sT}\big) - B^V_3 \big(C_{\aT_1} C_{\aT_2} + C_{\aT_1} C_{\sT}\big)}{\Gamma_f},
\eeq
\Lat{cf.} \autoref{eq:BfRhoGamma}, where we defined the coefficients
\begin{align}
	B^V_1 &= \frac{\alpha |g_{V \gamma}|^2 \big(\Mf^2 - M_V^2\big)^5}{24 \Mf^9}, & B^V_2 &= \frac{\alpha |g_{V \gamma}|^2 M_V^2 \big(\Mf^2 - M_V^2\big)^3 \big(\Mf^2 + M_V^2\big)}{96 \Mf^9}, \notag \\
	B^V_3 &= \frac{\alpha |g_{V \gamma}|^2 M_V^2 \big(\Mf^2 - M_V^2\big)^4}{24 \Mf^9}.
\end{align}
The generalization to the extended VMD representation would be straightforward, once applied to the isoscalar sector.

\section{$\boldsymbol{f_1\to e^+e^-}$}
\label{sec:f1ee}

As the discussion in \autoref{sec:tree_level} shows, in general the constraints from $e^+e^-\to e^+e^- f_1$, $f_1\to 4\pi$, and $f_1\to \rho\gamma$ do not suffice to reliably determine all three free VMD parameters, with the branching fraction of $f_1\to4\pi$ not able to provide any additional input at all due to significant contamination from decay channels not related to the TFFs. In this way, the evidence for the decay $f_1\to e^+e^-$ reported by the SND collaboration~\cite{Achasov:2019wtd} is extremely interesting as future improved measurements of the decay have the potential to overconstrain the system of $C_{\aT_1}$, $C_{\aT_2}$, and $C_{\sT}$, as we will demonstrate in \autoref{sec:pheno}. In this section, we provide the required formalism to extract information on the $f_1$ TFFs from its decay into $e^+e^-$; \Lat{cf.} also Ref.~\cite{Rudenko:2017bel}. 

\begin{figure}[t]
	\centering
	\includegraphics{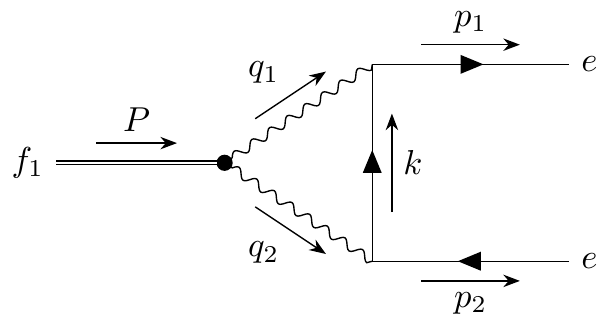}
	\caption{\LN{Feynman} diagram for the decay of the axial-vector meson $f_1$ into an electron--positron pair.}
	\label{fig:f1e+e-Decay}
\end{figure}

The \LN{Feynman} diagram for the one-loop process is depicted in \autoref{fig:f1e+e-Decay}. The general form of the amplitude is
\beq\label{eq:f1e+e-AmplitudeGeneral}
	\M(f_1 \to e^+ e^-) = e^4 \eps_\mu(P) \overline{u}^s(p_1) \gamma^\mu \gamma^5 A_1\big(\Mf^2,m_e^2 = 0,m_e^2 = 0\big) v^r(p_2),
\eeq
which implies
\beq
	\left|\M(f_1 \to e^+e^-)\right|^2 = \frac{4e^8 \Mf^2}{3} |A_1|^2
\eeq
for the spin-averaged amplitude squared and a decay width of
\beq\label{eq:f1e+e-DecayWidth}
	\Gamma(f_1 \to e^+ e^-) = \frac{64 \pi^3 \alpha^4 \Mf}{3} |A_1|^2.
\eeq
Here and in the following, the arguments of the reduced amplitude $A_1$ will be suppressed and we will work in the limit $m_e=0$. To extract 
$A_1$ from the full amplitude, we first consider the amplitude $\M(f_1 \to \gamma^* \gamma^*)$ and recast it into the more convenient form
\begin{align}
	\M(f_1 \to \gamma^* \gamma^*)
	&= \frac{\iu e^2}{\Mf^2} \eps^{\mu \nu \beta \gamma}\Big[\F_{\aT_1}(q_1^2,q_2^2)  \eps^*_\mu(q_1) \eps^*_\nu(q_2) \eps_\alpha(P) {q_1}_\beta {q_2}_\gamma (q_1^\alpha - q_2^\alpha) \\
	& - \frac{1}{2} \left[\F_{\aT_2}(q_1^2,q_2^2) + \F_{\sT}(q_1^2,q_2^2)\right]  \eps^*_\nu(q_2) \eps_\mu(P) {q_2}_\beta \left[{q_1}_\gamma \eps^*_\alpha(q_1) q_1^\alpha - \eps^*_\gamma(q_1) q_1^2\right] \notag \\
	& + \frac{1}{2} \left[\F_{\aT_2}(q_1^2,q_2^2) - \F_{\sT}(q_1^2,q_2^2)\right]  \eps^*_\nu(q_1) \eps_\mu(P) {q_1}_\beta \left[{q_2}_\gamma \eps^*_\alpha(q_2) q_2^\alpha - \eps^*_\gamma(q_2) q_2^2\right]\Big]. \notag
\end{align}
Inserting this amplitude into the QED loop, the full amplitude can be written as  
\begin{align}
\label{eq:f1e+e-AmplitudeFinal}
\M(f_1 \to e^+ e^-) &= \frac{4 \iu e^4}{\Mf^2} \eps_\alpha(P) P_\mu \overline{u}^s(p_1) \gamma_\beta \gamma^5 v^r(p_2) \int{\frac{\dfk}{(2\pi)^4} \frac{k^\mu k^\beta k^\alpha}{k^2 q_1^2 q_2^2} \F_{\aT_1}(q_1^2,q_2^2)} \\
	&+ \frac{\iu e^4}{\Mf^2} \eps_\beta(P) \overline{u}^s(p_1) \gamma_\mu \gamma^5 v^r(p_2)\notag\\
	&\quad\times \int{\frac{\dfk}{(2\pi)^4} \frac{k^\mu k^\beta}{k^2 q_1^2 q_2^2}} \Big[(q_2^2 - q_1^2) \F_{\aT_2}(q_1^2,q_2^2) 
	- (q_2^2 + q_1^2) \F_{\sT}(q_1^2,q_2^2)\Big] \notag \\
	&+ \frac{\iu e^4}{2\Mf^2}\eps_\mu(P) \overline{u}^s(p_1) \gamma^\mu \gamma^5 v^r(p_2)\notag\\
	&\quad\times \int{\frac{\dfk}{(2\pi)^4} \frac{\big[2q_1^2 q_2^2 + k^2(q_1^2+q_2^2)\big] \F_{\sT}(q_1^2,q_2^2) 
	+ k^2\big(q_1^2 - q_2^2\big) \F_{\aT_2}(q_1^2,q_2^2)}{k^2 q_1^2 q_2^2}},\notag
\end{align}
where we have used the on-shell condition for the fermions, neglected their masses, and written the loop integration in the most symmetric way. In particular, rewriting the TFF combinations as
\begin{align}
 (q_2^2 - q_1^2) \F_{\aT_2}(q_1^2,q_2^2) 
	- (q_2^2 + q_1^2) \F_{\sT}(q_1^2,q_2^2)&=-2q_1^2\F_2(q_1^2,q_2^2)+2q_2^2\F_3(q_1^2,q_2^2),\notag\\
	\big[2q_1^2 q_2^2 + k^2(q_1^2+q_2^2)\big] \F_{\sT}(q_1^2,q_2^2) 
	+ k^2\big(q_1^2 - q_2^2\big) \F_{\aT_2}(q_1^2,q_2^2)&=
	2(k^2+q_2^2)q_1^2\F_2(q_1^2,q_2^2)\notag\\
	&\quad-2(k^2+q_1^2)q_2^2 \F_3(q_1^2,q_2^2)
\end{align}
shows that the \LN{BL} limits that are not well-defined---see \autoref{eq:FFDVSV} and the subsequent comment---always appear suppressed by the respective on-shell virtuality, as expected from the form of the physical helicity amplitudes. We conclude that these integration regions will therefore be of minor importance.
Moreover, all remaining integrals are ultraviolet and infrared convergent by inspection of the parameterization of the form factors in \autoref{eq:VMDParametrization} and \autoref{extended_VMD}. However, inserting the (isovector) VMD expressions directly into the loop integral would produce unphysical imaginary parts, which can be avoided by using the spectral representations of \autoref{eq:FormFactorsSpectralRepresentation} and \autoref{extended_VMD_Spectral_Representation} instead, to ensure the correct analytic properties. 

We performed the remaining \LN{Passarino}--\LN{Veltman} reduction in two ways: first, in an automated way 
using \Lat{FeynCalc}~\cite{Mertig:1990an,Shtabovenko:2016sxi,Shtabovenko:2020gxv}, \Lat{FeynHelpers}~\cite{Shtabovenko:2016whf} (which collects \Lat{FIRE}~\cite{Smirnov:2014hma} and \Lat{Package-X}~\cite{Patel:2015tea}), and \Lat{LoopTools}~\cite{Hahn:1998yk}, and directly by introducing \LN{Feynman} parameters in \autoref{eq:f1e+e-AmplitudeFinal}. Decomposing the 
amplitude as
\begin{align}\label{eq:f1e+e-AmplitudeGeneralCalculated}
	\M(f_1 \to e^+ e^-) &= e^4 \eps_\mu(P) \overline{u}^s(p_1) \gamma^\mu \gamma^5 A_1 v^r(p_2), \\
	A_1 &= \big(D_1^{I=1} + D_1^{I=0}\big) C_{\aT_1} + \big(D_2^{I=1} + D_2^{I=0}\big) C_{\aT_2} + \big(D_3^{I=1} + D_3^{I=0}\big) C_{\sT} + D_\asym, \notag
\end{align}
the latter approach, in the minimal VMD ansatz, leads to the representation
\begin{align}
 D_{1/2}^{I=1}&=\frac{\Mrho^2\Mrhoprime^2}{16\pi^4 N_\aT \Mf^4}\int_{4\Mpi^2}^\infty\dx \int_{\sthr}^\infty\dy \int_0^1\dz\,\Im\big[P^\text{BW}_\rho(x)\big]\,\Im\big[P^\text{BW}_{\rho'}(y)\big]f_{1/2}(x,y,z,\Mf),\notag\\
 D_{3}^{I=1}&=\frac{\Mrho^4}{16\pi^4 N_\sT \Mf^4}\int_{4\Mpi^2}^\infty\dx \int_{4\Mpi^2}^\infty\dy \int_0^1\dz\,\Im\big[P^\text{BW}_\rho(x)\big]\,\Im\big[P^\text{BW}_{\rho}(y)\big]f_3(x,y,z,\Mf),
\end{align}
where
\begin{align}
f_1&=\frac{\bar x-\bar y}{\bar x \bar y}\bigg[\frac{\bar x z \log\frac{\Delta(\bar x,\bar y,z)}{-\bar x z}}{\Delta(\bar y,z)}-(1-z)\log\Delta(\bar x,\bar y,z)\bigg]\notag\\
&+\frac{z}{\bar x \bar y}\Big(\bar x\log(-\bar x z)-\bar y\log(-\bar y z)\Big)+\frac{(1-z)(1-3z)\log\frac{\Delta(\bar y,z)}{\Delta(\bar x,z)}}{2\bar x \bar y} - (x\leftrightarrow y),\notag\\
f_2&=\frac{\bar x-\bar y}{2\bar x \bar y}\bigg[\frac{\bar x z \log\frac{\Delta(\bar x,\bar y,z)}{-\bar x z}}{\Delta(\bar y,z)}+z\log\Delta(\bar x,\bar y,z)+\frac{1}{4}\bigg]
-\frac{3z-2}{2\bar x\bar y}\Big(\bar x \log\Delta(\bar x,z)-\bar y\log\Delta(\bar y,z)\Big)
\notag\\
&-\frac{z}{\bar x \bar y}\Big(\bar x\log(-\bar x z)-\bar y\log(-\bar y z)\Big)-\frac{(1-z)(1-3z)\log\frac{\Delta(\bar y,z)}{\Delta(\bar x,z)}}{2\bar x \bar y} - (x\leftrightarrow y),\notag\\
f_3&=-\frac{2z-1}{2\bar x\bar y \Delta(\bar y,z)^3}\bigg[2z^3\bar x^2\log\frac{\Delta(\bar x,\bar y,z)}{-\bar x z} \notag \\
&\qquad \qquad \qquad \quad \, \, +(1-z)\bar y \Delta(\bar y,z)\Big[2\bar x z+\Delta(\bar y,z)\Big(1-3z+2(\bar x+\bar y)\Big)\Big]\bigg]\notag\\
&+\frac{z(4z-2+\bar x(22z-5))}{4\bar x\bar y\Delta(\bar y,z)^2}\bigg[\bar x z \log\frac{\Delta(\bar x,\bar y,z)}{-\bar x z}+(1-z)\Delta(\bar y,z)\bigg]
-\frac{z^2(5+9\bar x)}{2\bar y}\frac{\log\frac{\Delta(\bar x,\bar y,z)}{-\bar x z}}{\Delta(\bar y,z)}\notag\\
&+\frac{(1-z)\Big[5(8z^2-7z+1)+18(2\bar x z+\bar y(1-z))\Big]\log \Delta(\bar x,\bar y,z)}{4\bar x\bar y}
+\frac{\bar x^2-1}{4\bar y}\log\frac{1-\bar x}{-\bar x}\notag\\
&+
\frac{\bar y^2-1}{4\bar x}\log\frac{1-\bar y}{-\bar y}
-\frac{3}{2\bar y}\log(-\bar x)-
\frac{3}{2\bar x}\log(-\bar y)
+\frac{7+19(\bar x+\bar y)+6(\bar x^2+\bar y^2)}{24\bar x\bar y},
\end{align}
with 
\beq
 \Delta(x,y,z)=z(1-z)-z x-(1-z) y,\qquad
 \Delta(x,z)=z -x,\qquad \bar x =\frac{x}{\Mf^2},
 \qquad \bar y =\frac{y}{\Mf^2},
\eeq
and the correct analytic continuation is defined by $x\to x-i\eps$, $y\to y-i\eps$ in the logarithms. Similar expressions apply
for the isoscalar parts and the extended VMD parameterization, the latter including the asymptotic contribution
\begin{align}
 D_\asym&=\frac{3F_{f_1}^\eff}{8\pi^2\Mf^3}\int_{\sm}^\infty\dx \int_0^1\dz\, f_\asym(x,z,\Mf),\\
 f_\asym&=\frac{z^4(1-z)^2}{2\bar x(\bar x -z)^4(z(1-z)-\bar x)^2}\bigg[(2-z)z^2\big(8-23z+27z^2-14z^3\big)\notag\\
 &-\bar x z\big(32-100 z+131 z^2-76z^3+14z^4\big)+\bar x^2\big(16-46z+51z^2-18z^3\big)\bigg]\notag\\
 &+\frac{z(1-z)}{2\bar x(\bar x -z)^3}\bigg[z^2\big(17-37z+37z^2-14z^3\big)+\bar x\big(2+11z-17z^2+10z^3\big)\notag\\
 &-3\bar x^2(2z+1)\bigg]
 -\frac{z^2(z(z+2)+2\bar x(5-2z)-9\bar x^2)}{2(\bar x-z)^4}\log\frac{z(1-z)-\bar x}{-\bar x z}.\notag
\end{align}
In all cases the numerical integration is performed with the \Lat{Cuhre} algorithm from the \Lat{Cuba} library~\cite{Hahn:2004fe}.

\begin{table}[t]
	\centering
	\begin{tabular}{ l  r  r}
	\toprule
	& $\Gammarhoprime^{(4\pi)}(q^2)$ & $\Gammarhoprime^{(\omega \pi,\pi \pi)}(q^2)$ \\\midrule
	$D_1^{\rho\rho'}\times 10^3$ & $(-0.126)^{+0.026}_{-0.031}+(-1.501)^{+0.099}_{-0.121}\,\iu$ & $(-0.173)^{+0.030}_{-0.034}+(-1.659)^{+0.107}_{-0.126}\,\iu$\\
	$D_2^{\rho\rho'}\times 10^3$ & $(-0.978)^{+0.030}_{-0.038} + 1.593^{-0.119}_{+0.144}\,\iu$ & $(-1.032)^{+0.030}_{-0.036}+1.755^{-0.129}_{+0.150}\,\iu$\\
	$D_3^{\rho\rho}\times 10^3$ & \multicolumn{2}{c}{$3.189 + 2.338\,\iu$}\\
	$D_3^{\rho\rho'}\times 10^3$ & $4.66^{-0.30}_{+0.37}+\,0.88^{-0.05}_{+0.06}\iu$ & $5.26^{-0.33}_{+0.39}+0.99^{-0.05}_{+0.06}\,\iu$\\
	$D_3^{\rho'\rho'}\times 10^3$ & $6.78^{-0.90}_{+1.19}+0.06^{+0.00}_{-0.00}\,\iu$ & $8.85^{-1.14}_{+1.45}+0.09^{+0.01}_{-0.01}\,\iu$\\
	$D_3^{\omega\omega}\times 10^3$ & \multicolumn{2}{c}{$3.835 + 3.193 \,\iu$}\\
	$D_3^{\phi\phi}\times 10^3$ & \multicolumn{2}{c}{$8.736 + 3.775 \,\iu$}\\
	$\bar D_\asym\times 10^3$ & \multicolumn{2}{c}{$0.146\quad 0.038\quad 0.019\quad 0.011$}\\
	\bottomrule
	\end{tabular} 
	\caption{Numerical results for the constants defined in \autoref{eq:D_i_constants} for the two $\rho'$ spectral functions $\Gammarhoprime(q^2) = \Gammarhoprime^{(4\pi)}(q^2)$, \autoref{eq:energyDependentWidthRhoPrime}, and $\Gammarhoprime(q^2) = \Gammarhoprime^{(\omega \pi,\pi \pi)}(q^2)$, \autoref{eq:energyDependentWidthRhoPrimeAlternative}. The uncertainties refer to the variation $\Gammarhoprime=(400\pm 60)\MeV$, see \autoref{appx:constants}, which gives the dominant parametric effect.
	$\bar D_\asym$ is given for the reference points $\sqrt{\sm} \in \{1.0,1.3,1.5,1.7\}\GeV$.}
	\label{tab:PVConstants}  
\end{table}

For the numerical analysis we further write the coefficients in \autoref{eq:f1e+e-AmplitudeGeneralCalculated} according to
\begin{align}\label{eq:D_i_constants}
	D_i &= D_i^{I=1} + D_i^{I=0}, \quad i=1,2,3, & D_\asym &= \frac{F_{f_1}^\eff \Mf^3}{\Mrho^4}\bar D_\asym, \notag\\
	D_i^{I=1} &= \frac{D_i^{\rho \rho'}}{N_\aT}, \quad D_i^{I=0} = 0, \quad i=1,2, &
D_3^{I=0} &= R^\omega D_3^{\omega \omega} +  R^\phi D_3^{\phi \phi},\notag\\ 
	D_3^{I=1}\big|_{\widetilde{\text{VMD}}} &= \frac{D_3^{\rho\rho}(1-\eps_1-\eps_2)+D_3^{\rho\rho'}\eps_1+D_3^{\rho'\rho'}\eps_2}{\widetilde{N}_\sT}, &
	D_3^{I=1}\big|_\text{VMD} &= \frac{D_3^{\rho\rho}}{N_\sT},
\end{align}
where the prefactor for $D_\asym$ is motivated from \autoref{eq:effectiveDecayConstantVMDIsoscalar} to ensure that the resulting dimensionless coefficients can be compared in a meaningful way. Our numerical results are shown in \autoref{tab:PVConstants}, including the uncertainties from the variation in $\Gammarhoprime$. Even after taking the change in the normalizations into account, see \autoref{tab:renormalizations}, these results show that the uncertainties due to the spectral shape and the width itself can lead to comparable effects.

\begin{table}[t]
	\centering
	\begin{tabular}{ l  r  r}
	\toprule
	& $\Gammarhoprime^{(4\pi)}(q^2)$ & $\Gammarhoprime^{(\omega \pi,\pi \pi)}(q^2)$ \\\midrule
	$D_1^{I=1}\times 10^3$ & $(-0.218)^{+0.033}_{-0.034}+(-2.601)^{+0.007}_{-0.007}\,\iu$ & $(-0.269)^{+0.032}_{-0.032}+(-2.583)^{+0.011}_{-0.010}\,\iu$\\
	$D_2^{I=1}\times 10^3$ & $(-1.695)^{-0.060}_{+0.062} + 2.760^{-0.033}_{+0.031}\,\iu$ & $(-1.606)^{-0.053}_{+0.054}+2.732^{-0.038}_{+0.036}\,\iu$\\
	$D_3^{I=1}\big\rvert_\text{VMD}\times 10^3$ & \multicolumn{2}{c}{$3.961 + 2.904\,\iu$}\\
	$D_3^{I=1}\big\rvert_{\widetilde{\text{VMD}}}\times 10^3$  & $2.163^{+0.121}_{-0.148} + 3.592^{-0.061}_{+0.077}\,\iu$ & $1.930^{+0.128}_{-0.147}+3.685^{-0.070}_{+0.085}\,\iu$\\
	$D_3^{I=0}\times 10^3$ & \multicolumn{2}{c}{$-0.95(30) - 0.24(13)\,\iu$}\\
	$D_\asym\times 10^3$ & \multicolumn{2}{c}{$0.125(12)\quad 0.032(3)\quad 0.017(2)\quad 0.009(1)$}\\
	\bottomrule
	\end{tabular}
	\caption{Coefficients from \autoref{eq:D_i_constants}, based on \autoref{tab:PVConstants} and the normalizations from \autoref{tab:renormalizations}. For the extended VMD version the result in general depends on the $\eps_{1/2}$; here, we show the special case for $C_{\aT_2}=0$. For $D_3^{I=0}$ the error is propagated from \autoref{eq:SU3RatiosCouplings} and for $D_\asym$ from \autoref{eq:effectiveDecayConstantLCSR}.}
	\label{tab:PVConstantsNormalized}
\end{table}

To be able to better compare the various contributions, we also show the coefficients including their normalizations, see \autoref{tab:PVConstantsNormalized}, where we used the value of \autoref{eq:effectiveDecayConstantLCSR} for the asymptotic contribution. These numbers show that the symmetric contribution still produces the largest  coefficient, but not by much. Accordingly, the $f_1\to e^+e^-$ decay proves sensitive to the antisymmetric TFFs, about which not much is known at present. For the extended VMD ansatz, this observations implies an important caveat regarding the numbers shown in the table, which have been produced under the assumption that $C_{\aT_2}=0$. In this case, one observes distinct differences between the two VMD versions, which can be traced back to the different weight given to the $\rho\rho'$ contribution. Finally, the real part of the isoscalar coefficient comes out larger than expected from \autoref{eq:ratioISIVContribution}. This is due to the fact that the loop integral is effectively regularized by the vector-meson mass, and the masses of $\omega$ and $\phi$ differ by a sufficient amount that the cancellation in \autoref{eq:isoscalar_cancellation} between the two contributions becomes less effective. The imaginary part of the loop integral is finite also in the infinite-mass limit, so that its size complies better with the expected isoscalar suppression.

Since the coupling constants are real, we use the decay width from \autoref{eq:f1e+e-DecayWidth} to obtain a branching ratio of
\begin{align}\label{eq:Bfee}
	B(f_1 \to e^+ e^-) &= \frac{E_1 C_{\aT_1}^2 + E_2 C_{\aT_2}^2 + E_3 C_{\sT}^2 + E_{1,2} C_{\aT_1} C_{\aT_2} + E_{1,3} C_{\aT_1} C_{\sT} + E_{2,3} C_{\aT_2} C_{\sT}}{\Gamma_f}\notag\\
	&+\frac{E_{1,\asym} C_{\aT_1}+E_{2,\asym} C_{\aT_2}+E_{3,\asym} C_{\sT}+E_\asym}{\Gamma_f},
\end{align}
where we defined
\begin{align}
	E_i &= \frac{64 \pi^3 \alpha^4 \Mf}{3} |D_i|^2,\quad i = 1,2,3, \\
	E_{i,j} &= \frac{128 \pi^3 \alpha^4 \Mf}{3} \Re[D_i D_j^*], \quad (i,j) = (1,2), (1,3), (2,3),\notag\\
	E_{i,\asym} &= \frac{128 \pi^3 \alpha^4 \Mf}{3} \Re[D_iD_\asym], \quad i = 1,2,3, \qquad 
	E_\asym=\frac{64 \pi^3 \alpha^4 \Mf}{3} |D_\asym|^2,\notag
\end{align}
and the terms involving $D_\asym$ are only included in the extended VMD representation. 

\begin{figure}[t]
	\centering
	\includegraphics[width=0.495\textwidth]{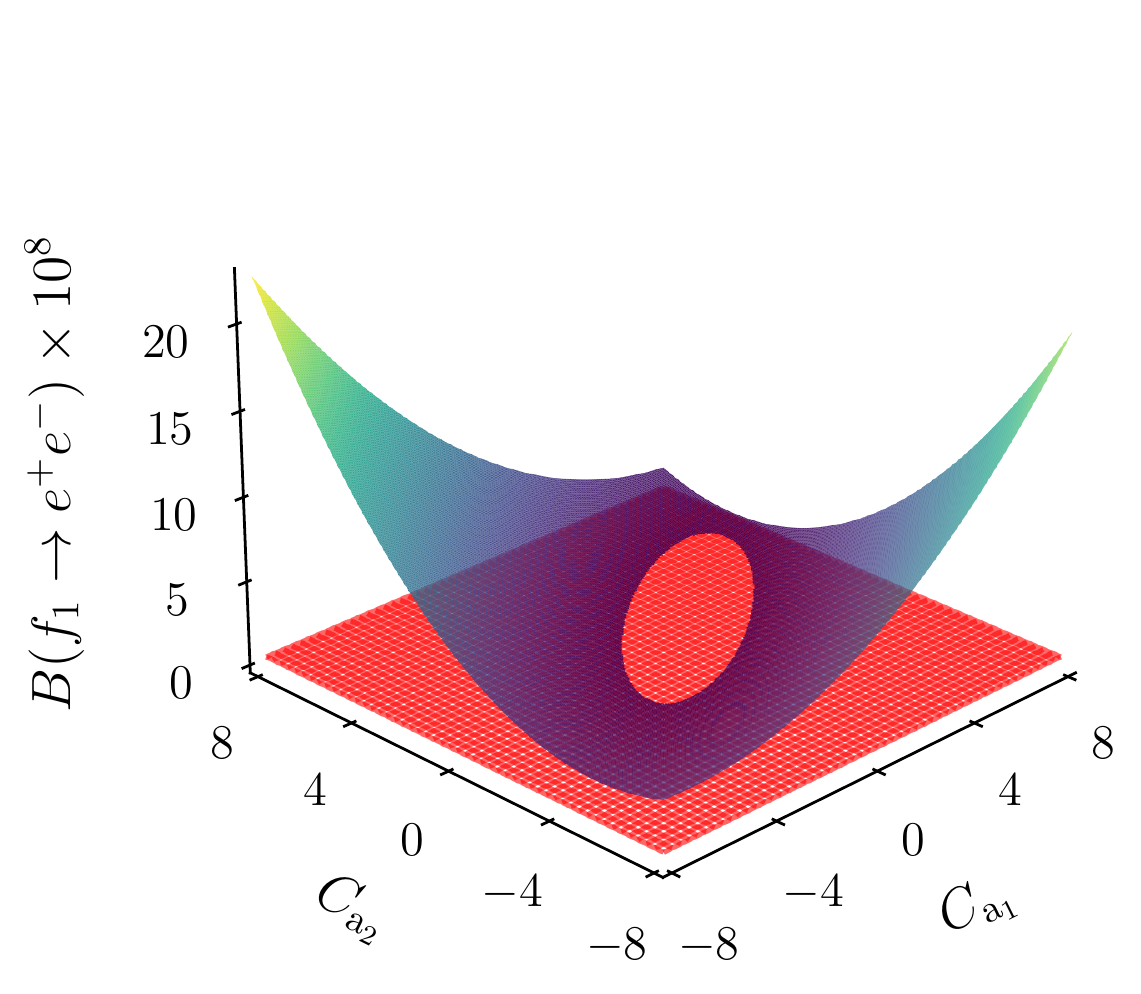}
	\hfill
	\includegraphics[width=0.495\textwidth]{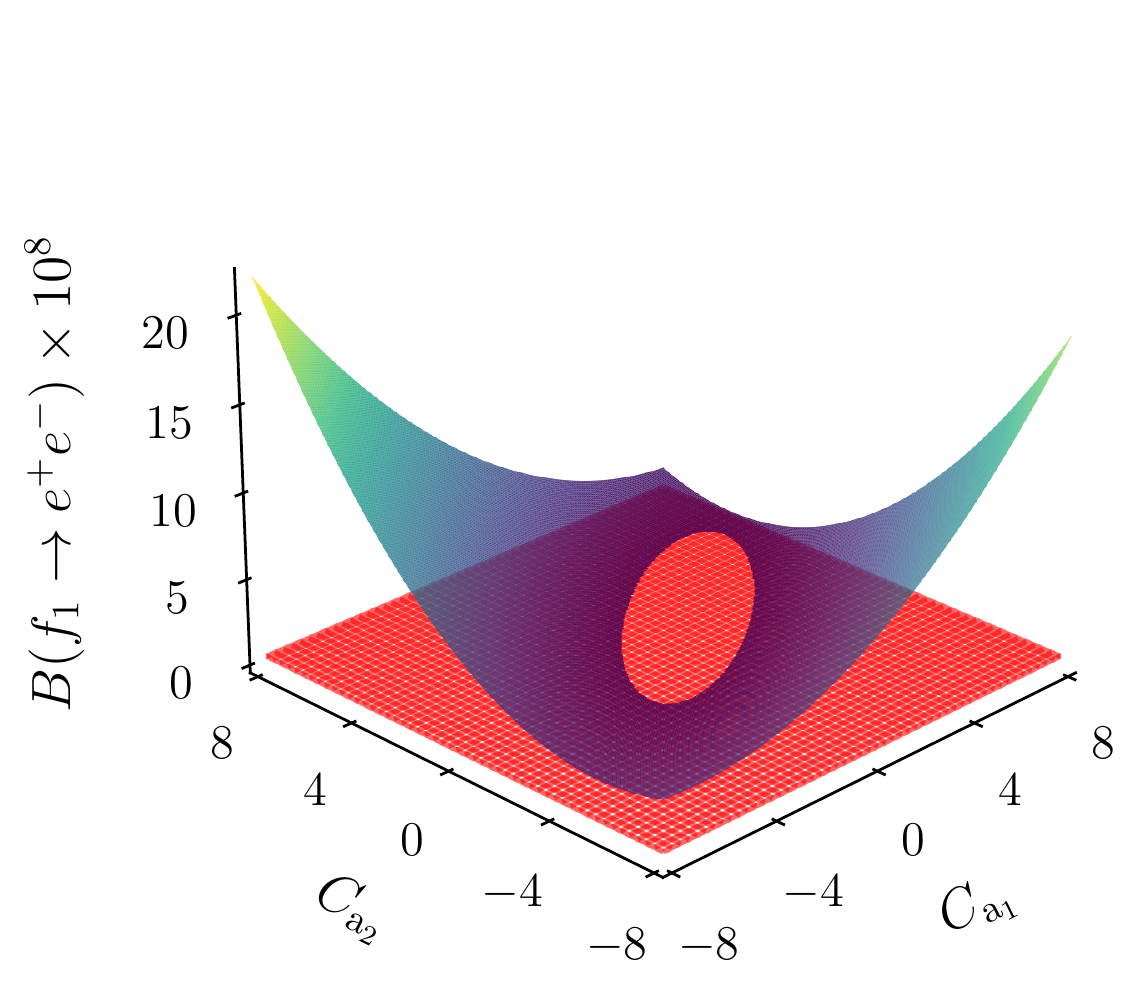}
	\\
	\includegraphics[width=0.495\textwidth]{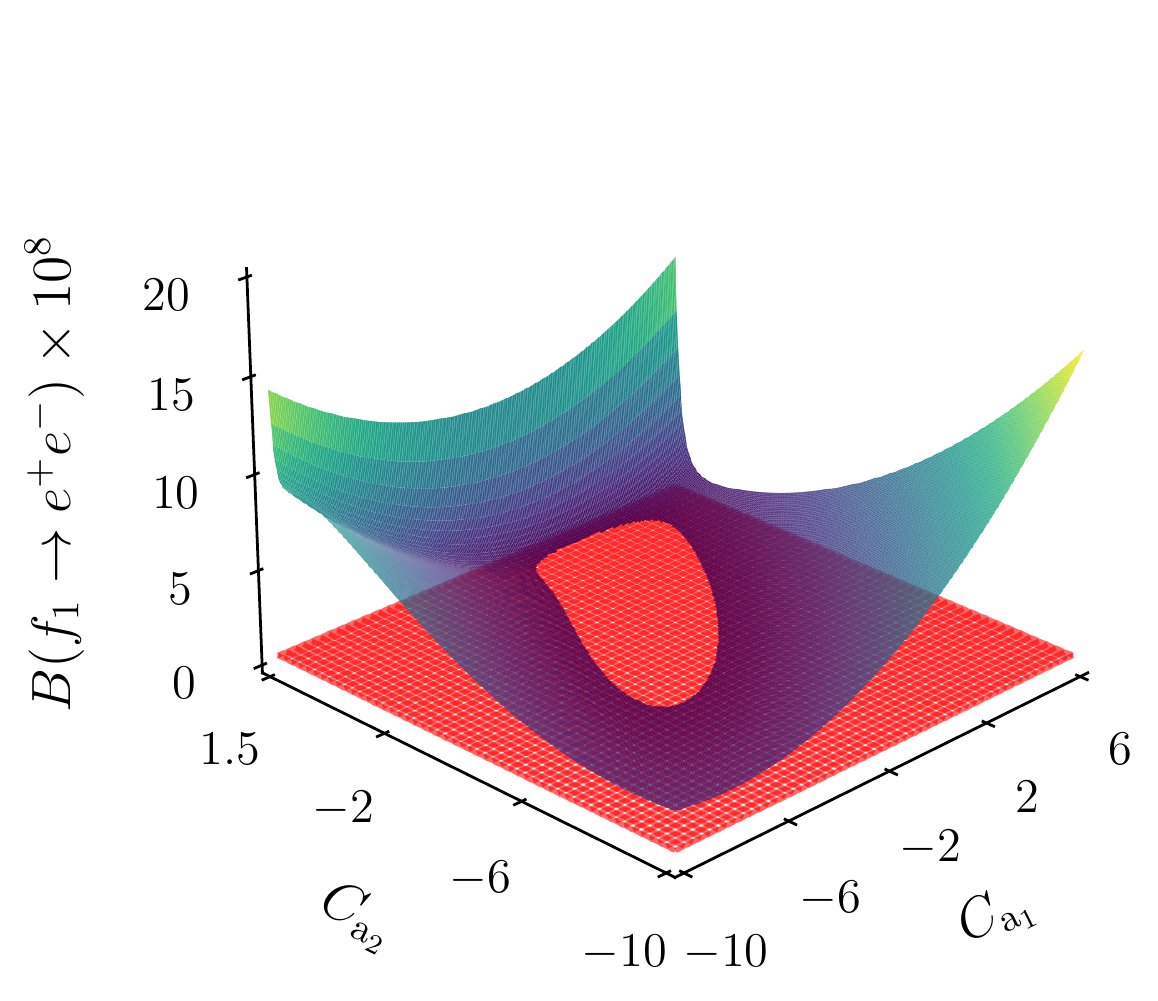}
	\hfill
	\includegraphics[width=0.495\textwidth]{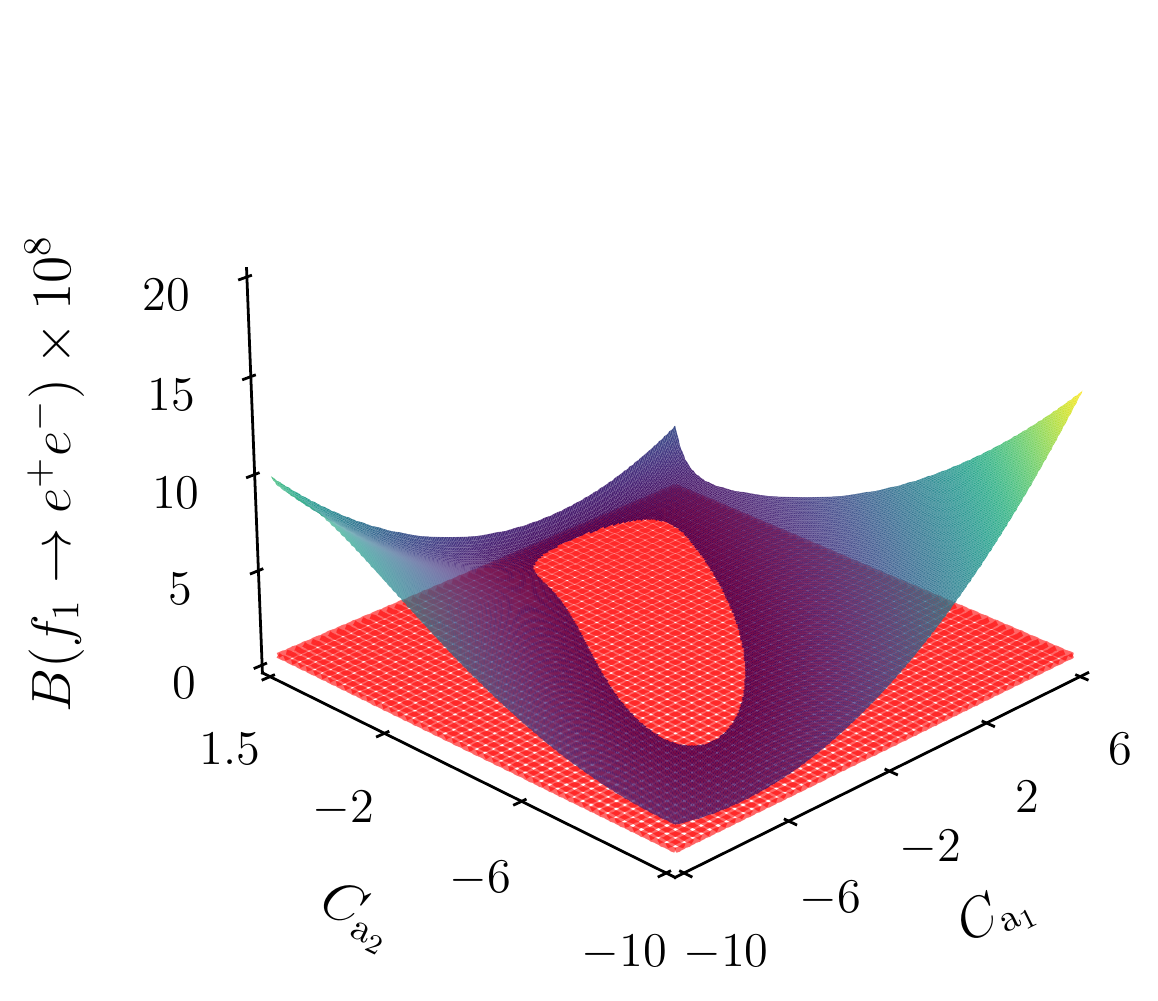}
	\caption{Surface plots of $B(f_1 \to e^+ e^-)$ (\textit{blue-yellow textured}), \autoref{eq:Bfee}, as obtained with the minimal (\textit{top}) and extended (\textit{bottom}) VMD parameterization (reference point $\sqrt{\sm} = 1.3 \GeV$ for the latter) and using the central value of $C_\sT = 0.93(11)$, \autoref{eq:CouplingCs}, $\Gammarhoprime(q^2) = \Gammarhoprime^{(4\pi)}(q^2)$ (\textit{left}), \autoref{eq:energyDependentWidthRhoPrime}, and $\Gammarhoprime(q^2) = \Gammarhoprime^{(\omega \pi,\pi \pi)}(q^2)$ (\textit{right}), \autoref{eq:energyDependentWidthRhoPrimeAlternative}, together with the central value of the measurement $B(f_1 \to e^+ e^-) = 5.1^{+3.7}_{-2.7} \times 10^{-9}$~\cite{Achasov:2019wtd} (\textit{red}).}
	\label{fig:B_f_e_e}
\end{figure}

Similarly to \autoref{eq:BfRhoGamma}, the solution of \autoref{eq:Bfee} in terms of the unknown couplings $C_{\aT_1}$ and $C_{\aT_2}$ represents an ellipse in the minimal VMD case, which, however, changes for the extended VMD representation, see \autoref{fig:B_f_e_e}. Here, we used the central value of $C_\sT = 0.93(11)$, \autoref{eq:CouplingCs}, to remove one unknown and set $\sqrt{\sm} = 1.3 \GeV$ for the asymptotic contribution~\cite{Hoferichter:2018dmo,Hoferichter:2018kwz}. In fact, the results in \autoref{tab:PVConstants}  and \autoref{tab:PVConstantsNormalized} show that $D_\asym$ remains small for a wide range of matching scales $\sm$, so that the details of the matching do not play a role in view of the present experimental uncertainties. For definiteness, we will continue to use $\sqrt{\sm} = 1.3 \GeV$ in the following, with the understanding that the matching can be refined once improved data become available, along the lines described in \autoref{appx:asymptotics}.

In order to solve for all couplings, we need to consider a combined analysis of all constraints, see \autoref{sec:pheno}. However, given that the biggest contribution tends to come from the symmetric term, see \autoref{tab:PVConstants}, it is instructive to study the case $C_{\aT_1}=C_{\aT_2}=0$ and consider the $f_1\to e^+e^-$ decay as an independent determination of $C_{\sT}$. For the minimal VMD ansatz we find 
\beq
\label{eq:Cs_VMD}
 C_{\sT}=1.7^{+0.6}_{-0.5},
\eeq
where the isoscalar contribution implies an increase by about $0.3(1)$. The extended variant gives\footnote{Due to the interference with the asymptotic contribution, there are, in principle, two solutions, which, however, are very close in magnitude.} 
\beq
\label{eq:Cs_VMD_tilde}
 C_{\sT}=1.9^{+0.8}_{-0.6},
\eeq
where the uncertainties from the dependence on the $\rho'$ spectral function, its width, and the asymptotic contribution, $\Delta C_{\sT}\lesssim 0.03$, are negligible compared to both the experimental error and the uncertainty from the isoscalar contribution.
Both values are larger than the L3 result given in 
\autoref{eq:CouplingCs}, indicating that indeed a significant 
contribution from the antisymmetric TFFs should be expected, which in view of the results from \autoref{tab:PVConstantsNormalized} is well possible with plausible values of $C_{\aT_{1/2}}$. Finally, the difference between \autoref{eq:Cs_VMD} and \autoref{eq:Cs_VMD_tilde} gives a first estimate  of the sensitivity to the chosen VMD ansatz.

\section{Combined phenomenological analysis}
\label{sec:pheno}

\begin{table}[t]
	\centering
	\begin{tabular}{l  r r r r}
	\toprule
		Reference & $B(f_1 \to \rho \gamma)$ & $r_{\rho \gamma}$& $B(f_1 \to \phi \gamma)$ & $B(f_1 \to e^+ e^-)$ \\
		\midrule
		VES~\cite{Amelin:1994ii} & $2.8(9) \perc$ & $3.9(1.3)$ & &\\
		PDG~\cite{Zyla:2020zbs} & $6.1(1.0) \perc$ & & &\\
		Our fit & $4.2(1.0) \perc$ & & &\\
		Serpukhov~\cite{Bityukov:1987bj,Zyla:2020zbs} &  & & $0.74(26)\times 10^{-3}$&\\
		SND~\cite{Achasov:2019wtd} & & & &$(5.1\substack{+3.7\\-2.7}) \times 10^{-9}$\\\bottomrule
	\end{tabular}
	\caption{Summary of the experimental measurements used in our global analysis. In addition, we use the L3 data for $e^+e^-\to e^+e^- f_1$, see \autoref{sec:L3}.}
	\label{tab:expCollected}
\end{table}

In this section, we perform a global analysis of the experimental constraints from $e^+e^-\to e^+e^- f_1$, $f_1\to \rho\gamma$, and $f_1\to e^+e^-$. We will also consider $f_1\to\phi\gamma$ due to its relation via $\Uthree$ symmetry, but not include $f_1\to 4\pi$ for the reasons stated in \autoref{sec:4pi} and \autoref{appx:f14pi}. 
Most of the input quantities follow in a straightforward way from the experimental references and the compilation in Ref.~\cite{Zyla:2020zbs}, see \autoref{tab:expCollected}, except for the branching fraction of the $\rho\gamma$ channel, for which the fit by the Particle Data Group (PDG) and the direct measurement by VES~\cite{Amelin:1994ii} disagree by $2.5\sigma$. 

The PDG fit proceeds in terms of the five branching fractions for  $f_1\to4\pi$, $a_0(980)\pi$ (excluding $a_0(980)\to K\bar K$), $\eta\pi\pi$ (excluding $a_0(980)\pi$), $K\bar K\pi$, and $\rho\gamma$, 
 including data on
\begin{enumerate}
	\item $\Gamma(f_1\to K\bar K\pi)/\Gamma(f_1\to 4\pi)$~\cite{Armstrong:1989zr,Armstrong:1989jk,Barberis:1997vf},
	\item $\Gamma(f_1\to 4\pi)/\Gamma(f_1\to\eta\pi\pi)$~\cite{Gurtu:1978yv,Bolton:1991nx},
	\item $\Gamma(f_1\to \rho\gamma)/\Gamma(f_1\to 4\pi)$~\cite{Coffman:1989nk},
	\item $\Gamma(f_1\to a_0(980)\pi \,[\text{excluding }\,K\bar K\pi])/\Gamma(f_1\to\eta\pi\pi)$~\cite{Gurtu:1978yv,Corden:1978cz,Dickson:2016gwc},
	\item $\Gamma(f_1\to K\bar K\pi)/\Gamma(f_1\to\eta\pi\pi)$~\cite{Corden:1978cz,Gurtu:1978yv,Barberis:1998by,Dickson:2016gwc,Campbell:1969cx,Defoix:1972vd},
	\item $\Gamma(f_1\to\eta\pi\pi)/\Gamma(f_1\to\rho\gamma)$~\cite{Dickson:2016gwc,Barberis:1998by,Armstrong:1991bj},
\end{enumerate}
however, with the notable exception of the constraint from Ref.~\cite{Amelin:1994ii}.\footnote{Reference~\cite{Amelin:1994ii} only quotes the final $\rho\gamma$ branching fraction, not $\Gamma(f_1\to\eta\pi\pi)/\Gamma(f_1\to\rho\gamma)$ as measured in the experiment, but the $\eta\pi\pi$ branching fraction from Ref.~\cite{Hikasa:1992je} is very close to the one from Ref.~\cite{Zyla:2020zbs}, rendering the systematic error from the conversion negligible.}
This fit has a reduced $\chi^2/\text{dof}=24.0/14=1.71$, reflecting the significant tensions in the data base. These tensions become exacerbated when including Ref.~\cite{Amelin:1994ii} in the fit, leading to a slightly smaller $\rho\gamma$ branching fraction of $5.3\perc$, with $\chi^2/\text{dof}=33.5/15=2.23$. The origin of the tensions can be traced back to the input for $\Gamma(f_1\to\eta\pi\pi)/\Gamma(f_1\to\rho\gamma)$, which is measured as $21.3(4.4)$~\cite{Dickson:2016gwc}, $10(1)(2)$~\cite{Barberis:1998by}, and $7.5(1.0)$~\cite{Armstrong:1991bj},\footnote{The latter value is given as $5.0(7)$ in Ref.~\cite{Armstrong:1991bj} for $\eta\pi^+\pi^-$, and has thus been increased by the isospin factor $3/2$ in the PDG listing.  There is also a limit $B(f_1\to\rho\gamma)<5\perc$ at $95\perc$ confidence level from Ref.~\cite{Bityukov:1991pk}, in tension with Refs.~\cite{Barberis:1998by,Armstrong:1991bj}.} with some additional sensitivity to the $\rho\gamma$ channel from $\Gamma(f_1\to \rho\gamma)/\Gamma(f_1\to 4\pi)$~\cite{Coffman:1989nk}.  

\begin{table}[t]
	\centering
	\begin{tabular}{l  r  r  r  r}
	\toprule
$f_1\to \phi\gamma$ & \multicolumn{2}{c}{No} & \multicolumn{2}{c}{Yes}\\
 & \multicolumn{1}{c}{Solution 1} & \multicolumn{1}{c}{Solution 2}
& \multicolumn{1}{c}{Solution 1} & \multicolumn{1}{c}{Solution 2}\\\midrule	
$\chi^2/\text{dof}$ & $2.72/2=1.36$ & $6.60/2=3.30$ & $8.67/3=2.89$ & $8.28/3=2.76$ \\
$p$-value & $0.26$ &$0.04$ & $0.03$ &$0.04$\\
$C_\sT$ & $0.97(13)$ &$1.01(18)$ & $0.95(18)$ & $0.99(17)$\\
$C_{\aT_1}$ & $-0.23(13)$ &$0.91(21)$ & $-0.09(14)$ & $0.80(17)$\\
$C_{\aT_2}$ & $0.51(21)$ &$0.53(39)$ & $0.17(25)$ & $0.34(30)$\\
$\rho_{\sT\aT_1}$ & $0.43$ &$0.41$ & $0.21$ & $0.31$\\
$\rho_{\sT\aT_2}$ & $-0.42$ &$-0.13$ & $-0.50$ & $-0.37$\\
$\rho_{\aT_1\aT_2}$ & $-0.44$ &$0.77$ & $-0.29$ & $0.66$\\
$B(f_1 \to e^+ e^-)\times 10^9$ & $2.7(6)$ &$0.7(3)$ & $1.8(6)$ & $0.7(3)$\\
$B(f_1\to \phi\gamma)\times 10^3$ & $2.5(1.3)$ & $1.5(1.1)$ & $1.3(8)$ & $1.1(7)$\\
$B(f_1\to \omega\gamma)\times 10^3$ & $5.6(1.7)$ & $4.4(2.2)$ & $2.7(1.3)$ & $3.3(1.4)$\\
\bottomrule
	\end{tabular}
	\caption{Best-fit results for the three VMD couplings $C_\sT$, $C_{\aT_1}$, and $C_{\aT_2}$ in the minimal VMD representation. Each fit includes the constraints from the normalization and slope of the TFF measured by L3 in $e^+e^-\to e^+e^- f_1$, from $B(f_1\to\rho\gamma)$, $r_{\rho\gamma}$, and $B(f_1\to e^+e^-)$. In addition, we show the variants including $B(f_1\to \phi\gamma)$ as a sixth constraint assuming $\Uthree$ symmetry. The uncertainties include the scale factor $S=\sqrt{\chi^2/\text{dof}}$. We also show the correlations $\rho_{ij}$ among the three couplings and the value of $B(f_1\to e^+e^-)$ preferred by each fit. 
	Since  the experimental uncertainties dominate by far in the case of $B(f_1\to e^+e^-)$, we only show the results for $\Gammarhoprime^{(\omega \pi,\pi \pi)}(y)$ and $\sqrt{\sm}=1.3\GeV$ and do not include the theory uncertainties discussed in detail in \autoref{sec:f1ee}. The uncertainties quoted for $B(f_1\to V\gamma)$ refer to the fit errors and $R^V$, but do not include any $\Uthree$ uncertainties.}
	\label{tab:solutionsCa1_Ca2_minimal}
\end{table}

The main reason why the fit prefers the $\rho\gamma$ branching fraction from Refs.~\cite{Barberis:1998by,Armstrong:1991bj} is that the $\chi^2$ minimization is set up in terms of $\Gamma(f_1\to\eta\pi\pi)/\Gamma(f_1\to\rho\gamma)$, not the inverse quantity, as would be canonical given that $\Gamma(f_1\to\rho\gamma)$ is the smallest of the fit components and could thus be treated perturbatively. Using $\Gamma(f_1\to\rho\gamma)/\Gamma(f_1\to\eta\pi\pi)$ instead in the minimization gives a similar $\chi^2/\text{dof}=24.9/14=1.78$, but reduces the $\rho\gamma$ branching fraction to $4.9(9)\perc$ (including the scale factor from Ref.~\cite{Zyla:2020zbs}), close to the naive average of Refs.~\cite{Coffman:1989nk,Dickson:2016gwc,Barberis:1998by,Armstrong:1991bj} when taking the respective normalization channel from the fit. Including in addition the measurement from Ref.~\cite{Amelin:1994ii}, we find $\chi^2/\text{dof}=28.6/15=1.91$ and
\begin{align}
B(f_1\to 4\pi)&=33.4(1.8)\perc && [32.7(1.9)\perc],\notag\\
B(f_1\to a_0(980)\pi\, [\text{excluding}\, a_0(980)\to K\bar K])&=38.6(4.2)\perc && [38.0(4.0)\perc],\notag\\
B(f_1\to \eta\pi\pi\,[\text{excluding}\,a_0(980)\pi])&=14.6(4.1)\perc && [14.0(4.0)\perc],\notag\\
B(f_1\to K\bar K\pi)&=9.2(4)\perc && [9.0(4)\perc],\notag\\
 B(f_1\to\rho\gamma)&=4.3(8)\perc\, && [6.1(1.0)\perc],
\end{align}
where  the results of the PDG fit are indicated in brackets (for better comparison the same channel-specific scale factors have been applied as in Ref.~\cite{Zyla:2020zbs}). Finally, the limit from Ref.~\cite{Bityukov:1991pk} tends to further reduce the average a little, which together with a slightly increased scale factor when including Refs.~\cite{Bityukov:1991pk,Amelin:1994ii} leads us to quote
\beq
B(f_1\to\rho\gamma)=4.2(1.0)\perc
\eeq
as our final average, which we will use in the subsequent analysis, see \autoref{tab:expCollected}. While our main argument in favor of this
procedure is the avoidance of a fit  bias towards the larger $\rho\gamma$ branching fractions, one may also compare to theoretical expectations. The models considered in Refs.~\cite{Dickson:2016gwc,Babcock:1976hr,Ishida:1988uw,Lutz:2008km,Osipov:2018ejk} in general do prefer smaller $\rho\gamma$ branching fractions, but the spread among the models is too large to make that comparison conclusive.    

\begin{table}[t]
	\centering
	\begin{tabular}{l  r  r  r  r}
	\toprule
$f_1\to \phi\gamma$ & \multicolumn{2}{c}{No} & \multicolumn{2}{c}{Yes}\\
 & \multicolumn{1}{c}{Solution 1} & \multicolumn{1}{c}{Solution 2}
& \multicolumn{1}{c}{Solution 1} & \multicolumn{1}{c}{Solution 2}\\\midrule	
$\chi^2/\text{dof}$  & $2.25/2=1.12$ & $4.40/2=2.20$ & $4.01/3=1.34$ & $7.53/3=2.51$\\
$p$-value & $0.33$ & $0.11$ & $0.26$ & $0.06$\\
$C_\sT$  & $1.00(10)$ & $1.02(14)$ & $1.00(11)$ &$1.02(15)$\\
$C_{\aT_1}$  & $-0.18(12)$ & $0.85(14)$ & $-0.19(12)$ &$0.85(15)$\\
$C_{\aT_2}$  & $1.03(36)$ & $1.17(32)$ & $-0.20(29)$ & $0.13(47)$\\
$\rho_{\sT\aT_1}$  & $0.10$ & $0.86$ & $0.10$ & $0.86$\\
$\rho_{\sT\aT_2}$  & $0.00$ & $0.21$ & $-0.34$ &$-0.32$\\
$\rho_{\aT_1\aT_2}$  & $0.08$ & $0.19$ & $0.18$ &$-0.27$\\
$\eps_1$ & $2.59(1.33)$ & $3.00(1.15)$ & $-1.79(1.01)$ & $-0.64(1.65)$\\
$B(f_1 \to e^+ e^-)\times 10^9$  & $5.1(3.3)$ & $5.1(4.7)$ & $1.5(4)$ & $0.3(4)$\\
$B(f_1\to \phi\gamma)\times 10^3$ & $4.4(2.4)$ & $3.4(2.0)$ & $0.8(6)$ & $0.8(7)$\\
$B(f_1\to \omega\gamma)\times 10^3$ & $9.1(3.1)$ & $6.8(2.2)$ & $1.9(1.0)$ & $3.3(1.1)$\\
\bottomrule
	\end{tabular}
	\caption{Same as \autoref{tab:solutionsCa1_Ca2_minimal}, but for the extended VMD case, including the resulting parameter $\eps_1$.}
	\label{tab:solutionsCa1_Ca2_extended}
\end{table}

\begin{figure}[t]
	\centering
	\includegraphics[width=0.496\textwidth]{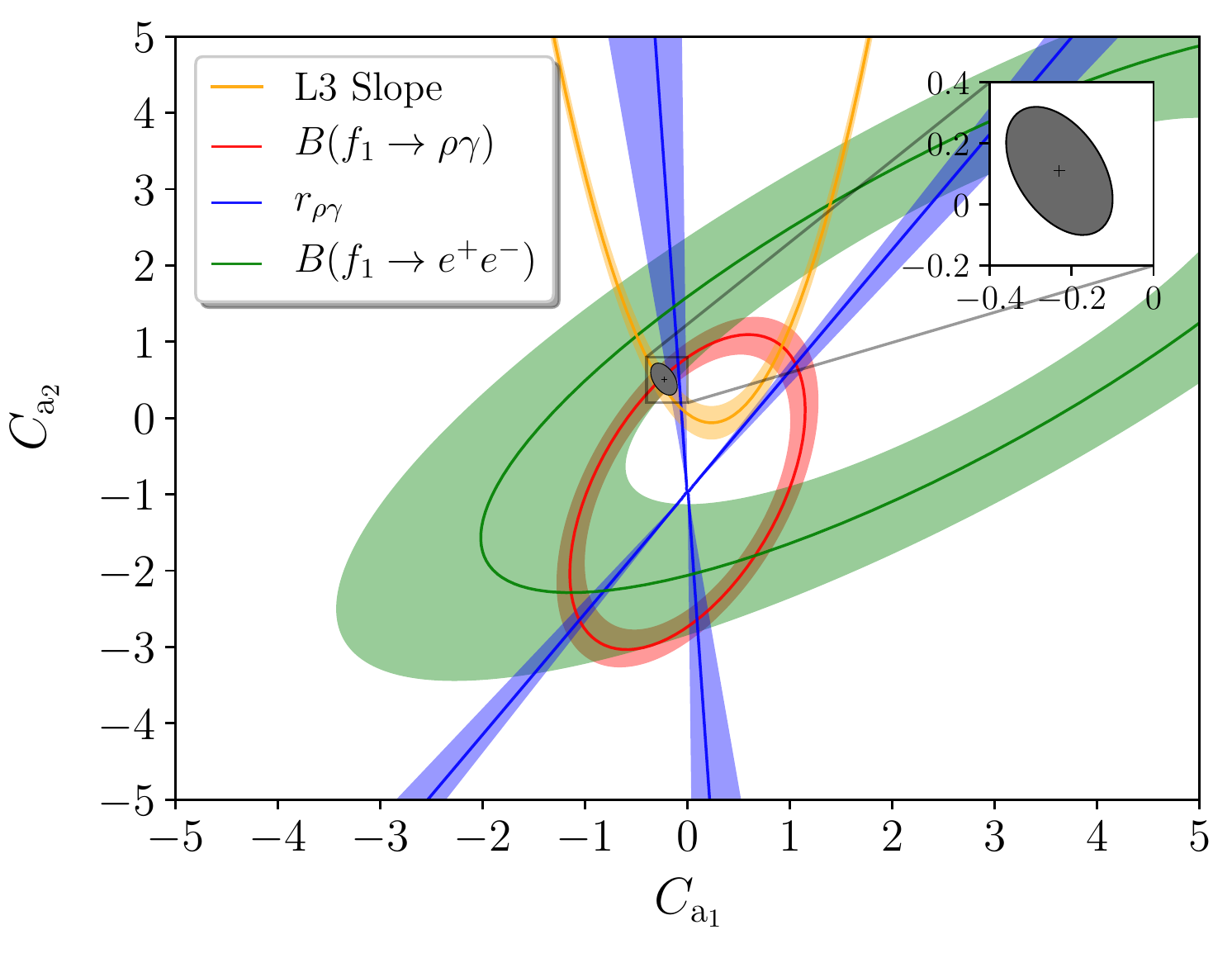}
	\includegraphics[width=0.496\textwidth]{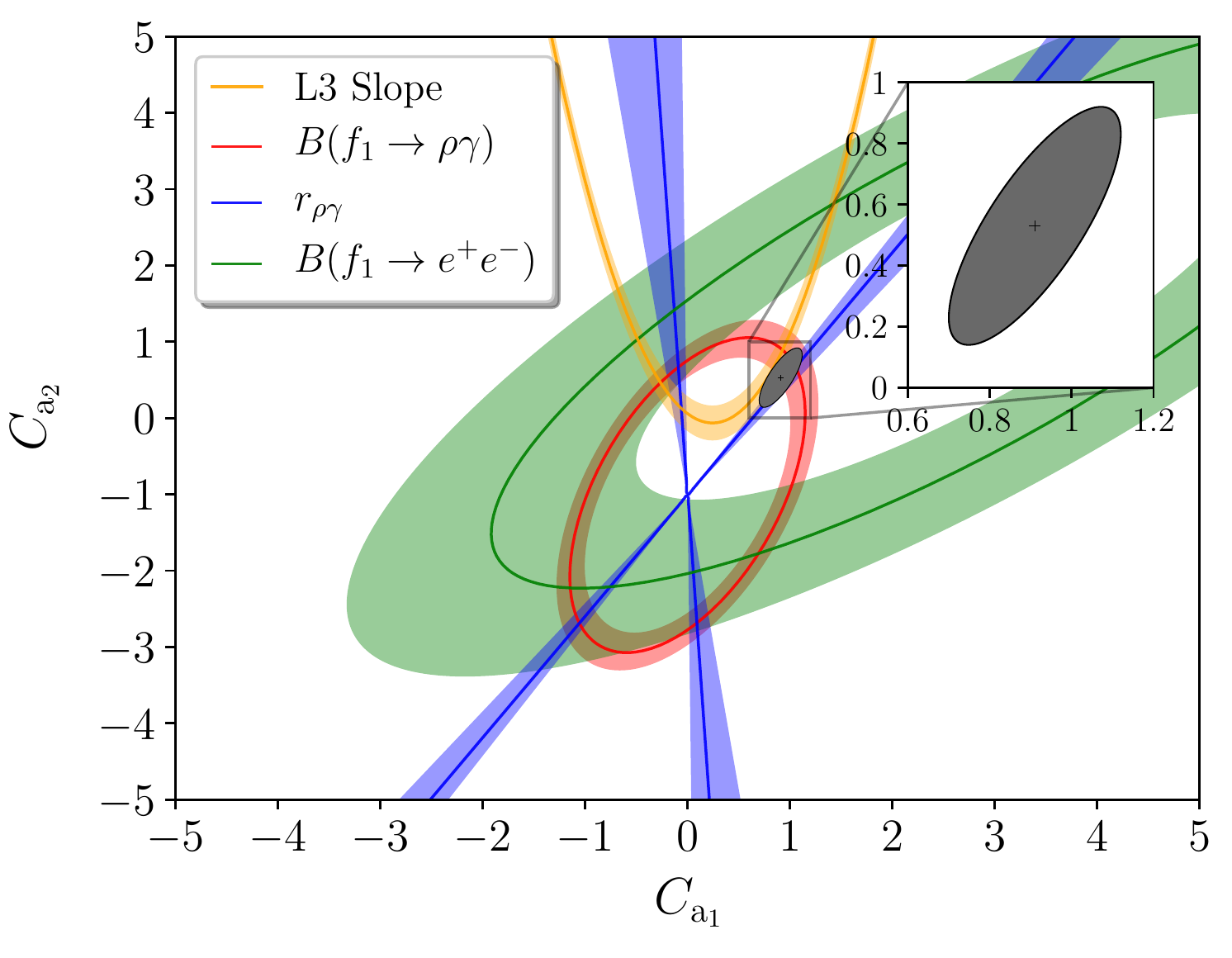}
	\includegraphics[width=0.496\textwidth]{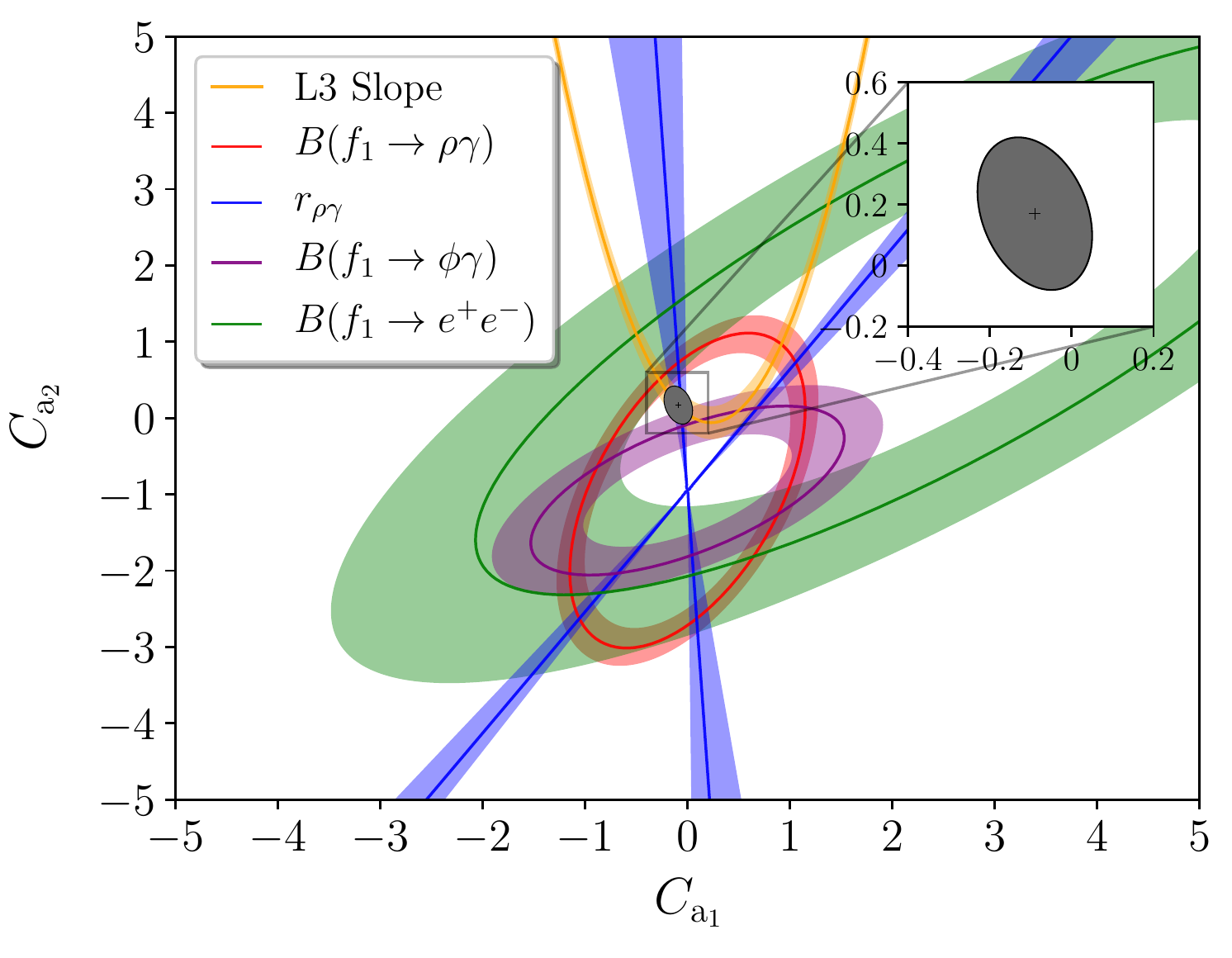}
	\includegraphics[width=0.496\textwidth]{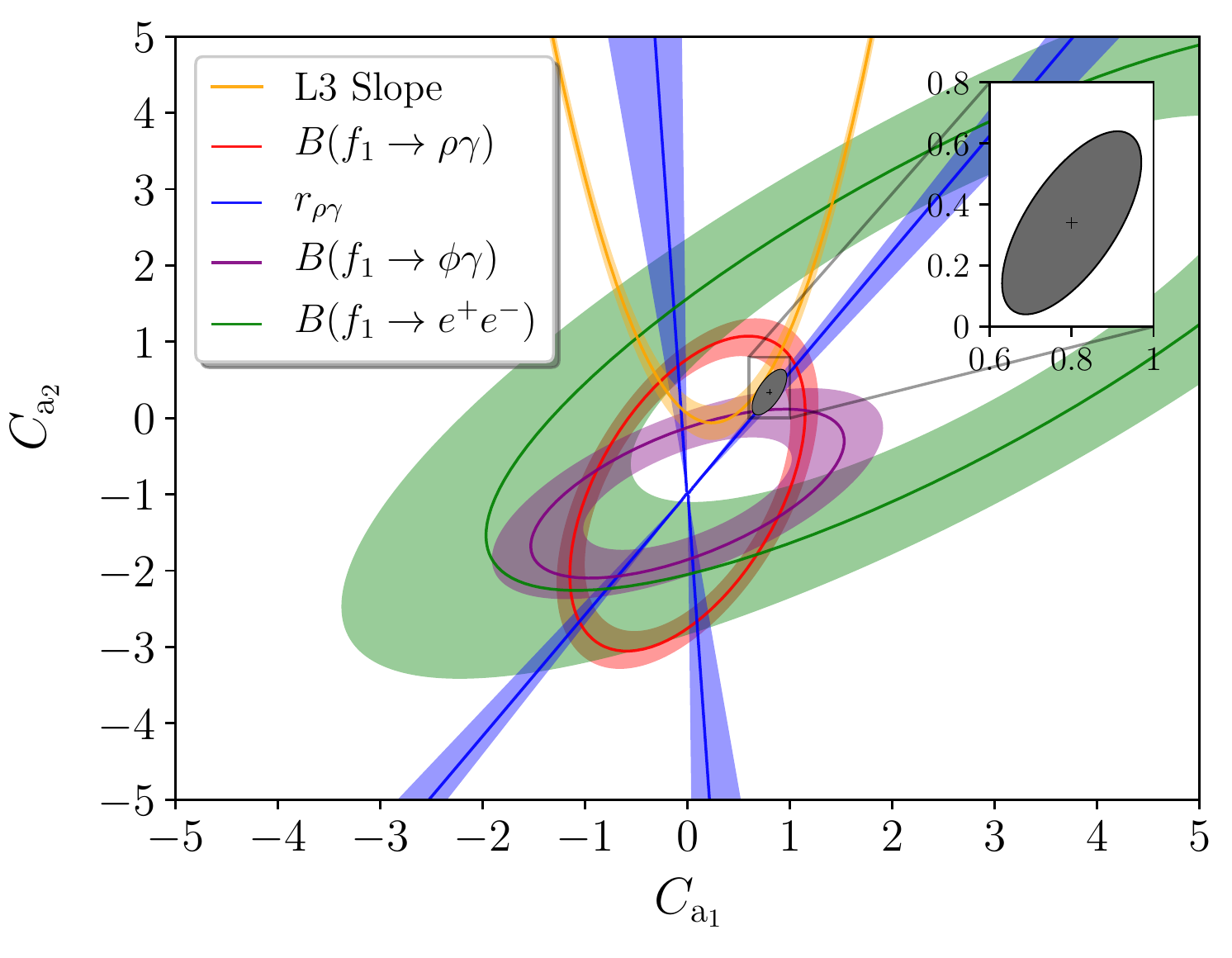}
	\caption{Contours in the $C_{\aT_1}$--$C_{\aT_2}$ plane for the best-fit value of $C_\sT$ in the minimal VMD representation: for Solution 1 (\textit{left}) and Solution 2 (\textit{right}),  without (\textit{upper}) and including (\textit{lower}) the constraint from $B(f_1\to\phi\gamma)$. The best-fit region is indicated by the $\Delta \chi^2=1$ ellipse (inflated by the scale factor).}
	\label{fig:solutionMinimalVMD}
\end{figure}

\begin{figure}[t]
	\centering
	\includegraphics[width=0.496\textwidth]{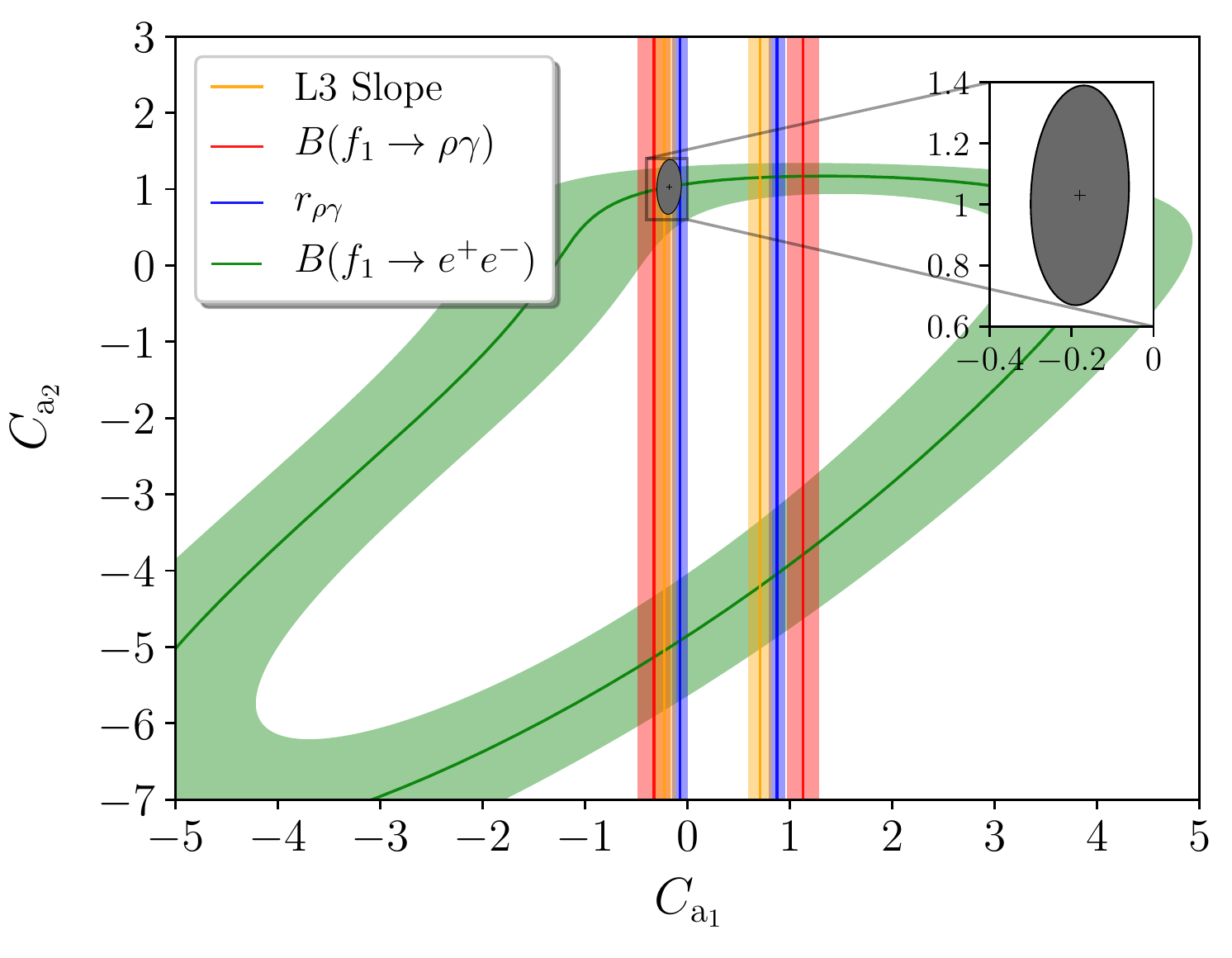}
	\includegraphics[width=0.496\textwidth]{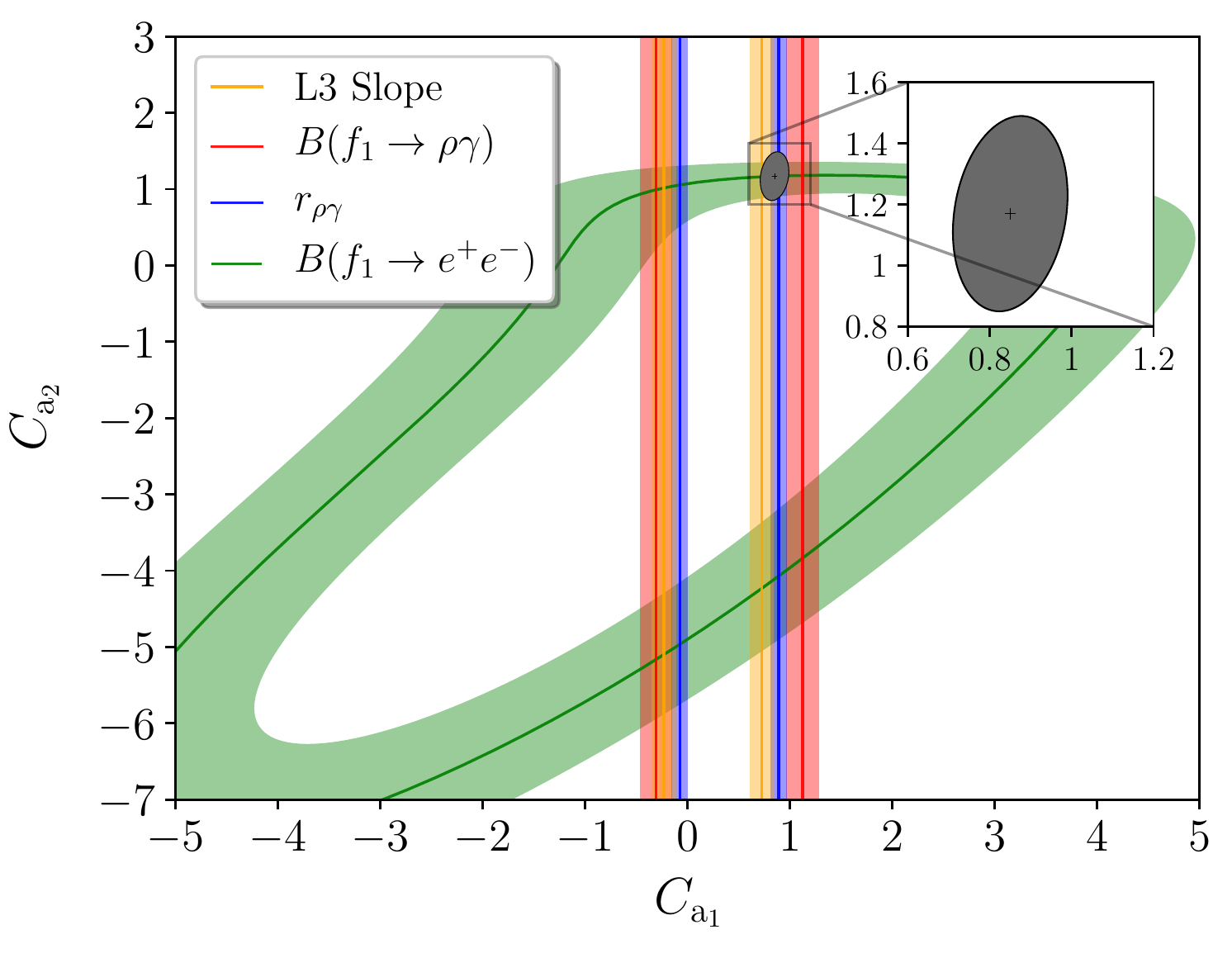}
	\includegraphics[width=0.496\textwidth]{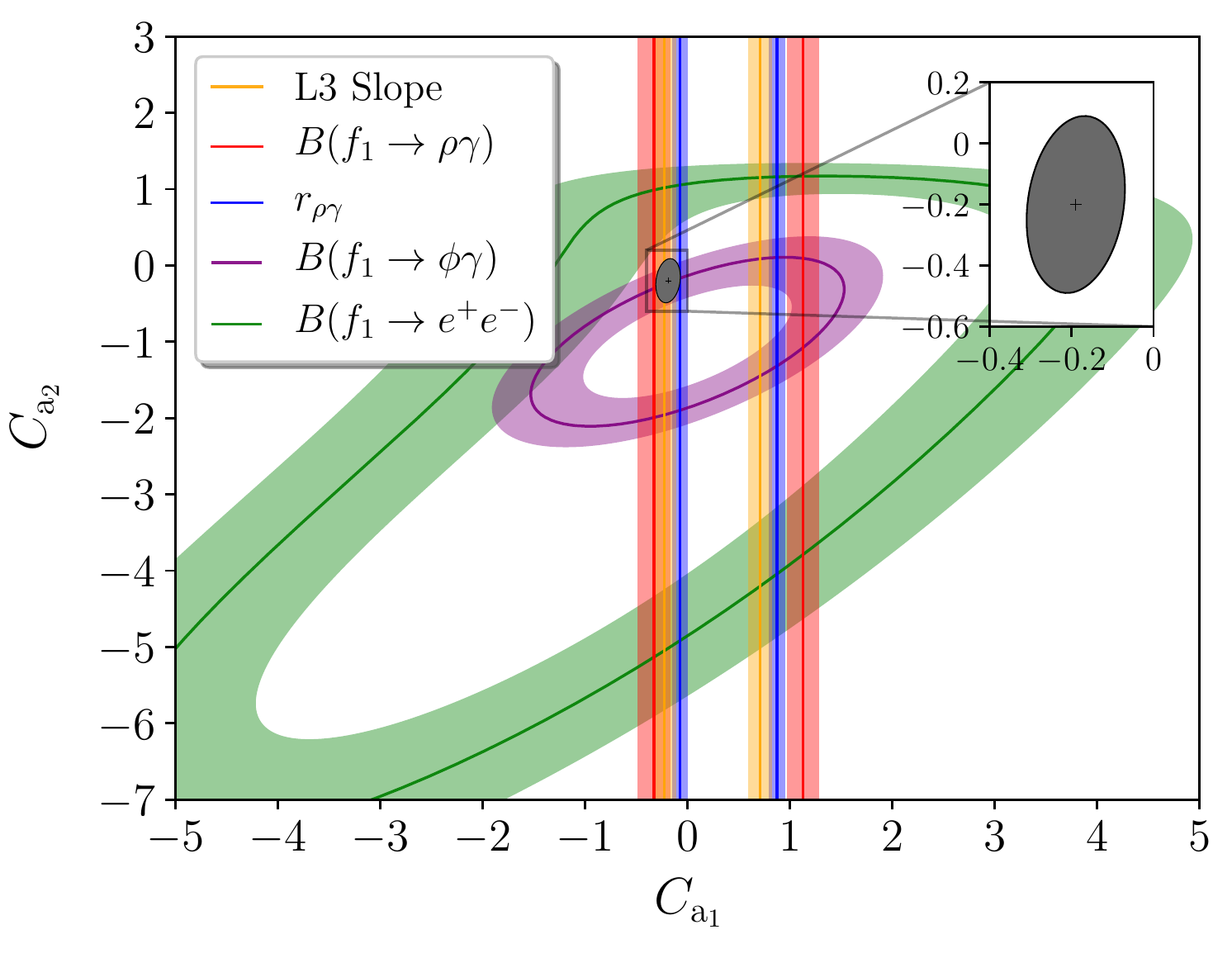}
	\includegraphics[width=0.496\textwidth]{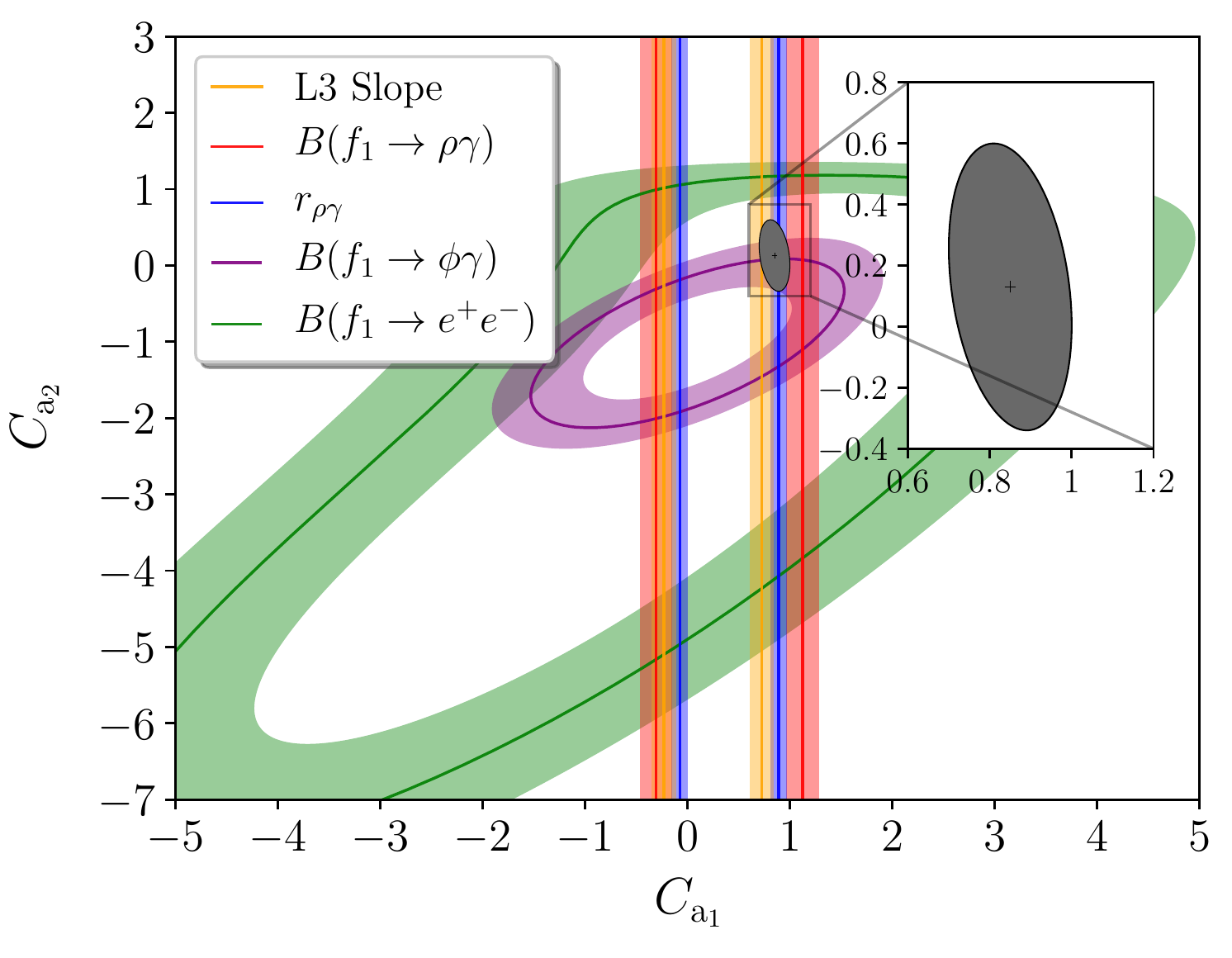}
	\caption{Contours in the $C_{\aT_1}$--$C_{\aT_2}$ plane for the best-fit value of $C_\sT$ in the extended VMD representation: for Solution 1 (\textit{left}) and Solution 2 (\textit{right}),  without (\textit{upper}) and including (\textit{lower}) the constraint from $B(f_1\to\phi\gamma)$. The best-fit region is indicated by the $\Delta \chi^2=1$ ellipse (inflated by the scale factor). We do not consider equivalent solutions with very large negative $C_{\aT_2}$, as arise without the $B(f_1\to\phi\gamma)$ constraint. Further local minima when including $B(f_1\to \phi\gamma)$ mirror the indicated Solutions 1 and 2 on the lower branch of the ellipse, but display a worse $\chi^2/\text{dof}$ and are thus discarded.}
	\label{fig:solutionExtendedVMD}
\end{figure}

The results of the global analysis are shown in \autoref{tab:solutionsCa1_Ca2_minimal} and \autoref{tab:solutionsCa1_Ca2_extended}, restricted to the parameterization $\Gammarhoprime^{(\omega \pi,\pi \pi)}(y)$ due to the dominant experimental uncertainties. The latter are propagated as given in \autoref{tab:expCollected}, except for $B(f_1\to \phi\gamma)$, for which we use $B(f_1\to \phi\gamma)/(R^\phi)^2=3.0(1.6)\perc$ as  data point in the minimization, including the uncertainty on $R^\phi$ from \autoref{eq:SU3RatiosCouplings}.
As a side result, \autoref{tab:solutionsCa1_Ca2_minimal} and \autoref{tab:solutionsCa1_Ca2_extended} also contain predictions for the branching fraction of the yet unmeasured decay $f_1\to\omega\gamma$.
The outcome in the four cases considered---minimal and extended VMD representations each with and without the constraint from $B(f_1\to\phi\gamma)$---is illustrated in \autoref{fig:solutionMinimalVMD} and \autoref{fig:solutionExtendedVMD}. In all cases the parameter $C_\sT$ is by far best constrained, its value hardly changes compared to the L3 reference point given in \autoref{eq:CouplingCs}, with a slight preference for a small upward shift. The main distinctions concern the couplings $C_{\aT_1}$ and $C_{\aT_2}$, with qualitative differences between the two VMD scenarios. In each case, however, we find two sets of solutions, corresponding to a small negative value of $C_{\aT_1}$ (Solution 1) or a sizable positive one (Solution 2), respectively, both of which are shown in the tables and figures. In most cases, Solution 1 is strongly preferred, the exception being the minimal VMD fit including $B(f_1\to \phi\gamma)$, in which case Solution 2 displays a slightly better fit quality.  

\begin{figure}[t]
	\centering
	\includegraphics[width=\textwidth]{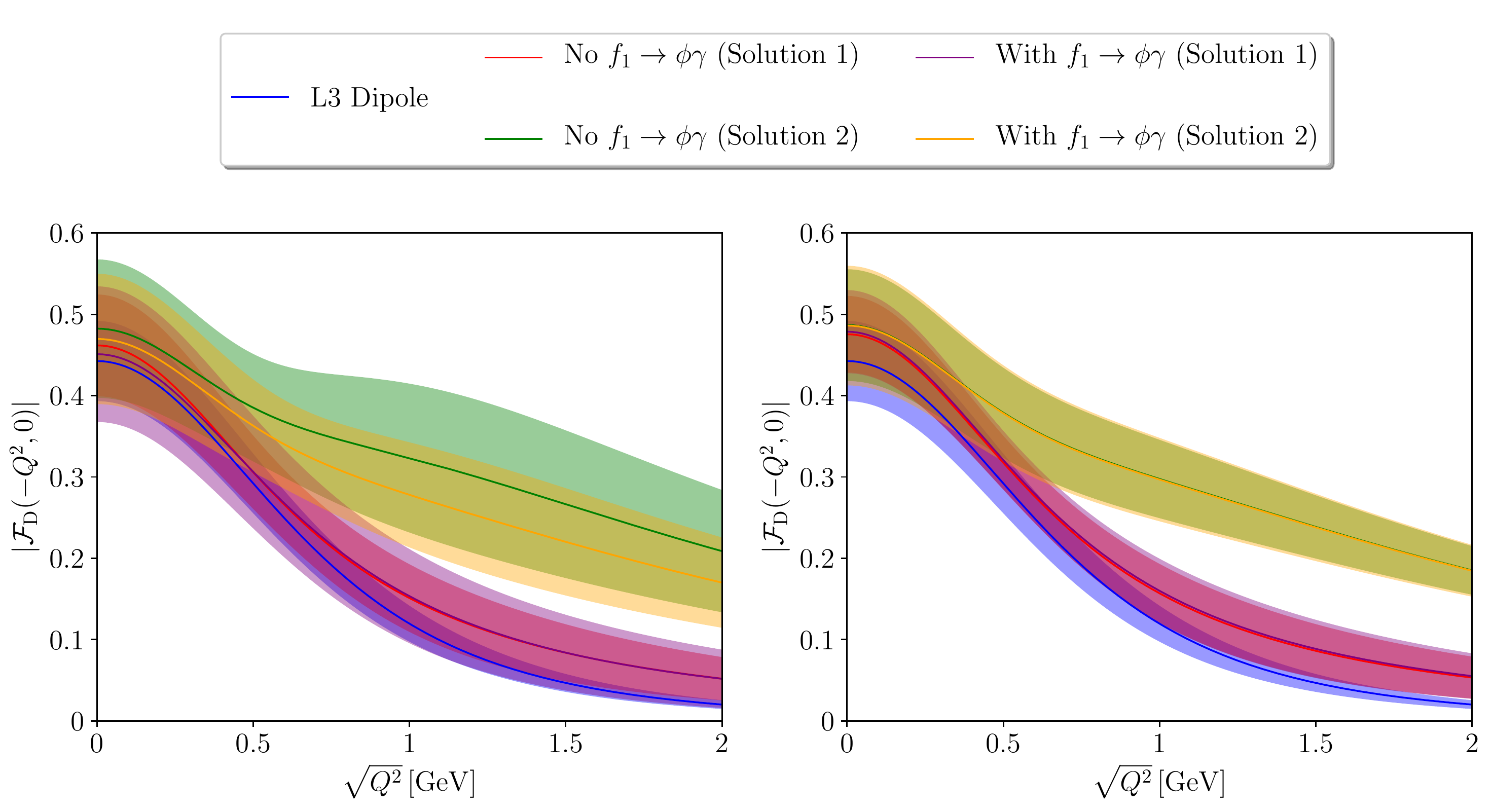}
	\caption{Comparison of the fit solutions for the effective form factor probed in $e^+e^-\to e^+e^- f_1$ to the L3 measurement~\cite{Achard:2001uu}, according to \autoref{L3_matching}, for the minimal VMD representation (\textit{left}) and the extended one (\textit{right}). The L3 dipole band includes the uncertainties on $\lvert\F_\text{D}(0,0)\rvert$ and $\Lambda_\text{D}$ as given by \autoref{eq:L3}, added in quadrature; ours propagate the uncertainties from \autoref{tab:solutionsCa1_Ca2_minimal} and \autoref{tab:solutionsCa1_Ca2_extended}, respectively.}
	\label{fig:L3Comparison} 
\end{figure}

In the minimal VMD representation, all constraints are sensitive to $C_{\aT_2}$, but especially once including $B(f_1\to\phi\gamma)$ there is significant tension among the different bands. 
In Solution 2, 
the region preferred by all constraints but $B(f_1\to e^+e^-)$, which thus dominate the fit, would imply a much smaller value of $B(f_1\to e^+e^-)$ than reported by SND~\cite{Achasov:2019wtd}, while Solution 1 is better in line with the SND result. 
An improved measurement of $B(f_1\to e^+e^-)$ could therefore differentiate between these scenarios. 
In addition, we compare the resulting relevant form factor combination to the L3 dipole fit---see \autoref{sec:L3}---in \autoref{fig:L3Comparison}.
While some tension is expected due to the singly-virtual asymptotic behavior of $\F_{\aT_1}(q_1^2,q_2^2)$, see \autoref{tab:FFLimits}, the resulting curves for Solution 2 start to depart from the L3 band already around $Q=0.5\GeV$, which further disfavors this set of solutions. 

In the extended VMD representation, the dependence on $C_{\aT_2}$ disappears in all observables apart from $B(f_1\to e^+e^-)$ and, potentially, $B(f_1\to \phi\gamma)$. Accordingly, in the fit without the latter, the value of $C_{\aT_2}$ is solely determined by $B(f_1\to e^+e^-)$, and the best-fit value of this branching fraction thus coincides with the input. There is good consistency among the other constraints, as reflected by a reduced $\chi^2$ around unity. In this case, an improved measurement of $B(f_1\to e^+e^-)$ could thus be interpreted as a determination of $C_{\aT_2}$. Once $B(f_1\to \phi\gamma)$ is included, one obtains an additional constraint on $C_{\aT_2}$, which, however, needs to be treated with care. First, the uncertainties on $R^\phi$ have been included in the fit, but in addition there are $\Uthree$ uncertainties that are difficult to quantify. Moreover, the isoscalar contributions have been treated in their minimal variant throughout, but if excited $\omega'$ and $\phi'$ states were included, the dependence on $C_{\aT_2}$ would again change, even disappear in a scenario similar to the extended VMD representation for the isovector contributions. Since the fit including $B(f_1\to \phi\gamma)$ favors a value of $B(f_1\to e^+e^-)$ smaller than SND (for Solution 1 similar in size to the ones for Solution 1 in the minimal VMD case), an improved measurement of $B(f_1\to e^+e^-)$ would also allow one to differentiate between these scenarios. In addition to the worse $\chi^2$, Solution 2 is again disfavored by the comparison to L3, see \autoref{fig:L3Comparison}.    

In contrast, for Solution 1 of both the minimal and the extended VMD fit departures from the L3 dipole only arise around $Q=1\GeV$, which implies agreement with all but the last data point of Ref.~\cite{Achard:2001uu} (centered around $Q=1.8\GeV$, where the curves still agree within uncertainties). In fact, a large part of the pull is a result of the slightly increased value of $C_{\sT}$ from the global fit, while the impact of the asymptotic behavior of $\F_{\aT_1}(-Q^2,0)$ remains small.
Finally, we observe that most extended VMD fits require a substantial $\rho'$ contribution, as reflected by the large values of $\eps_1$ shown in \autoref{tab:solutionsCa1_Ca2_extended}. In fact, for the fit without $B(f_1\to \phi\gamma)$ it even exceeds the coefficient of the $\rho$ contribution, which could be considered an indication that smaller values of $B(f_1\to e^+e^-)$ are preferred. We also implemented a variant of the extended VMD fit in which $\eps_1$ was allowed to float freely, but this did not improve the fit quality, with a resulting $\eps_1$ consistent with the ones imposed via \autoref{eq:eps1VMD}.

\section{Summary and outlook}
\label{sec:summary}

In this paper, we performed a comprehensive analysis of the TFFs of the axial-vector resonance $f_1(1285)$, motivated by its contribution to HLbL scattering in the anomalous magnetic moment of the muon. Our study 
is based on all available constraints from $e^+e^-\to e^+e^- f_1$, $f_1\to 4\pi$, $f_1\to \rho\gamma$, $f_1\to\phi\gamma$, and $f_1\to e^+e^-$, all of which are sensitive to different aspects of the $f_1\to\gamma^*\gamma^*$ transition. Since the amount of data is limited, a completely model-independent determination of all three TFFs is not feasible at present, leading us to consider parameterizations motivated by vector meson dominance. To assess the sensitivity to the chosen parameterization, we constructed two variants, a minimal one that produces non-vanishing results for all TFFs, and an extension that improves the asymptotic behavior by matching to short-distance constraints. In each case this leaves  three coupling constants as free parameters, $C_\sT$, $C_{\aT_1}$, and $C_{\aT_2}$, for the symmetric and the two antisymmetric TFFs, in terms of which the analysis is set up. 

As a first step, we derived master formulae for all processes in terms of these couplings and performed cross checks when analyzing each process in terms of the dominant coupling $C_{\sT}$. This reveals that the decay $f_1\to 4\pi$ does not provide further information on the TFFs, as the mechanism $f_1 \to a_1 \pi \to \rho \pi \pi \to 4\pi$ likely dominates with respect to $f_1\to\rho\rho\to4\pi$, and only the latter can be related to the $f_1$ TFFs.
The process is thus discarded in the subsequent analysis. For the remaining observables we performed detailed uncertainty estimates, including the subleading isoscalar contributions, the properties of the $\rho'$ meson and its spectral function, and the matching to short-distance constraints. In all cases we conclude that the dominant uncertainties are currently of experimental origin. 

Combining all constraints in a global fit, we found that the symmetric coupling $C_\sT$ is by far best determined, with substantial differences in $C_{\aT_1}$ and $C_{\aT_2}$ among the different scenarios, see \autoref{tab:solutionsCa1_Ca2_minimal}, \autoref{tab:solutionsCa1_Ca2_extended}, \autoref{fig:solutionMinimalVMD}, and \autoref{fig:solutionExtendedVMD} for our central results. 
Out of two sets of solutions---Solution 1 with a small negative value of $C_{\aT_1}$, Solution 2 with a sizable positive one---the former is in general preferred by the fit, with Solution 2 further disfavored by the comparison to space-like $e^+e^-\to e^+e^- f_1$ data, see \autoref{fig:L3Comparison}. 
In the case of the minimal VMD representation, we observed some tension between $B(f_1\to e^+e^-)$ and the remaining constraints especially when including $B(f_1\to \phi\gamma)$ in the fit, leading to a preference for a branching fraction below the value 
 recently reported by the SND collaboration.  
In the extended parameterization, the dependence on $C_{\aT_2}$ drops out in all observables but $B(f_1\to e^+e^-)$ and, potentially, $B(f_1\to \phi\gamma)$, but limited information about the isoscalar sector together with necessary $\Uthree$ assumptions render the latter constraint less reliable. 
While the $f_1\to \phi\gamma$ branching fraction seems to prefer a smaller value of $B(f_1\to e^+e^-)$ (similar to the minimal VMD fit), we conclude that the parameter that controls its size, $C_{\aT_2}$, is largely unconstrained at the moment, and would thus profit most from an improved measurement of $B(f_1\to e^+e^-)$. 

In general, new measurements of $B(f_1\to e^+e^-)$---as possible in the context of $e^+e^-\to\text{hadrons}$ energy scans at SND and CMD-3---would be highly beneficial to further constrain the $f_1$ TFFs, given that the resulting constraints are complementary to other observables, in particular, providing sensitivity to doubly-virtual kinematics and the antisymmetric TFFs. 
Apart from a more reliable determination of $C_{\aT_2}$, one could also validate and, if necessary, refine the underlying VMD assumptions. Furthermore, improved measurements of $e^+e^-\to e^+e^- f_1$ would be valuable to further constrain the singly-virtual TFFs---in particular, the asymptotic behavior of $\F_{\aT_1}(q^2,0)$---ideally adding new data points above $1\GeV$ and being analyzed using the full momentum dependence given in \autoref{L3_matching}, to avoid the corresponding limitation in the interpretation of the L3 data. Such analyses are possible at BESIII~\cite{Ablikim:2019hff} and Belle II~\cite{Kou:2018nap}. 
To go beyond VMD parameterizations, the energy dependence in the (dispersively improved) \LN{Breit}--\LN{Wigner} propagators would need to be constrained by data, which would require differential information on $f_1$ decays.  
At the moment, our analysis summarizes the combined information on the $f_1$ TFFs that can be extracted from the available data in terms of simple parameterizations, which we expect to become valuable for forthcoming estimates of the axial-vector contributions to HLbL scattering.  


\acknowledgments 
We thank Gilberto Colangelo, Stefan Leupold, and Peter Stoffer for valuable discussions and comments on the manuscript.  
Financial support by the DFG through the funds provided to the Sino--German Collaborative
Research Center TRR110 ``Symmetries and the Emergence of Structure in QCD''
(DFG Project-ID 196253076 -- TRR 110) and the SNSF (Project No.\ PCEFP2\_181117) is gratefully acknowledged. 


\appendix

\section{Asymptotic behavior including mass effects}  
\label{appx:asymptotics}
In this appendix, we generalize the considerations of Refs.~\cite{Hoferichter:2018dmo,Hoferichter:2018kwz} regarding a double-spectral representation of \LN{BL} scaling to include mass effects that arise from the kinematic variables in the denominator. Starting from
\begin{align}
	\F_2(q_1^2,q_2^2) &= - F_\Ax^\eff \Maxial^3 \frac{\partial}{\partial q_1^2}\int_0^1\du \frac{\phi(u)}{u q_1^2 + (1-u) q_2^2 - u (1-u) \Maxial^2} + \Order(1/q_i^6), \notag \\
	\F_3(q_1^2,q_2^2) &= F_\Ax^\eff \Maxial^3 \frac{\partial}{\partial q_2^2}\int_0^1\du \frac{ \phi(u)}{u q_1^2 + (1-u) q_2^2 - u (1-u) \Maxial^2} + \Order(1/q_i^6),
\end{align}
see \autoref{eq:FFAsymptotic}, we see that the asymptotic behavior of the axial-vector TFFs can still be deduced from the simpler pseudoscalar case, which leads us to study the generic integral
\beq\label{eq:I_asymptotic}
	\tilde{I}(q_1^2,q_2^2,m^2)= \int_0^1\du \frac{ \phi(u)}{u q_1^2 + (1-u) q_2^2 - u (1-u) m^2}, 
\eeq
which, in the case $q_1^2=q_2^2=-Q^2$, evaluates to
\begin{align}
	 \tilde{I}(-Q^2,-Q^2,m^2)&=-\frac{6}{Q^2 \xi}\bigg[1-\frac{2}{\sqrt{\xi(4+\xi)}}\log\frac{\sqrt{4+\xi}+\sqrt{\xi}}{\sqrt{4+\xi}-\sqrt{\xi}}\bigg]\notag\\
 &=-\frac{1}{Q^2}\bigg(1-\frac{\xi}{5}+\frac{3}{70}\xi^2-\frac{\xi^3}{105}+\Order\big(\xi^4\big)\bigg),\qquad \xi=\frac{m^2}{Q^2}. 
\end{align}
Given the large masses of the axial-vector mesons, $m = m_\Ax$, such corrections in $\xi$ may become relevant and \autoref{eq:I_asymptotic} defines a convenient test case to study their impact.

As a first step, we observe that \autoref{eq:I_asymptotic} can still be formulated as a single dispersion relation~\cite{Khodjamirian:1997tk} via the transformation $x=-\frac{1-u}{u}\big(q_2^2- m^2u\big)$,
\begin{align}
\label{eq:I_single_disp}
	\tilde{I}(q_1^2,q_2^2,m^2)&=\frac{1}{\pi}\int_0^\infty \dx \frac{\Im \tilde{I}(x,q_2^2,m^2)}{x-q_1^2},\notag\\
	\Im \tilde{I}(x,y,m^2)&=\frac{3\pi}{m^4}\bigg(\frac{(x-y)^2-m^2(x+y)}{\sqrt{\lambda(x,y,m^2)}}-x+y\bigg),
\end{align}
where $y=q_2^2$ has been assumed to be space-like. Analytic continuation in $q_2^2$ then allows one to rewrite the imaginary part in \autoref{eq:I_single_disp} in terms of another dispersion relation, leading to
\begin{align}
\label{eq:I_double_disp}
 	\tilde{I}(q_1^2,q_2^2,m^2)&=\frac{1}{\pi^2}\int_0^\infty \dx\int_0^\infty \dy\frac{\tilde{\rho}(x,y,m^2)}{(x-q_1^2)(y-q_2^2)}\notag\\
	 &=-\frac{6}{m^2}\bigg[1+\frac{q_1^2}{m^2}\log\bigg(1-\frac{m^2}{q_1^2}\bigg)+\frac{q_2^2}{m^2}\log\bigg(1-\frac{m^2}{q_2^2}\bigg)\bigg]\notag\\
	 &+\frac{q_1^2 q_2^2}{\pi^2}\int_0^\infty \dx\int_0^\infty \dy\frac{\tilde{\rho}(x,y,m^2)}{x(x-q_1^2)y(y-q_2^2)},
\end{align}
with double-spectral function
\beq
	\tilde{\rho}(x,y,m^2)=\frac{3\pi}{m^4}\frac{(x-y)^2-m^2(x+y)}{\sqrt{-\lambda(x,y,m^2)}}\theta\big(-\lambda(x,y,m^2)\big).
\eeq
Restricting the integration in $x$, $y$ should then allow one to isolate the asymptotic contributions while keeping the leading mass corrections. In the subtracted version, the singly-virtual limits become explicit since
\beq
	-\frac{6}{m^2}\bigg[1+\frac{q_i^2}{m^2}\log\bigg(1-\frac{m^2}{q_i^2}\bigg)\bigg]=\int_0^1\du \frac{\phi(u)}{u q_i^2-u(1-u)m^2}.
\eeq

Further, to make connection with the massless limit of \autoref{eq:asym_double_spectral}, which amounts to 
\beq
	\tilde{I}(q_1^2,q_2^2,m^2) \overset{m \to 0}{\to} I(q_1^2,q_2^2) \to -3q_1^2q_2^2\int_{\sm}^\infty\frac{\dx}{(x-q_1^2)^2(x-q_2^2)^2},
\eeq
see \autoref{eq:double_spectral_integral}, we first note that this variant had  been constructed in such a way that the singly-virtual contributions are removed, suggesting a matching in the limit $q_1^2=q_2^2=-Q^2$, in which
\beq
	-3q_1^2q_2^2\int_{\sm}^\infty\frac{\dx}{(x-q_1^2)^2(x-q_2^2)^2}
	= -\frac{1}{Q^2}\bigg[1-3\frac{\sm}{Q^2}+6\bigg(\frac{\sm}{Q^2}\bigg)^2 -10 \bigg(\frac{\sm}{Q^2}\bigg)^3+\Order\bigg(\bigg(\frac{\sm}{Q^2}\bigg)^4\bigg)\bigg].
\eeq
To evaluate \autoref{eq:I_double_disp} in the same limit, we symmetrize the integration to $v=x+y$, $w=x-y$ and introduce a step function $\theta(v-\vm)$. In these variables, the $w$ integration extends between $w_\pm =\pm\sqrt{2m^2 v-m^4}$, which shows that in the massless limit the double-spectral density indeed collapses to a $\delta$ function, see \autoref{eq:double_spectral}. For $q_1^2=q_2^2=-Q^2$, the $w$ integration can be performed analytically, leading to
\begin{align}
	 \tilde{I}(-Q^2,-Q^2,m^2)&=\frac{6}{m^4}\int_{\vm}^\infty \dv
 	\Bigg(\frac{(v+2Q^2)^2-m^2 v}{(v+2Q^2)\sqrt{(v+2Q^2)^2-2m^2 v+m^4}}-1\Bigg)\\
	 &=-12 Q^2\int_{\vm}^\infty \dv\frac{v+Q^2}{(v+2Q^2)^4}+\Order\big(m^2\big)\notag\\
	 &= 
	 -\frac{1}{Q^2}\bigg[1-3\frac{\vm}{4Q^2}+6\bigg(\frac{\vm}{4Q^2}\bigg)^2 -8 \bigg(\frac{\vm}{4Q^2}\bigg)^3+\Order\bigg(\bigg(\frac{\vm}{Q^2}\bigg)^5\bigg)\bigg]
	 +\Order\big(m^2\big).\notag
\end{align}
The first three terms in the expansion thus match upon the identification $\vm=4\sm$.

\section{Phenomenological Lagrangians}
\label{appx:SU3}

In this appendix, we define the Lagrangians used for  
the $\rho \gamma$, $\rho \pi \pi$, and $\rho \omega \pi$ couplings
and discuss the information that can be extracted 
for their  $\rho'$ analogs. In particular, we derive 
estimates for the branching ratios $B(\rho' \to \pi \pi)$ and $B(\rho' \to \omega \pi)$, which are necessary inputs for the construction of the energy-dependent width $\Gammarhoprime^{(\omega \pi,\pi \pi)}(q^2)$ in \autoref{eq:energyDependentWidthRhoPrimeAlternative}.

For the coupling of photons to the vector mesons $\{\rho,\omega,\phi,\rho',\ldots\}$, we use the effective interaction Lagrangian~\cite{Klingl:1996by}
\beq\label{eq:lagrangianVGamma}
\Lagrangian_{\text{V}\gamma} = -\frac{e}{2} F^{\mu \nu} \left(\frac{\rho_{\mu \nu}}{g_{\rho \gamma}} + \frac{\omega_{\mu \nu}}{g_{\omega \gamma}} + \frac{\phi_{\mu \nu}}{g_{\phi \gamma}}+ \frac{\rho'_{\mu \nu}}{g_{\rho' \gamma}} + \ldots\right), 
\eeq
where $F^{\mu \nu} = \partial^{\mu} A^{\nu}-\partial^\nu A^\mu$ is the electromagnetic field strength tensor with the photon field $A^\mu$, $\{\rho^{(\prime)}_{\mu \nu},\rho^{(\prime)}_\mu\}$, $\{\omega_{\mu \nu},\omega_\mu\}$, and $\{\phi_{\mu \nu},\phi_\mu\}$ are the respective vector meson equivalents, and the ellipsis refers to excited isoscalar vector mesons that we omit from the following discussion for simplicity. The couplings of the three ground-state vector mesons are linked via $\SUthree$ symmetry according to $g_{\rho \gamma} : g_{\omega \gamma} : g_{\phi \gamma} = 1 : 3 : 3/\sqrt{2}$~\cite{Klingl:1996by}, with the sign of $g_{\phi\gamma}$ adjusted according to \autoref{vector_SU3}.
In the following, we neglect complex phases associated with actual pole residues (which are known to be tiny~\cite{Hoferichter:2017ftn}), and work with the phase convention $\text{sgn}\,g_{\rho \gamma} = +1$.
From the Lagrangian, the partial decay width of the vector mesons into $e^+ e^-$ follows as
\beq
	\Gamma(\text{V} \to e^+ e^-) = \frac{4\pi \alpha^2}{3 \lvert g_{\text{V} \gamma} \rvert^2} \left(1 + \frac{2m_e^2}{m_\text{V}^2}\right) \sqrt{m_\text{V}^2 - 4m_e^2}.
\eeq
For the $\rho$ meson, one can solve for the coupling and insert the (experimental) value $\Gamma(\rho \to e^+ e^-) = 7.04 \keV$~\cite{Zyla:2020zbs} to find
\beq\label{eq:couplingRhoGamma}
	 g_{\rho \gamma}  = 4.96.
\eeq
This value agrees well with the residue $\lvert g_{\rho \gamma} \rvert=4.9(1)$ extracted from the pion vector form factor~\cite{Hoferichter:2017ftn}, and is also close to the expectation from $\SUthree$ symmetry, $ g_{\rho \gamma}^{\SUthree} = g_{\omega \gamma} /3=5.6$, where $ g_{\omega \gamma} $ can be similarly extracted from $\Gamma(\omega \to e^+ e^-) = B(\omega \to e^+ e^-) \Gamma_\omega = 0.625 \keV$~\cite{Zyla:2020zbs},
\beq\label{eq:couplingOmegaGamma}
	 g_{\omega \gamma}  = 16.7.
\eeq
Furthermore, one can use $\Gamma(\phi \to e^+ e^-) = 1.27 \keV$ to solve for the coupling of the $\phi$ meson, yielding
\beq\label{eq:couplingPhiGamma}
	 g_{\phi \gamma}  = 13.38.
\eeq

For the VMD application considered in this work, we 
also need a formulation in which the coupling of photons to vector mesons is momentum independent, with the respective vector meson considered 
on shell.
Such a coupling can be formally defined via the Lagrangian
\beq\label{eq:lagrangianVGammaVMD}
	\widetilde{\Lagrangian}_{\text{V}\gamma} = e A^\mu \big(\widetilde{g}_{\rho \gamma} \rho_\mu + \widetilde{g}_{\omega \gamma} \omega_\mu + \widetilde{g}_{\phi \gamma} \phi_\mu + \widetilde{g}_{\rho' \gamma} \rho'_\mu + \ldots\big), 
\eeq
where matching the amplitudes resulting from \autoref{eq:lagrangianVGamma} and \autoref{eq:lagrangianVGammaVMD} for on-shell mesons determines
\beq\label{eq:couplingRhoGammaVMD}
	\widetilde{g}_{\text{V}\gamma} = \frac{m_\text{V}^2}{g_{\text{V} \gamma}}.
\eeq
In particular, we carry over the sign convention for the coupling constants $\widetilde{g}_{\text{V}\gamma}$ from $g_{\text{V} \gamma}$ above.

In order to describe the coupling of (uncharged) isovector vector mesons to two pions, we employ the effective interaction Lagrangian~\cite{Klingl:1996by}
\beq\label{eq:lagrangianRhoPiPi}
	\Lagrangian_{\rho^{(\prime)} \pi \pi} = \big(g_{\rho \pi \pi} \rho_\mu + g_{\rho' \pi \pi} \rho'_\mu \big) \big(\pi^+ \partial^\mu \pi^- - \pi^- \partial^\mu \pi^+\big),
\eeq
where $\pi^\pm$ denote the pion fields of definite charge and the coupling to two neutral pions is forbidden by \LN{Bose} symmetry.
We find the decay width for $\rho^{(\prime)} \to \pi^+ \pi^-$ 
\beq\label{eq:decayWidthRhoPiPi}
	\Gamma\left(\rho^{(\prime)} \to \pi^+ \pi^-\right) = \frac{M_{\rho^{(\prime)}} \lvert g_{\rho^{(\prime)} \pi \pi} \rvert^2}{48 \pi} \left(1 - \frac{4\Mpi^2}{M_{\rho^{(\prime)}}^2}\right)^{3/2}.
\eeq
A VMD ansatz for the pion vector form factor,\footnote{Strictly speaking, this form is based on \autoref{eq:lagrangianVGammaVMD}, not \autoref{eq:lagrangianVGamma}, but the difference essentially amounts to a constant that does not affect the relative signs.}
\beq \label{eq:FpiV-VMD}
	F_\pi^\text{V}(q^2) \approx \frac{g_{\rho \pi \pi}}{g_{\rho \gamma}} \frac{\Mrho^2}{\Mrho^2 - q^2-i\sqrt{q^2}\Gammarho(q^2)} + \frac{g_{\rho' \pi \pi}}{g_{\rho' \gamma}} \frac{\Mrhoprime^2}{\Mrhoprime^2 - q^2-i\sqrt{q^2}\Gamma_{\rho'}(q^2)} ,
\eeq
dictates $g_{\rho \pi \pi}$ to have the same sign as $g_{\rho\gamma}$, 
so that under the assumption $\Gamma(\rho \to \pi^+ \pi^-) = \Gammarho$~\cite{Zyla:2020zbs}, we obtain
\beq\label{eq:couplingRhoPiPi}
	 g_{\rho \pi \pi}  = 5.98,
\eeq
again close to the actual residue $\lvert g_{\rho \pi \pi} \rvert=6.01^{+0.04}_{-0.07}$~\cite{GarciaMartin:2011jx}.  

Finally, starting from the anomalous interaction Lagrangian $\Lagrangian_{\text{V} \Phi}^{(3)}$ given in Ref.~\cite{Klingl:1996by}, we write down the Lagrangian that describes the coupling of the neutral isovector vector mesons to $\omega\pi^0$,
\beq\label{eq:lagrangianRhoOmegaPi}
\Lagrangian_{\rho^{(\prime)} \omega \pi} = \frac{\eps^{\mu \nu \alpha \beta}}{2}  (\partial_\beta \pi^0) \Big\{ g_{\rho \omega \pi}\big[(\partial_\mu \rho_\nu) \omega_\alpha + (\partial_\mu \omega_\nu) \rho_\alpha\big]
+ g_{\rho' \omega \pi}\big[(\partial_\mu \rho'_\nu) \omega_\alpha + (\partial_\mu \omega_\nu) \rho'_\alpha\big] \Big\}.
\eeq
 The corresponding $\rho' \to \omega \pi$ decay width is given by
\beq\label{eq:decayWidthRhoOmegaPi}
	\Gamma(\rho' \to \omega \pi) = \frac{\lvert g_{\rho' \omega \pi} \rvert^2}{96\pi M_{\rho'}^3} \lambda\big(M_{\rho'}^2, \Momega^2, \Mpi^2\big)^{3/2}.
\eeq

In the following, we estimate the couplings $\lvert g_{\rho' \gamma} \rvert$, $\lvert g_{\rho' \pi \pi} \rvert$, and $\lvert g_{\rho' \omega \pi} \rvert$, as well as the relevant relative signs in these.
One purpose is the construction of the energy-dependent width $\Gammarhoprime^{(\omega \pi,\pi \pi)}(q^2)$ in \autoref{eq:energyDependentWidthRhoPrimeAlternative}, which---besides the shape of the decay widths $\Gamma(\rho' \to \pi \pi)$ and $\Gamma(\rho' \to \omega \pi)$---requires the branching ratios $B(\rho' \to \pi \pi)$ and $B(\rho' \to \omega \pi)$ as input.
In addition, this allows us to assess the relative importance of $\rho'$
contributions in $f_1\to\gamma^*\gamma^*$ versus those in $f_1\to4\pi$.

Analyses of the pion vector form factor using improved variants of \autoref{eq:FpiV-VMD} suggest a $\rho'$ contribution relative to the dominant $\rho$ therein of an approximate strength~\cite{Fujikawa:2008ma,Roig:2011iv,Schneider:2012ez} 
\beq\label{eq:ratioCouplings1}
	\frac{g_{\rho' \pi \pi}/g_{\rho' \gamma}}{g_{\rho \pi \pi}/g_{\rho \gamma}} \approx -\frac{1}{10}.
\eeq
On the other hand, the $\omega \to \pi \gamma^*$ TFF~\cite{Schneider:2012ez,Achasov:2016zvn}
can be approximated in a VMD picture according to
\beq \label{eq:omegaTFF}
	f_{\omega \pi}(q^2) \approx \frac{g_{\rho \omega \pi}}{g_{\rho \gamma}} \frac{\Mrho^2}{\Mrho^2 - q^2-i\sqrt{q^2}\Gammarho(q^2)} + \frac{g_{\rho' \omega \pi}}{g_{\rho' \gamma}} \frac{\Mrhoprime^2}{\Mrhoprime^2 - q^2-i\sqrt{q^2}\Gamma_{\rho'}(q^2)} .
\eeq
The asymptotic behavior $f_{\omega \pi}(q^2) = \Order(q^{-4})$~\cite{Farrar:1975yb,Vainshtein:1977db,Lepage:1979zb,Lepage:1980fj} implies a superconvergence sum-rule constraint on the couplings of \autoref{eq:omegaTFF} according to
\beq\label{eq:ratioCouplings2}
	\frac{g_{\rho' \omega \pi}/g_{\rho' \gamma}}{g_{\rho \omega \pi}/g_{\rho \gamma}} = -\frac{\Mrho^2}{\Mrhoprime^2}\approx -\frac{1}{4}, 
\eeq
which is consistent with the experimental analysis of Ref.~\cite{Achasov:2016zvn}.
From the experimental width $\Gamma(\omega \to \pi \gamma) = 0.71 \MeV$~\cite{Zyla:2020zbs} and the corresponding formula~\cite{Schneider:2012ez}
\beq
	\Gamma(\omega \to \pi \gamma) = \frac{\alpha(\Momega^2 - \Mpi^2)^3}{24 \Momega^3} \lvert f_{\omega \pi}(0)\rvert^2,
\eeq
we furthermore obtain the normalization $\lvert f_{\omega \pi}(0) \rvert = 2.3 \GeV^{-1}$ and thus
\beq\label{eq:couplingRhoOmegaPi}
	 g_{\rho \omega \pi} \approx 15.4 \GeV^{-1}
\eeq
when combined with \autoref{eq:ratioCouplings2}, choosing a positive sign convention for $f_{\omega \pi}(0)$.  Moreover, from \autoref{eq:ratioCouplings1} and \autoref{eq:ratioCouplings2} one deduces the ratio
\beq
	\frac{g_{\rho' \omega \pi}}{g_{\rho' \pi \pi}} \approx 6.4 \GeV^{-1},
\eeq
so that under the assumption $\Gammarhoprime \approx \Gamma(\rho' \to \pi \pi) + \Gamma(\rho' \to \omega \pi)$---neglecting another significant contribution from $\rho' \to a_1 \pi$ ($a_1 \to 3\pi$)\footnote{References~\cite{Kozyrev:2019ial,Akhmetshin:1998df} show that $e^+e^-\to a_1\pi$, the second-largest subchannel of $e^+e^-\to 4\pi$ beyond $e^+e^-\to \omega\pi$, is already
important at the $\rho'$. Adding the $a_1\pi$ channel will decrease the $g_{\rho'\pi\pi}$ and $g_{\rho'\omega\pi}$ couplings in parallel, with the ratio of
branching fractions $B(\rho' \to \pi \pi)/B(\rho' \to \omega \pi)$ kept fixed, but they then will not add up to $100\perc$  anymore.}---one can use  \autoref{eq:decayWidthRhoPiPi} and \autoref{eq:decayWidthRhoOmegaPi} to obtain
\begin{align}
	\lvert g_{\rho' \pi \pi}\rvert &\approx 1.60, & \lvert g_{\rho' \omega \pi}\rvert &\approx 10.3 \GeV^{-1}.
\end{align}
The branching ratios then become
\begin{align}\label{eq:rhoPrimeBranchingRatios}
	B(\rho' \to \pi \pi) &\approx 6 \perc, & B(\rho' \to \omega \pi) &\approx 94 \perc,
\end{align}
and, for completeness, the $\rho' \gamma$ coupling is estimated as
\beq
	\lvert g_{\rho' \gamma}\rvert \approx 13.3.
\eeq
The estimate \autoref{eq:rhoPrimeBranchingRatios} agrees with the expectation that the $\rho'$ should be largely inelastic, and the resulting spectral function in \autoref{eq:energyDependentWidthRhoPrimeAlternative} thus essentially defines an
estimate of the $4\pi$ channel dominated by $\omega\pi$. We stress that these considerations should only be considered rough estimates, the main point being to define another plausible variant that allows us to assess the sensitivity of our results to the assumptions made for the $\rho'$ spectral function. 
Finally, for our analysis of $f_1\to4\pi$ including effects of the $\rho'$, we require the ratio of coupling constants
\beq
\label{def_kappa}
\frac{g_{\rho' \pi \pi}\times g_{\rho' \gamma}}{g_{\rho \pi \pi}\times g_{\rho \gamma}} \approx -0.7.
\eeq

\section{Comparison to the literature}
\label{appx:literature}

In this appendix, we briefly compare the basis of \LN{Lorentz} structures and TFFs as well as the parameterization of the latter for the $f_1$ used in this work to the previous analysis of Refs.~\cite{Rudenko:2017bel,Milstein:2019yvz}.
Since the TFFs are not (anti-)symmetrized in Ref.~\cite{Rudenko:2017bel}, we use the basis introduced in \autoref{sec:lorentz_decomposition} for our comparison, that is, in particular, the structures from \autoref{eq:structures}.
When using \autoref{eq:amplitude} to translate the amplitude $\M(f_1 \to \gamma^* \gamma^*)$ from Ref.~\cite{Rudenko:2017bel} to the tensor matrix element given in \autoref{eq:MDecomposition}, we find the structures to be related by
\begin{align}
	T_{1\,\Rud}^{\mu \nu \alpha}(q_1,q_2) &= -T_1^{\mu \nu \alpha}(q_1,q_2), \notag \\
	T_{2\,\Rud\,}^{\mu \nu \alpha}(q_1,q_2) &= -T_3^{\mu \nu \alpha}(q_1,q_2), \notag \\
	T_{3\,\Rud}^{\mu \nu \alpha}(q_1,q_2) &= T_2^{\mu \nu \alpha}(q_1,q_2),
\end{align}
and the TFFs to be linked via
\begin{align}
	\F_{1}^{\Rud}(q_1^2,q_2^2) &= -4\pi \F_1(q_1^2,q_2^2), \notag \\
	\F_{2}^{\Rud}(q_1^2,q_2^2) &= -4\pi \F_3(q_1^2,q_2^2), \notag \\
	\F_{3}^{\Rud}(q_1^2,q_2^2) &= 4\pi \F_2(q_1^2,q_2^2).
\end{align}
While the structures are thus identical to ours except for two global signs and a permutation, the additional factor of $4\pi$ in the TFFs appears due to the fact that the fine-structure constant $\alpha$ is used
in the definition of their matrix element instead of the factor $e^2$.  
The symmetry properties of the TFFs in their basis are given by $\F_{1}^{\Rud}(q_2^2,q_1^2) = -\F_{1}^{\Rud}(q_1^2,q_2^2)$ and $\F_{2}^{\Rud}(q_2^2,q_1^2) = \F_{3}^{\Rud}(q_1^2,q_2^2)$, where an (anti-)symmetrization similar to \autoref{eq:FFNewBasis} would of course be straightforward.
Moreover, the two-photon decay width, \autoref{eq:twoPhotonDecayWidth}, becomes
\beq
	\widetilde{\Gamma}_{\gamma \gamma}^{\Rud} = \frac{\alpha^2}{192\pi} \Maxial \lvert \F_{2}^{\Rud}(0,0) \rvert^2 = \frac{\alpha^2}{192\pi} \Maxial \lvert \F_{3}^{\Rud}(0,0) \rvert^2.
\eeq

The strategy that is used in Ref.~\cite{Rudenko:2017bel} to determine the explicit parameterization of the TFFs in accord with a VMD model is, in fact, quite different from our approach---the model does not correspond to a strict VMD ansatz.
Instead of proposing a VMD-like parameterization for the form factors $\F_{i}^{\Rud}(q_1^2,q_2^2)$ as we did in \autoref{eq:VMDParametrization},  three form factors $h_i(q_1^2,q_2^2)$ are introduced, based on which an amplitude $\M(f_1 \to {\rho^0}^* {\rho^0}^*)$ is constructed by replacing $\F_{i}^{\Rud}(q_1^2,q_2^2) \to h_i(q_1^2,q_2^2)$ in $\M(f_1 \to \gamma^* \gamma^*)$; analogously, two complex coupling constants $g_1$ and $g_2$ are introduced to construct an amplitude $\M(f_1 \to \rho \gamma)$.
We disagree that such complex couplings are allowed since the resulting imaginary parts need to reflect the actual analytic structure of the amplitude. Moreover, the explicit form of the  
 $h_i(q_1^2,q_2^2)$, to account for an off-shell dependence of the $\rho$ mesons, introduces unphysical kinematic singularities. 
 
By employing a $\rho\gamma$ coupling similar to the one we introduced by means of \autoref{eq:lagrangianVGamma}, the form factors $\F_{i}^{\Rud}(q_1^2,q_2^2)$ and $h_i(q_1^2,q_2^2)$ are then related to each other in Ref.~\cite{Rudenko:2017bel}, where the latter can further be linked to the coupling constants $g_1$ and $g_2$. Using the $\rho\gamma$ coupling in the convention of the present work, the form factors are found to be
\begin{align}\label{eq:FFRudenko}
	\F_{1}^{\Rud}(q_1^2,q_2^2) &= \frac{e g_1 (\Mrho^2 - \iu \Mrho \Gammarho)(q_2^2 - q_1^2)}{g_{\rho \gamma} (q_1^2 - \Mrho^2 + \iu \Mrho \Gammarho)(q_2^2 - \Mrho^2 + \iu \Mrho \Gammarho)}, \notag \\
	\F_{2/3}^{\Rud}(q_1^2,q_2^2) &= -\frac{e g_2 \Mrho^2 (\Mrho^2 - \iu \Mrho \Gammarho)}{g_{\rho \gamma} (q_1^2 - \Mrho^2 + \iu \Mrho \Gammarho)(q_2^2 - \Mrho^2 + \iu \Mrho \Gammarho)},
\end{align}
the width $\Gammarho$ being the (energy-independent) total width of the $\rho$ meson, as opposed to our energy-dependent parameterization of \autoref{eq:energyDependentWidthRho} and \autoref{eq:energyDependentWidthRho_barrier_factors}.
Moreover, the magnitude of the couplings $g_1$ and $g_2$ is determined in Ref.~\cite{Rudenko:2017bel} by making use of experimental data on $f_1 \to \rho \gamma$, see \autoref{sec:vmd}; the relative phase between these coupling constants remains undetermined, despite using, in addition, input from $f_1 \to 4\pi$.

By rewriting \autoref{eq:FFRudenko} as
\beq
	\F_{1}^{\Rud}(q_1^2,q_2^2) = \frac{e g_1 (\Mrho^2 - \iu \Mrho \Gammarho)}{g_{\rho \gamma} (q_1^2 - \Mrho^2 + \iu \Mrho \Gammarho)} - \frac{e g_1 (\Mrho^2 - \iu \Mrho \Gammarho)}{g_{\rho \gamma} (q_2^2 - \Mrho^2 + \iu \Mrho \Gammarho)},
\eeq
one observes that $\F_{1}^{\Rud}(q_1^2,q_2^2)$ does not correspond to 
a VMD ansatz in the strict sense, but rather arises from two diagrams, each being composed of one direct photon coupling and one VMD-like $\rho$ coupling. As we argued in \autoref{sec:vmd}, an actual VMD representation of the antisymmetric TFFs requires the introduction of a second multiplet. 
Further, \autoref{eq:FFRudenko} shows that the second and third TFFs are parameterized symmetrically, \Lat{i.e.}, the antisymmetric part is neglected.
In either case, we believe that the $f_1\to 4\pi$ decay does not allow one to extract information on the $f_1$ TFFs, for the reasons described in \autoref{sec:4pi} and \autoref{appx:f14pi}.

Finally, we would like so stress that, in addition to using complex couplings, energy-independent widths are problematic when inserted into the $f_1\to e^+e^-$ loop integral, leading to imaginary parts below the respective thresholds and thus distorting the analytic structure. Given, in addition, the appearance of kinematic singularities and different high-energy behavior, it is difficult to compare our phenomenological results to the ones of Refs.~\cite{Rudenko:2017bel,Milstein:2019yvz}. 

\section{$\boldsymbol{f_1\to a_1 \pi \to \rho \pi \pi \to 4\pi}$}
\label{appx:f14pi}

In order to investigate whether the intermediate state $a_1(\to \rho \pi)\pi$ can account for the discrepancy in the branching ratio of $f_1 \to 4\pi$ found in \autoref{sec:4pi}, we use the effective interaction Lagrangians
\begin{align}
	\Lagrangian_{f_1 a_1 \pi} &= \frac{g_{f_1 a_1 \pi}}{2}\eps^{\mu \nu \alpha \beta}  (\partial_\beta \pi^\mp) \big[(\partial_\mu {a_1^\pm}_\nu) {f_1}_\alpha + (\partial_\mu {f_1}_\nu) {a_1^\pm}_\alpha\big],\notag \\
	\Lagrangian_{a_1 \rho \pi} &= g_{a_1 \rho \pi} \big[{a_1^-}_\mu \rho^\mu \pi^+ - {a_1^+}_\mu \rho^\mu \pi^-\big], \label{eq:a1Lagrangians}
\end{align}
where $\Lagrangian_{f_1 a_1 \pi}$ is constructed in analogy to \autoref{eq:lagrangianRhoOmegaPi} and $\Lagrangian_{a_1 \rho \pi}$ represents the simplest Lagrangian possible, the relative sign originating from isospin symmetry.
Before constructing an amplitude for $f_1\to a_1 \pi \to \rho \pi \pi \to 4\pi$, we will in the following estimate the couplings $g_{f_1 a_1 \pi}$ and (the magnitude of) $g_{a_1 \rho \pi}$.

For the estimate of $g_{f_1 a_1 \pi}$, we start from the observation that the \LN{Wess}--\LN{Zumino}--\LN{Witten} anomaly~\cite{Wess:1971yu,Witten:1983tw,Zyla:2020zbs}
\beq\label{eq:WZW}
	F_{\pi^0 \gamma^* \gamma^*}(0,0) = \frac{1}{4 \pi^2 F_\pi}=  0.2745(3)\GeV^{-1}
\eeq
is largely saturated by the VMD ansatz
\begin{align}\label{eq:anomaly_VMD}
	F_{\pi^0 \gamma^* \gamma^*}(0,0) &\approx \frac{g_{\rho \omega \pi}}{g_{\rho \gamma} g_{\omega \gamma}} \left[\frac{\Mrho^2 \Momega^2}{(\Mrho^2 - q_1^2 - i \sqrt{q_1^2} \Gammarho(q_1^2))(\Momega^2 - q_2^2)} + (q_1 \leftrightarrow q_2)\right]\Bigg\rvert_{q_1^2 = 0 = q_2^2} \notag \\
	&= \frac{2 g_{\rho \omega \pi}}{g_{\rho \gamma} g_{\omega \gamma}} = \frac{2 g_{\rho \omega \pi} F_\rho F_\omega}{\Mrho \Momega}\approx 0.37\GeV^{-1},
\end{align}
where we used \autoref{eq:couplingRhoGamma}, \autoref{eq:couplingOmegaGamma}, and \autoref{eq:couplingRhoOmegaPi}. The decay constants  of the $\rho$ and $\omega$ meson,
\beq
	\braket{0 | j^\mu_\text{em}(0) | V(p,\lambda_V)} = F_V M_V \epsilon_\mu(p), \quad V = \rho, \omega,
\eeq
are related to our previous notation by $g_{V\gamma}=M_V/F_V$. This rough agreement suggests that an estimate of the axial-vector analogs can be obtained in a similar manner, leading to the axial-vector-meson-dominance ansatz
\beq
	F_{\pi^0 \gamma^* \gamma^*}(0,0) \approx \frac{2 g_{f_1 a_1 \pi} F_{a_1} F_{f_1}}{\Ma \Mf},
\eeq
with the corresponding decay constants defined by
\beq
	\braket{0 | \bar{q}(0) \gamma_\mu \gamma_5 \mathcal{Q}q(0) | A(p,\lambda_A)} = F_A m_A \epsilon_\mu(p), \quad A = a_1, f_1.
\eeq
Comparing the two parameterizations results in
\beq\label{eq:couplingf1a1Pi}
	\frac{g_{f_1 a_1 \pi}}{g_{\rho \omega \pi}} \approx \frac{F_\rho F_\omega}{\Mrho \Momega} \frac{\Ma \Mf}{F_{a_1} F_{f_1}} \approx 1.3,
\eeq
where we used $F_{a_1}=168(7)\MeV$, $F_{f_1}=87(7)\MeV$~\cite{Yang:2007zt,Hoferichter:2020lap}.

An estimate of $\lvert g_{a_1 \rho \pi}\rvert$ is  obtained by calculating the decay width of $a_1 \to \rho \pi$ and matching to the experimental width under the assumption $\Gamma(a_1 \to \rho \pi) = \Gamma_{a_1}$, taking into account that $\Gamma(a_1 \to \rho \pi) = \Gamma(a_1^{\pm} \to \rho^{\pm} \pi^0) + \Gamma(a_1^{\pm} \to \rho^0 \pi^{\pm}) = 2 \Gamma(a_1^{\pm} \to \rho^0 \pi^{\pm})$ for the charged channel. We find\footnote{Note that, in addition to the expected $S$-wave phase space, the Lagrangian $\Lagrangian_{f_1 a_1 \pi}$ also produces a numerically small $P$-wave contribution proportional to $\lvert \mathbf{p}_\rho \rvert^3$, which---strictly speaking---would only vanish when performing a partial-wave projection. Given the uncertainties inherent in the $f_1\to a_1\pi\to \rho \pi \pi \to 4\pi$ estimate presented here, especially in view of the width and spectral shape of the $a_1$, a more refined treatment is not warranted, and we simply remove these terms in \autoref{eq:decaywidtha1}.}
\begin{align}\label{eq:decaywidtha1}
	\Gamma(a_1 \to \rho \pi) = \frac{\lvert g_{a_1 \rho \pi}\rvert^2}{8 \pi} \frac{\lvert \mathbf{p}_\rho \rvert}{\Ma^2} \left( 1 + \frac{\lvert\mathbf{p}_\rho\rvert^2}{3\Mrho^2} \right) \to \frac{\lvert g_{a_1 \rho \pi}\rvert^2}{8 \pi} \frac{\lvert \mathbf{p}_\rho \rvert}{\Ma^2},
\end{align}
where $\lvert\mathbf{p}_\rho\rvert = \sqrt{\lambda(\Ma^2,\Mrho^2,\Mpi^2)}/(2\Ma)$ is the magnitude of the three-momentum in the center-of-mass frame, yielding
\beq\label{eq:couplinga1RhoPi}
	\lvert g_{a_1 \rho \pi}\rvert = ( 3.7 \ldots 5.7 ) \GeV,
\eeq
where the given variation is due to the width of the $a_1$.

\begin{figure}[t]
	\centering
	\includegraphics{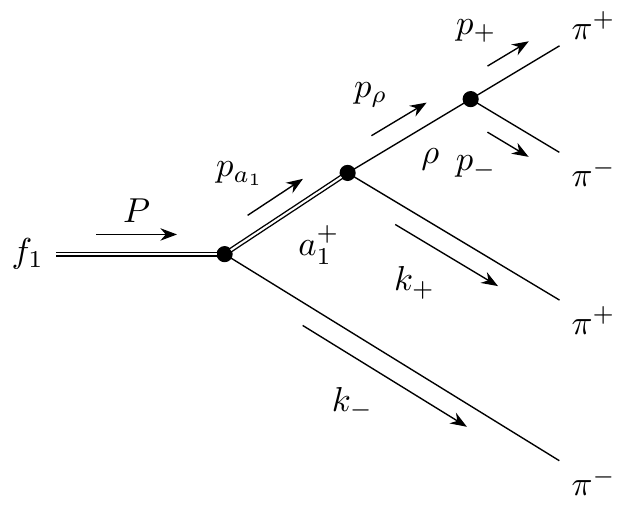}
	\qquad \qquad
	\includegraphics{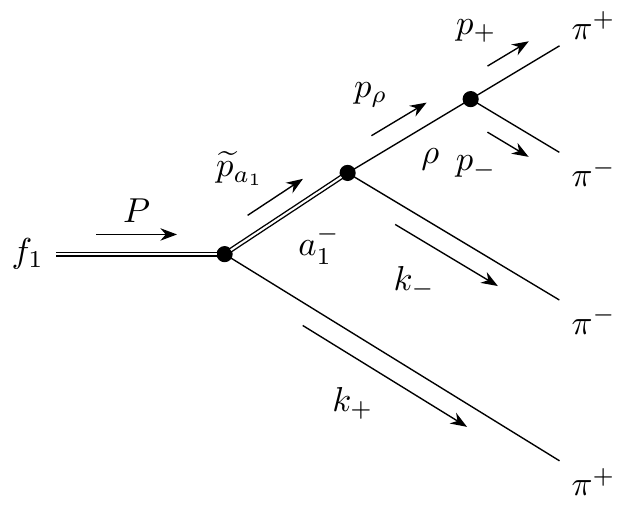}
	\caption{\LN{Feynman} diagrams for $f_1 \to \pi^+ \pi^- \pi^+ \pi^-$ via $a_1 \pi$. Since the two $\pi^+$ and $\pi^-$ are respectively indistinguishable, there exist eight diagrams in total, that is four diagrams with $a_1^+$ (\textit{left}) and four diagrams with $a_1^-$ (\textit{right}), which are obtained by permuting the momenta appropriately.}
	\label{fig:fapi}
\end{figure} 

The amplitude for $f_1 \to a_1 \pi \to \rho \pi \pi \to 4 \pi$ can be constructed with \autoref{eq:a1Lagrangians} and \autoref{eq:lagrangianRhoPiPi}, where eight diagrams have to be taken into account, see \autoref{fig:fapi}, leading to
\begin{align}
	\M_{a_1 \pi}(f_1 \to \pi^+ \pi^- \pi^+ \pi^-) &= \frac{g_{f_1 a_1 \pi} g_{a_1 \rho \pi} g_{\rho \pi \pi}}{\big(p_{a_1}^2 - \Ma^2 + i \sqrt{p_{a_1}^2} \Gamma_{a_1}(p_{a_1}^2)\big)\big(p_\rho^2 - \Mrho^2 + i \sqrt{p_\rho^2} \Gammarho(p_\rho^2)\big)} \notag \\
	& \hspace{-1.2cm}\times\eps^\mu(P) \eps_{\mu \nu \alpha \beta} \left[ 2 k_-^\nu p_-^\alpha p_+^\beta + k_-^\nu k_+^\alpha (p_+ - p_-)^\beta \right] + (p_- \leftrightarrow k_-) + (p_+ \leftrightarrow k_+) \notag \\
	& \hspace{-1.2cm} + (p_+ \leftrightarrow k_+, p_- \leftrightarrow k_-) - (k_+ \leftrightarrow k_-) - (k_+ \leftrightarrow k_-, p_- \leftrightarrow k_-) \notag \\
	& \hspace{-1.2cm} - (k_+ \leftrightarrow k_-, p_+ \leftrightarrow k_+) - (k_+ \leftrightarrow k_-, p_+ \leftrightarrow k_+, p_- \leftrightarrow k_-),
\end{align}
with the momenta defined as in \autoref{fig:fapi} and the pions on shell, $p_\pm^2 = \Mpi^2 = k_\pm^2$.
For the energy-dependent width of the $a_1$ meson, we choose an ansatz based on \autoref{eq:decaywidtha1},
\begin{align}
	\Gamma_{a_1}(q^2) &= \theta\big(q^2 - (\Mrho + \Mpi)^2\big) \frac{\gamma_{a_1 \to \rho \pi}(q^2)}{\gamma_{a_1 \to \rho \pi}(\Ma^2)} \Gamma_{a_1}, & \gamma_{a_1 \to \rho \pi}(q^2) &= \frac{\sqrt{\lambda(q^2,\Mrho^2,\Mpi^2)}}{(q^2)^{3/2}},
\end{align}
and the energy-dependent width $\Gammarho(q^2)$ is as specified in \autoref{eq:energyDependentWidthRho_barrier_factors}.
The decay width and thus branching ratio can then be calculated via the four-body phase-space integration of
\beq
	\mathrm{d}\Gamma_{a_1 \pi}(f_1 \to \pi^+\pi^-\pi^+\pi^-) = \frac{1}{2\Mf}\left\lvert\M_{a_1 \pi}(f_1 \to \pi^+\pi^-\pi^+\pi^-)\right\rvert^2 \mathrm{d}\Phi_4(P,p_+,p_-,k_+,k_-).
\eeq
Although we could proceed in complete analogy to \autoref{sec:4pi}, it is instructive to write the differential four-body phase space differently from \autoref{eq:phase_space_recursion}, namely in the form~\cite{Zyla:2020zbs}
\beq
	\mathrm{d}\Phi_4(P,p_+,p_-,k_+,k_-) = \mathrm{d}\Phi_2(p_\rho;p_+,p_-) \mathrm{d}\Phi_2(p_{a_1};p_\rho,k_+) \mathrm{d}\Phi_2(P;p_{a_1},k_-) \frac{\mathrm{d}p_{a_1}^2}{2\pi} \frac{\mathrm{d}p_\rho^2}{2\pi},
\eeq
where $\mathrm{d}\Phi_2(P;p_{a_1},k_-)$, $\mathrm{d}\Phi_2(p_{a_1};p_\rho,k_+)$, and $\mathrm{d}\Phi_2(p_\rho;p_+,p_-)$ are the respective two-body phase spaces of the subsystems $\{a_1(p_{a_1})\pi^-(k_-)\}$, $\{\rho(p_\rho)\pi^+(k_+)\}$, and $\{\pi^+(p_+)\pi^-(p_-)\}$. 
As argued in \autoref{sec:4pi}, each two-body phase space can be evaluated in the corresponding center-of-mass frame and we have to perform an explicit \LN{Lorentz} transformation from the center-of-mass frames of $\{a_1(p_{a_1})\pi^-(k_-)\}$ and $\{\pi^+(p_+)\pi^-(p_-)\}$ into the one of $\{\rho(p_\rho)\pi^+(k_+)\}$ in order to evaluate all the scalar products appearing in $\lvert\M_{a_1 \pi}(f_1 \to \pi^+\pi^-\pi^+\pi^-)\rvert^2$.\footnote{As in \autoref{sec:4pi}, the decay rate involves an additional symmetry factor of $S=1/(2!)^2$ because of the two pairs of indistinguishable particles in the final state.}
We perform the phase space integration numerically with the \Lat{Cuhre} algorithm from the \Lat{Cuba} library~\cite{Hahn:2004fe}, obtaining
\beq
	\Gamma_{a_1 \pi}(f_1 \to \pi^+\pi^-\pi^+\pi^-) = \lvert g_{f_1 a_1 \pi}\rvert^2 \lvert g_{a_1 \rho \pi}\rvert^2 \lvert g_{\rho \pi \pi}\rvert^2 \times (3.27 \ldots 2.46)\times 10^{-9}\GeV.
\eeq
Combining the above result with $\lvert g_{f_1 a_1 \pi}\rvert \approx 1.3 \times 15.4 \GeV^{-1}$, $\lvert g_{a_1 \rho \pi} \rvert = (3.7 \ldots 5.7) \GeV$, and $\lvert g_{\rho \pi \pi} \rvert = 5.98$, \autoref{eq:couplingf1a1Pi}, \autoref{eq:couplingRhoOmegaPi}, \autoref{eq:couplinga1RhoPi}, and \autoref{eq:couplingRhoPiPi}, we find the branching ratio to be given by
\beq
\label{eq:estimate1}
	B_{a_1 \pi}(f_1 \to \pi^+ \pi^- \pi^+ \pi^-) \approx (2.8 \ldots 5.0) \perc,
\eeq
in fair agreement with the experimental value $B(f_1 \to \pi^+ \pi^- \pi^+ \pi^-) = 10.9(6) \perc$~\cite{Zyla:2020zbs}.
We also considered the variant of this estimate obtained when further approximating the decay $f_1\to a_1 \pi \to \rho \pi \pi \to 4\pi$ by $f_1\to a_1 \pi \to \rho \pi \pi$, assuming that the $\rho$ decays into two charged pions only:
\beq
	\Gamma_{a_1 \pi}(f_1 \to \rho \pi \pi) = \lvert g_{f_1 a_1 \pi}\rvert^2 \lvert g_{a_1 \rho \pi}\rvert^2 \times (2.40 \ldots 2.06)\times 10^{-7}\GeV,
\eeq
and
\beq
\label{eq:estimate2}
	B_{a_1 \pi}(f_1 \to \rho \pi \pi) \approx (5.8\ldots 11.8) \perc,
\eeq
leading to a result closer to the experimental branching fraction, which indicates that $\rho$ dominance in this decay mode is again subject to sizable corrections.
In both estimates, given that the VMD saturation of the anomaly, \autoref{eq:anomaly_VMD}, actually overpredicts
the expected value, \autoref{eq:WZW}, a somewhat smaller value of $\lvert g_{f_1 a_1 \pi}\rvert$ may be favored. 

We stress that the estimates presented here are merely supposed to give an indication for why the VMD description of $f_1 \to 4\pi$ in \autoref{sec:4pi} is in serious disagreement with the experimental branching ratio, \Lat{i.e.}, we do not claim to have a reliable prediction for $B_{a_1 \pi}(f_1 \to \pi^+ \pi^- \pi^+ \pi^-)$, as, in particular, the uncertainty in assuming an axial-vector saturation of the anomaly is difficult to quantify.
Still, the arguments leading to \autoref{eq:estimate1} and \autoref{eq:estimate2} should make plausible that  
the intermediate state $a_1 \pi$ can indeed cover the experimental branching ratio to a large degree, thus rendering the $f_1 \to 4\pi$ decay unsuitable for extracting information on the $f_1$ TFFs.

\section{Constants and parameters}
\label{appx:constants}

\begin{table}[t]
	\centering
	\begin{tabular}{ l  c  r r }
	\toprule
		Quantity & Variable & Value & Reference\\
		\midrule
		Mass pion & $\Mpi$ & $139.57 \MeV$ & \cite{Zyla:2020zbs}\\
		Mass $a_1(1260)$ & $\Ma$ & $1230(40) \MeV$ & \cite{Zyla:2020zbs}\\
		Mass $f_1(1285)$ & $\Mf$ & $1281.9(5) \MeV$ & \cite{Zyla:2020zbs}\\
		Mass $f_1(1420)$ & $\Mfprime$ & $1426.3(9) \MeV$ & \cite{Zyla:2020zbs}\\
		Mass $\omega(782)$ & $\Momega$ & $782.65(12) \MeV$ & \cite{Zyla:2020zbs}\\
		Mass $\phi(1020)$ & $\Mphi$ & $1019.461(16) \MeV$ & \cite{Zyla:2020zbs}\\
		Mass $\rho(770)$ (charged) & $\Mrho$ & $775.11(34) \MeV$ & \cite{Zyla:2020zbs}\\
		Mass $\rho(1450)$ & $\Mrhoprime$ & $1465(25) \MeV$ & \cite{Zyla:2020zbs}\\
		Total width $a_1(1260)$ & $\Gamma_{a_1}$ & $(250 \ldots 600) \MeV$ & \cite{Zyla:2020zbs}\\
		Total width $f_1(1285)$ & $\Gamma_{f_1}$ & $22.7(1.1) \MeV$ & \cite{Zyla:2020zbs}\\
		Total width $f_1(1420)$ & $\Gamma_{f_1'}$ & $54.5(2.6)\MeV$& \cite{Zyla:2020zbs}\\
		Total width $\rho(770)$ (charged) & $\Gammarho$ & $149.1(8) \MeV$& \cite{Zyla:2020zbs}\\
		Total width $\rho(1450)$ & $\Gammarhoprime$ & $400(60) \MeV$& \cite{Zyla:2020zbs}\\
		\midrule
		Mass $\rho(770)$ (charged) & $\Mrho$ & $774.9(6) \MeV$&\cite{Fujikawa:2008ma}\\
		Mass $\rho(1450)$ (charged) & $\Mrhoprime$ & $1428(30) \MeV$&\cite{Fujikawa:2008ma}\\
		Total width $\rho(770)$ (charged) & $\Gammarho$ & $148.6(1.8) \MeV$&\cite{Fujikawa:2008ma}\\
		Total width $\rho(1450)$ (charged) & $\Gammarhoprime$ & $413(58) \MeV$&\cite{Fujikawa:2008ma}\\
\midrule
Mass $\rho(770)$ (neutral) & $\Mrho$ & $775.02(35) \MeV$&\cite{Lees:2012cj}\\
		Mass $\rho(1450)$ (neutral) & $\Mrhoprime$ & $1493(15) \MeV$&\cite{Lees:2012cj}\\
		Total width $\rho(770)$ (neutral) & $\Gammarho$ & $149.59(67) \MeV$&\cite{Lees:2012cj}\\
		Total width $\rho(1450)$ (neutral) & $\Gammarhoprime$ & $427(31) \MeV$&\cite{Lees:2012cj}\\
		\bottomrule
	\end{tabular}
	\caption{Selected masses and decay widths from Ref.~\cite{Zyla:2020zbs}, in comparison to the $\rho(770)$ and $\rho(1450)$ parameters from Refs.~\cite{Fujikawa:2008ma,Lees:2012cj}.}
	\label{tab:constants}
\end{table}

In this appendix, we collect the particle masses and decay widths used throughout this work, see \autoref{tab:constants}. Isospin-breaking effects can be safely neglected, in particular, the pion mass is identified with the mass of the charged pion.
Some comments are in order, however, regarding the treatment of broad resonances, most notably the $\rho(1450)$ and, to a lesser extent, the $\rho(770)$. Especially for the former, the quoted masses and widths are strongly reaction dependent, as referring to \LN{Breit}--\LN{Wigner} parameters, not to the model-independent pole parameters. We thus need to make sure that we use determinations that apply to the channels that we consider here. Since the main application concerns the description of multi-pion decay channels in the VMD propagators, both for the $\rho(770)$ and the $\rho(1450)$, it appears most natural to consider reactions that provide access to both resonances, which points towards $\tau\to\pi\pi\nu_\tau$ from Ref.~\cite{Fujikawa:2008ma} and $e^+e^-\to \pi\pi$ from Ref.~\cite{Lees:2012cj}. In particular, this allows us to see if there are relevant systematic differences between the charged and neutral channel. For the $\rho(770)$, the mass parameter agrees well between all channels, but while there is also good agreement between Refs.~\cite{Fujikawa:2008ma,Lees:2012cj} for the width, the compilation from Ref.~\cite{Zyla:2020zbs} quotes a significantly lower value for the neutral channel. Accordingly, we will use its $\rho(770)$ parameters from the charged channel in our analysis. Regarding the $\rho(1450)$, the mass from Ref.~\cite{Zyla:2020zbs} lies half-way  between Refs.~\cite{Fujikawa:2008ma,Lees:2012cj}, with a width that agrees well with both channels within uncertainties. We will therefore take over the recommended parameters for the $\rho(1450)$.


\end{document}